\documentclass[11pt]{amsart}
\usepackage{hyperref}
\usepackage{subfigure,caption2}

\usepackage{amssymb}
\usepackage[dvips]{graphicx}
\usepackage[rflt]{floatflt}
\numberwithin{equation}{section}
\newcommand{\smallpagebreak}{{\par\vspace{2 mm}\noindent}}

\newcommand{\demo}{\par\noindent{\it Proof.\/} \ }

\newcommand{\dsize}{\textstyle}
\newcommand{\D}{\displaystyle}

\newcommand{\R}{{\mathbb R}}

\newcommand{\Z}{{\mathbb Z}}
\newcommand{\N}{{\mathbb N}}
\newcommand{\C}{{\mathbb C}}

\newcommand{\re}{{\rm Re}\,}
\newcommand{\im}{{\rm Im}\,}
\newcommand{\ind}{\text{ind}\,}
\newcommand{\res}{{\rm res}\, }
\newcommand{\Const}{{\rm Const}}
\theoremstyle{plain}
\newtheorem{Th}{Theorem}[section]
\newtheorem{Le}{Lemma}[section]
\newtheorem{Pro}{Proposition}[section]
\newtheorem{Cor}{Corollary}[section]
\theoremstyle{definition}

\newtheorem{Def}{Definition}[section]
\title{Geometric tools of the adiabatic complex WKB method}
\author{Alexander Fedotov} \author{Fr{\'e}d{\'e}ric Klopp}
\address[Alexander Fedotov]{Departement of Mathematical Physics, St
  Petersburg State University, 1, Ulia\-novskaja, 198904 St
  Petersburg-Petrodvorets, Russia}
\email{ \href
{mailto:fedotov@mph.phys.spbu.ru}{fedotov@mph.phys.spbu.ru}}
\address[Fr{\'e}d{\'e}ric Klopp]{D{\'e}partement de Math{\'e}matique, Institut
  Galil{\'e}e, U.R.A 7539 C.N.R.S, Universit{\'e} de Paris-Nord, Avenue J.-B.
  Cl{\'e}ment, F-93430 Villetaneuse, France}
\email{\href
{mailto:klopp@math.univ-paris13.fr}{klopp@math.univ-paris13.fr}}
\thanks{F.K.'s research was partially supported by the program RIAC
  160 at Universit{\'e} Paris 13 and by the FNS 2000 ``Programme Jeunes
  Chercheurs''. Both authors also acknowledge support from the
  Mittag-Leffler Institute where this work was essentially completed.}
\begin{document}
\begin{abstract}
  The paper is devoted to the description of the main geometric and
  analytic tools of a complex WKB method for adiabatic problem. We
  illustrate their use by numerous examples.
  \vskip.5cm
  \par\noindent   \textsc{R{\'e}sum{\'e}.}
  L'article est consacr{\'e} {\`a} la description des principaux outils
  g{\'e}om{\'e}triques et analytiques d'une m{\'e}thode WKB complexe pour des
  probl{\`e}mes adiabatiques. Nous illustrons leur utilisation par de
  nombreux exemples.
\end{abstract}
\setcounter{section}{-1}
\maketitle
\section{Introduction}
\label{sec:intro}
\noindent In this paper, we study the asymptotic behavior of
solutions of the one-dimensional  Schr{\"o}dinger equation
\begin{equation}
  \label{eq:single}
  -\frac{d^2\psi}{dx^2}\,(x)+V(x)\,\psi(x)+W(\varepsilon
  x)\,\psi(x)=E\,\psi(x),\quad x\in\R,
\end{equation}
where $\varepsilon$ is a small positive parameter, and $V(x)$ a real
valued periodic function, $V(x+1)=V(x)$. We also assume that $V\in
L^2_{loc}$ and that $\zeta\mapsto 
W(\zeta)$ is real analytic in some
neighborhood of $\R\subset\C$.
\smallpagebreak The term $W(\varepsilon x)$ can be regarded as an
adiabatic perturbation of the periodic potential $V(x)$. The analysis
of perturbed periodic Schr{\"o}dinger equations is a classical topic of
mathematical physics. For example, in solid state physics, such
equations models behavior of electrons in crystals placed in an
external field (\cite{Bu:84,Bu:87}); in astrophysics, they model
periodic motions perturbed by the presence of massive objects
(\cite{MR82d:58038}). As in solid state physics, so in astrophysics,
the perturbations can often be regarded as very regular and slow
varying with respect to the underlying periodic system. This naturally
leads to an equation of the form~\eqref{eq:single}.
\subsection{Asymptotic methods}
\label{sec:asymptotic-methods}
The classical WKB methods are used for the analysis of equations of
the form 
\begin{equation}
  \label{eq:20}
  -\frac{d^2\psi}{dx^2}+W(\varepsilon x)\psi(x)=E\psi(x).
\end{equation}
The potential $W(\varepsilon\cdot)$ can be regarded as an adiabatic
perturbation of the free operator $\D-\frac{d^2}{dx^2}$.
In~\eqref{eq:single}, $\D-\frac{d^2}{dx^2}$ is replaced by the periodic
Schr{\"o}dinger operator
\begin{equation}
  \label{Ho}
   H_0=-\frac{d^2}{dx^2}(x)+V(x),\quad\quad  V(x+1)=V(x),\quad x\in\R.
\end{equation}
In~\cite{Bu:84}, to study solutions of~\eqref{eq:single}, V.~Buslaev
has suggested an analog the classical real WKB method. Both these
methods do not allow to control important exponentially small effects
(e.g. over barrier tunneling coefficients, exponentially small
spectral gaps). To study these effect for~\eqref{eq:20}, one can use
the classical complex WKB method. And, in~\cite{Fe-Kl:98c}, we
have developed an analog thereof to study such exponentially small
effects for equation~\eqref{eq:single}.
\smallpagebreak In our method (as in the classical complex WKB
method), one assumes that the adiabatic perturbation $W(\cdot )$ is
analytic and one tries to make the ``slow'' variable complex. But,
in~\eqref{eq:single}, as $V$ can be rather singular, one has to
``decouple'' the ``slow'' and the ``fast'' variables. We do this by
introducing an additional parameter, say $\zeta$, so that
equation~\eqref{eq:single} takes the form
\begin{equation}
  \label{family}
  -\frac{d^2}{dx^2}\psi(x)+(V(x)+W(\varepsilon
  x+\zeta))\psi(x)= E\psi(x), \quad x\in\R.
\end{equation}
The idea of our method is to study solutions of~\eqref{family} on the
complex plane of $\zeta$ and, then, to recover information on their
behavior in $x$ along the real line.
\smallpagebreak There is a natural condition that can be imposed on
solutions of~\eqref{family} so as to relate their behavior in $x$ to
their behavior in $\zeta$:
\begin{equation}
  \label{consistency}
  \psi(x+1,\zeta)=\psi(x,\zeta+\varepsilon)\quad\forall\zeta.
\end{equation}
We call it the {\it consistency} condition. On the complex plane of
$\zeta$, there are certain {\it canonical} domains where the solutions
satisfying the consistency condition have simple asymptotic behavior
(see section~\ref{sec:Canonical-domains} and Theorem~\ref{T5.1}):
\begin{equation}
  \label{simple-ass}
  \psi(x,\zeta)=e^{\dsize \frac{i}\varepsilon\int_{\zeta_0}^\zeta\kappa
  d\zeta}\left(\Psi(x, E-W(\zeta))+o(1)\right),\quad \varepsilon\to0.
\end{equation}
Here, $\Psi$ and $\kappa$ are a Bloch (Floquet) solution and the Bloch
quasi-momentum (see sections~\ref{SSS:BS} and~\ref{SS3.2}) of the
``unperturbed'' periodic equation
\begin{equation}
  \label{PSE}
   -\frac{d^2\Psi}{dx^2}+V(x)\Psi=\mathcal{E}\,\Psi,\quad
   \mathcal{E}=E-W(\zeta),
   \quad x\in\R.
\end{equation}
\smallpagebreak Having constructed solutions having simple asymptotic
behavior on a given canonical domain, one studies them outside this
domain using the transfer matrix techniques as in the classical
complex WKB method. The new asymptotic method has already been
successfully applied to study spectral properties of quasi-periodic
equations. In~\cite{Fe-Kl:02, Fe-Kl:01b, Fe-Kl:01c, Fe-Kl:03a}, using
this method, we have obtained a series of new results. However, trying
to proceed as in the classical complex WKB method, one meets numerous
technical problems which makes the computations very long. In this
paper, we present a new geometric approach replacing or simplifying
most of these computations.
\subsection{Canonical domains} 
\label{sec:comp-with-class}
Canonical domains are defined in terms of $\kappa(\zeta)$, the {\it
  complex momentum}. This function satisfies
\begin{equation}
  \label{IEC}
  \mathcal{E}(\kappa)+W(\zeta)=E,
\end{equation}
where $\mathcal{E}$ is the dispersion law of the periodic
operator~\eqref{Ho}. In the classical case, i.e. for $\D
H_0=-\frac{d^2}{dx^2}$, relation~\eqref{IEC} takes the form
$\kappa^2+W(\zeta)=E$. The properties of the complex momentum in the
adiabatic case are discussed in section~\ref{S4}. \\
Canonical domains are unions of canonical curves connecting two given
points in $\C$ (``two points'' condition). A canonical curve is
roughly a smooth vertical curve (i.e. intersecting the lines $\im\zeta
=0$ at non-zero angles) along which the function $\im\int^\zeta(\kappa
-\pi)d\zeta$ decreases, and the function $\im\int^\zeta\kappa d\zeta$
increases for increasing $\im\zeta$ (see
section~\ref{sec:Canonical-domains} for the precise definition).\\
Recall that, in the classical case, in the definition of the canonical
domains, there is no "two points" condition, and the canonical lines
are characterized by a growth condition on the function $\im\int^\zeta
\kappa d\zeta$. In our case, the ``verticality'' condition arises as
the periodicity $V(x+1)=V(x)$ singles out the ``horizontal'' direction
of the real line.\\
The basic fact of our method (established in~\cite{Fe-Kl:98c}) is
that, on any canonical domain, we can construct a solution with the
standard behavior~\eqref{simple-ass} (see Theorem~\ref{T5.1}).  It is
analytic in $\{Y_1<\im\zeta<Y_2\}$, the smallest strip containing the
canonical domain.
\subsection{The new geometric approach and its strategy}
\label{sec:new-geom-appr}
When applying the classical complex WKB method, one first describes
``maximal'' canonical domains; then, to get the global asymptotics of
a solution having simple asymptotic behavior on a given canonical
domain, one expresses it in terms of the solutions having simple
behavior on the other canonical domains. Therefore, one computes the
``transfer'' matrices relating basis of solutions having simple
asymptotic behavior on different overlapping canonical domains.\\
In the case of adiabatic perturbations of the periodic Schr{\"o}dinger
operator, the definition of the canonical domains contains more
conditions. In result, even ``maximal'' canonical domains are
generally quite ``small'' in the $\re\zeta$-direction. Moreover,
``maximal'' canonical domains become rather difficult to find. So,
when computing the transfer matrices relating solutions with simple
asymptotic behavior on two given different canonical domains, one has
to consider a ``long'' chain of auxiliary overlapping canonical
domains and to make many additional computations.\\
Fortunately, it appears that a solution having simple asymptotic
behavior~\eqref{simple-ass} on a canonical domain $K_0$ still has this
behavior on domains which can be much larger than the maximal
canonical domain containing $K_0$.  Domains where a consistent
solution $f$ has the simple asymptotic behavior~\eqref{simple-ass} are
called {\it continuation diagrams of $f$}. In this paper, we describe
an elementary geometric approach to computing continuation diagrams.
\smallpagebreak Instead of trying to find ``maximal'' canonical
domains, we begin by constructing a ``thin'' canonical domain. We use
the following simple observation (see Lemma~\ref{LCD}): any canonical
line is contained in a {\it local} canonical domain ``stretched''
along the canonical line. To construct a canonical line, we use
segments of some ``elementary'' curves described in
section~\ref{pre-cl} (see also
Proposition~\ref{pro:pcl:1}).\\
The main part of the work then consists in studying asymptotic
behavior of the solution constructed with Theorem~\ref{T5.1} outside
the local canonical domain. It appears that there are three general
principles allowing to compute the continuation diagram. We call these
principles the {\it main continuation tools}.
\smallpagebreak So, to construct a solution with simple asymptotics on
a large (not necessarily canonical) domain, we begin with a local
canonical domain, and then, step by step, at each step applying one of
the three continuation tools, we ``extend'' the continuation diagram,
``continuing'' (i.e.  justifying) the simple asymptotics of $f$ to a
larger domain.
\subsection{The main continuation tools}
\label{sec:main-cont-tools}
There are three continuation tools: the Rectangle Lemma,
Lemma~\ref{Rectangle}, the Adjacent Canonical Domain Principle,
Proposition~\ref{AddCD} and the Stokes Lemma, Lemma~\ref{st-lm}.  The
first two principles were formulated and proved in~\cite{Fe-Kl:02}
and~\cite{Fe-Kl:01b}. The Stokes Lemma is proved in the present paper.
We now briefly explain the respective roles of these tools and show
how they complement one another when computing the continuation
diagram.
\smallpagebreak{\bf The Rectangle Lemma.} Roughly, the Rectangle Lemma
says that a solution $f$ has the standard asymptotic
behavior~\eqref{simple-ass} along a horizontal line (i.e. a line
$\im\zeta=\Const$) as long as the leading term of its asymptotics is
growing along that line. This result is in agreement with the standard
WKB heuristics saying that the asymptotics of a solution stays valid
as long as its leading term is defined and increasing.
\smallpagebreak The leading term of the asymptotics contains the
exponential factor $\exp(\dsize \frac{i}\varepsilon\int^\zeta\kappa
d\zeta)$. For small $\varepsilon$, this factor determines the size of
the solution. If $\im\kappa>0$ in some domain $D$, then, $f$ is
increasing to the left; if $\im\kappa<0$ in $D$, then, $f$ is
increasing to the right. The Rectangle Lemma (Lemma~\ref{Rectangle})
is formulated in terms of the sign of the imaginary part of $\kappa$.
\smallpagebreak Let $\gamma$ be the canonical line used to construct
the solution $f$ locally. If, along a segment of $\gamma$, \ 
$\im\kappa>0$ (resp. $\im\kappa<0$), then, $f$ keeps its simple
behavior in a domain contiguous to $\gamma$ on its left (resp. right)
side.
\smallpagebreak A natural obstacle for ``continuation'' by means of
the Rectangle Lemma is a vertical line where $\im\kappa=0$. So,
usually, the domains where one justifies~\eqref{simple-ass} by means
of the Rectangle Lemma are curvilinear rectangles (or unions thereof).
\smallpagebreak{\bf The Adjacent Canonical Domain Principle.} Let $\gamma_0$ be
a curve canonical with respect to $\kappa_0$, some branch of the
complex momentum. The Adjacent Canonical Domain Principle,
Proposition~\ref{AddCD}, says that, if a solution $f$ has the simple
behavior~\eqref{simple-ass} in a domain {\it adjacent} to a canonical
curve $\gamma_0$ then, $f$ keeps its simple behavior in any domain
canonical with respect to $\kappa_0$ and {\it enclosing} $\gamma_0$.
\smallpagebreak The Adjacent Canonical Domain Principle is used to bypass the
vertical curves which are obstacles for the use of the Rectangle
Lemma. These can be either segments of the canonical line used to
start the construction of $f$ or vertical lines along which
$\im\kappa=0$. In both cases, the obstacles are curves canonical with
respect to some branch of the complex momentum.
\smallpagebreak By means of the Adjacent Canonical Domain Principle, one
justifies the standard behavior in $A$, a domain the boundary of which
contains the curve $\gamma_0$ and the lines beginning at the ends of
$\gamma_0$ defined by equations of the form $\im\int^\zeta\kappa_0
d\zeta=\Const$ and $\im\int^\zeta(\kappa_0-\pi) d\zeta=\Const$. Often
these two lines intersect one another, and the domain $A$ has the
shape of a curvilinear triangle. Otherwise, one considers domains $A$
of the form of a curvilinear trapezium; the fourth curve bounding such
a trapezium is one more canonical curve.  The precise description of
these two possible situations is the subject of the Trapezium Lemma,
Lemma~\ref{parallelogram-le}.
\smallpagebreak The trapezium shaped domains are used to avoid the
construction of ``maximal'' canonical domains {\it enclosing}
$\gamma_0$ as this can be rather tricky. As the fourth boundary of the
trapezium shaped domains, one usually chooses a curve which can be
bypassed either by means of the other continuation tools or by
applying The Adjacent Canonical Domain Principle once more.
\smallpagebreak{\bf The Stokes Lemma.} Lemma~\ref{st-lm} is akin to
the results of the classical complex WKB method on the behavior of
solutions in a neighborhood of a Stokes line where, instead of
decreasing, they start to increase, see~\cite{Fe:93}. \\
Consider $\zeta_0$ a branch point of the complex momentum. Assume
$W'(\zeta_0)\ne 0$. As in the classical complex WKB method, such a
point gives rise to three Stokes lines (i.e. lines starting at such a
branch point defined by $\im\int^\zeta(\kappa-\kappa(\zeta_0))
d\zeta=0$).\\
Let $\sigma$ be one of these lines that moreover is vertical. Consider
$V$, a neighborhood of $\sigma$ (more precisely, of a segment of
$\sigma$ containing only one branch point, namely, $\zeta_0$). Assume
that $V$ is so small that the Stokes lines divide it into three
sectors (see Fig.~\ref{stokes:fig:1}). Let $S_1$ and $S_3$ be the
sectors adjacent to $\sigma$, and let $S_2$ be the last sector.
Roughly, the Stokes Lemma says that, if $f$ has the standard behavior
inside $S_1\cup S_2$ and decreases as $\zeta\in S_1\cup S_2$
approaches $\sigma$ along the lines $\re\zeta=\Const$, then, $f$ has
the standard behavior in $V\setminus\sigma$.\\
In result, to get the leading term of the asymptotics of $f$ in the
sector $S_3$, one analytically continues this term from $S_1\cup S_2$
to $S_3$ inside $V\setminus \sigma$, i.e. around the branch point
$\zeta_0$ avoiding the line $\sigma$.\\
The Stokes Lemma complements the Adjacent Canonical Domain Principle.
Recall that the Adjacent Canonical Domain Principle allows to bypass
vertical curves where $\im\kappa=0$. The ends of the curves on which
$\im\kappa=0$ are branch points of the complex momentum. The Stokes
lines beginning at these points usually form the upper and the lower
boundaries of the domains where one justifies the standard behavior by
means of the Adjacent Canonical Domain Principle. The Stokes Lemma,
Lemma~\ref{st-lm}, allows us to justify the standard behavior beyond
these lines by ``going around'' the branch points.
\smallpagebreak{\bf On the choice of the initial canonical line.}
For our construction to be successful, we have to make a suitable
choice for the canonical line we start with. The idea is that this
line should be close to the curve where the constructed solution is
minimal: inside the continuation diagram, the factor
$\left|\exp(\dsize \frac{i}\varepsilon\int_{\zeta_0}^\zeta \kappa
  d\zeta)\right|$ has to increase as $\zeta$ moves away from this
curve (along the lines $\im\zeta=\Const$). To achieve this, one builds
the canonical line of segments of curves where $\im\kappa=0$ and of
segments of curves close to Stokes lines. In section~\ref{ex:lcd}, we
construct a canonical line of such curves. In
section,~\ref{ex:cont-diag}, we give a detailed example of the
computation of a continuation diagram of a solution constructed on a
canonical domain enclosing such a canonical line.
\subsection{Two-Waves Principle}
\label{sec:two-wave-principle}
Recall that a continuation diagram is a domain where $f$, a given
solution of~\eqref{family} satisfying~\eqref{consistency}, has the
simple behavior~\eqref{simple-ass}. In domains next to the
continuation diagram, the leading term of the asymptotics of the
solution is of the form
\begin{equation}
  \label{gen-as-rep}
  A_+\,e^{\dsize \frac{i}\varepsilon\int^\zeta\kappa
    d\zeta}\Psi_+(x, E-W(\zeta))+ A_-\,e^{\dsize
    -\frac{i}\varepsilon\int^\zeta\kappa d\zeta}\Psi_-(x, E-W(\zeta))
\end{equation}
with coefficients $A_\pm$ that depend non trivially on $E$. This
dependence makes it impossible to describe the solution by only one
of the terms in~\eqref{gen-as-rep} uniformly in $E$ and $\zeta$.\\
When studying a solution in domains adjacent to the continuation
diagram one meets many different cases. In this paper, we discuss only
one typical case. One encounters it when studying the solution in the
domains ``adjacent'' to the local canonical domain where the
construction of the solution was started. The precise geometrical
situation is described in section~\ref{sec:2waves}; the behavior of
the solution is governed by the Two-Waves Principle, Lemma~\ref{tw:1},
see also comments in section~\ref{tw:comments}.
\smallpagebreak Note that, in the case of the Two-Waves Principle, one
of the coefficients $A_\pm$ rapidly oscillates as a function of $E$
(for $\varepsilon\to 0$) and ``periodically'' vanishes. Its zeros are
described by a Bohr-Sommerfeld like condition. Recall that, when
starting, we try to construct solutions along the lines where they are
minimal. The values of $E$ for which one of the coefficients
in~\eqref{gen-as-rep} vanishes can be regarded as some sort of
``resonances''; when $E$ takes a ``resonant'' value, the solution
becomes minimal along a new curve.
\subsection{Examples}
\label{example}
In all the examples we have treated so far
(\cite{Fe-Kl:01b,Fe-Kl:01c,Fe-Kl:03a, Fe-Kl:03b}), we have seen that,
for a suitable choice for the initial canonical line, the continuation
diagram of the solution can be effectively computed by means of the
continuation tools described above. In the present paper, instead of
trying to formulate and prove this observation as a general statement,
we illustrate all our constructions by detailed examples. In these
examples, we assume that $W(\zeta)=\alpha\cos\zeta$, that all the gaps
of the periodic operator~\eqref{Ho} are open, and that the energy $E$
satisfies
\begin{description}
\item[({\bf C})] $[E_{2n}-\delta, E_{2n+1}+\delta]\subset
  [E-\alpha,E+\alpha]\subset(E_{2n-1},E_{2n+2})$,
\end{description}
where $[E_{2n-1}, E_{2n}]$ and $[E_{2n+1}, E_{2n+2}]$ are two
neighboring spectral bands of the periodic operator~\eqref{Ho}. This
case is of special interest in the sense that it will illustrate the
use of all our tools. From the quantum physicist's point of view, this
is the case when $[E_{2n-1}, E_{2n}]$ and $[E_{2n+1}, E_{2n+2}]$, the
spectral bands of $H_0$, interact due ``through'' the adiabatic
perturbation. In this case, one can observe several new interesting
spectral phenomena, see~\cite{Fe-Kl:03b,Fe-Kl:03a}. The examples we
consider in the present paper are used to study these effects
(see~\cite{Fe-Kl:03b}).
\subsection{The structure of the paper}
\label{sec:structure-paper}
In this text, we describe general constructions and results step by
step, illustrating each step with examples. More or less long proofs
of general results are postponed until the end of the paper.
\smallpagebreak Throughout the paper, we shall use a number of well
known facts on the periodic Schr{\"o}dinger operator~\eqref{Ho}. They are
described in section~\ref{S3}. In subsection~\ref{sec:Omega}, we also
introduce an analytic object defined in terms of the periodic
operator; it is playing an important role for the adiabatic
constructions.\\
In section~\ref{S4}, we define and study the complex momentum and
related objects (e.g. Stokes lines). We complete this section
(subsection~\ref{ex:st-ln}) with the analysis of the complex momentum
and the Stokes line for $W(\zeta)=\alpha\cos\zeta$.\\
In section~\ref{standard-behavior}, we introduce the concept of
standard behavior and define canonical lines and canonical domains; we
also formulate Theorem~\ref{T5.1} on the solutions having standard
behavior on a given canonical domain.\\
In section~\ref{local-constructions}, we define local canonical
domains and explain how to build canonical lines from segments of
``elementary curves''. Having presented general results, in
subsection~\ref{ex:lcd}, as an example, we construct a canonical line
using this method.\\
Section~\ref{sec:m-t} is devoted to the main continuation principles
and related objects. The Trapezium Lemma,
Lemma~\ref{parallelogram-le}, is proved in section~\ref{proof:T-l}.
The Stokes Lemma, Lemma~\ref{st-lm}, is proved in
section~\ref{Principles-proof}.\\
In section~\ref{ex:cont-diag}, we give a detailed example of the
computation of a continuation diagram.\\
Section~\ref{sec:2waves} is devoted to the Two-Waves Principle. In
subsection~\ref{ex:tw}, on a detailed example, we show how to use it.
The proof of the Two-Waves principle can be found in
section~\ref{tw:proof}.


%
\section{Periodic Schr{\"o}dinger operators}
\label{S3}
\noindent We first formulate well known results used throughout the
paper. Their proofs can be found, for example,
in~\cite{Eas:73,Ma-Os:75,McK-Tr:75,Ti:58}. In the end of the section,
we discuss a meromorphic function constructed in terms of the periodic
operator. This function plays an important role for the adiabatic
constructions.\\
Recall that the potential $V$ in~\eqref{Ho} is assumed to be a
$1$-periodic, real valued, $L^2_{loc}$-function.
\subsection{Gaps and bands}
\label{sec:gaps-bands}
The spectrum of the periodic operator~\eqref{Ho} is absolutely
continuous and consists of intervals of the real axis $[E_1,\,E_2]$,
$[E_3,\,E_4]$, $\dots$, $[E_{2n+1},\,E_{2n+2}]$, $\dots$, such that
\begin{gather*}
  E_1<E_2\le E_3<E_4\dots E_{2n}\le E_{2n+1}<E_{2n+2}\le \dots\,,\\
  E_n\to+\infty,\quad n\to+\infty.
\end{gather*}
The points $E_{j}$, \ $j=1,2,3\dots$, are the eigenvalues of the
differential operator~\eqref{Ho} acting on $L^2([0,2])$ with periodic
boundary conditions. The intervals defined above are called the {\it
  spectral bands}, and the intervals $(E_2,\,E_3)$, $(E_4,\,E_5)$,
$\dots$, $(E_{2n},\,E_{2n+1})$, $\dots$, are called the {\it spectral
  gaps}. If $E_{2n}<E_{2n+1}$, we say that the $n$-th gap is {\it
  open}.
\subsection{Bloch solutions}
\label{SSS:BS}
Let $\psi$ be a solution of the equation
\begin{equation}
  \label{PSO}
  -\frac{d^2}{dx^2}\psi\,(x)+ V\,(x)\psi\,(x)=E\psi\,(x), \quad x\in\R,
\end{equation}
satisfying the relation $\psi\,(x+1)=\lambda\,\psi\,(x)$ for all
$x\in\R$ with $\lambda\in\C$ independent of $x$. Such a solution is
called a {\it Bloch} solution, and the number $\lambda$ is the {\it
  Floquet multiplier}. Let us discuss the analytic properties of Bloch
solutions as functions of the spectral parameter.
\smallpagebreak Consider $\mathcal{S}_\pm $, two copies of the complex
plane of energies cut along the spectral bands. Paste them together to
get a Riemann surface with square root branch points.
We denote this surface by $\mathcal{S}$.\\
There exists a Bloch solution $E\mapsto\psi(x,E)$ of
equation~\eqref{PSO} meromorphic on $\mathcal S$. We normalize it by
the condition $\psi(0,E)\equiv 1$.  The poles of this solution are
located in the spectral gaps. More precisely, for each spectral gap,
there is one and only one pole projecting into this gap. This pole is
located either on $\mathcal{S}_+$ or on $\mathcal{S}_-$. The position
of the pole is independent of $x$.
\smallpagebreak Except at the edges of the spectrum (i.e. the branch
points of $\mathcal{S}$), the two branches of $\psi$ are linearly
independent solutions of~\eqref{PSO}.
\smallpagebreak Finally, we note that, in the spectral gaps, both
branches of $\psi$ are real valued functions of $x$, and, on the
spectral bands, they differ only by complex conjugation.
\subsection{The Bloch quasi-momentum}
\label{SS3.2}
Consider the Bloch solution $\psi(x,E)$. The corresponding Floquet
multiplier $\lambda\,(E)$ is analytic on $\mathcal{S}$. Represent it
in the form $\lambda(E)=\exp(ik(E))$. The function $k(E)$ is the {\it
  Bloch quasi-momentum}.
\smallpagebreak The Bloch quasi-momentum is an analytic
multi-valued function of $E$. It has the same branch points as
$\psi(x,E)$.\\
Let $D$ be a simply connected domain containing no branch point of
the Bloch quasi-momentum. In $D$, one can fix an analytic
single-valued branch of $k$, say $k_0$. All the other
single-valued branches of $k$ that are analytic in $E\in D$ are
related to $k_0$ by the formulae
\begin{equation}\label{eq:55}
   k_{\pm ,l}(E)=\pm k_0(E)+2\pi l,\quad l\in\Z.
\end{equation}
Consider $\C_+$ the upper half of the complex plane. On $\C_+$, one
can fix a single valued analytic branch of the quasi-momentum
continuous up to the real line. We can and do fix it by the condition
$ -ik(E+i0)>0$ for $E<E_1$. We call this branch the {\it main branch}
of the Bloch quasi-momentum and denote it by $k_p$.\\
The function $k_p$ conformally maps $\C_+$ onto the first quadrant of
the complex plane cut at compact vertical slits beginning at the
points $\pi l$, $l\in \N$. It is monotonically increasing along the
spectral bands so that $[E_{2n-1}, E_{2n}]$, the $n$-th spectral band,
is mapped on the interval $[\pi(n-1), \pi n]$. Inside any open gap,
$\re k_p(E)$ is constant, and $\im k_p(E)$ is positive and has only
one non-degenerate maximum. If the $n$th gap is open, in this gap, one
has $\re k_p(E)=\pi n$.\\
All the branch point of $k_{p}$ are of square root type: let $E_l$
be a branch point; then, in a sufficiently small neighborhood of
$E_l$, the function $k_p$ is analytic in $\sqrt{E-E_l}$, and
\begin{equation}
  \label{sqrt}
  k_{p}(E)-k_{p}(E_l)=c_l\sqrt{E-E_l}+O(E-E_l),\quad c_l\not=0.
\end{equation}
Finally, we note that the main branch can be analytically continued on
the complex plane cut only along the spectral gaps of the periodic
operator.
\subsection{A meromorphic function $\mathbf{\omega}$ and the
  differential $\mathbf{\Omega}$}
\label{sec:Omega}
We now define a meromorphic function on $\mathcal{S}$, the Riemann
surface associated to the periodic operator~\eqref{Ho}. First, we have
to recall more facts and to introduce some notations.
\subsubsection{Periodic component of the Bloch solution}
\label{sec:peri-comp-bloch}
At a given energy $E$, the Bloch solution $\psi(x,E)$ can be
represented in the form
\begin{equation}
  \label{PCBS:1}
  \psi(x,E)=e^{\dsize ik(E)x} p (x,E),
\end{equation}
where $k(E)$ is the Bloch quasi-momentum of $\psi(x,E)$ at $E$, and
the function $p(x,E)$ is $1$-periodic in $x$. The function $p$ is
called the {\it periodic component} of $\psi$ with respect to
the branch $k(E)$.\\
Note that, as $k(E)$ is defined modulo $2\pi$, the function $p(x,E)$
is defined up to the factor $e^{2\pi i m x}$,\ $m\in \Z$.  The
branches $p$ and $k$ are related by
\begin{equation}\label{k->p}
k\to k+2\pi m\quad \Longleftrightarrow p\to e^{-2\pi i m x}\,p.
\end{equation}
\subsubsection{Notations}
\label{sec:notations}
For $E\in \mathcal{S}$, let $\hat E$ be the other point in
$\mathcal{S}$ having the same projection on $\C$ as $E$. We let
\begin{equation}\label{hats}
  \hat \psi(x,E)=\psi(x,\hat E),\quad \hat k(E)=-k(E),\quad\hat
  p(x,E)=e^{-\dsize i\hat k(E)\,x}\,\hat \psi(x,E).
\end{equation}
The function $\hat \psi$ is one more Bloch solution of the periodic
Schr{\"o}dinger equation that, for $E$ outside $\{E_l\}$, is linearly
independent of $\psi(x,E)$. The function $\hat k$ is its
quasi-momentum, and $\hat p$ its periodic component.
\subsubsection{The sets $P$ and $Q$}
\label{sec:sets-p-q}
Introduce two discrete sets on $\mathcal S$. Let $P$ be the set of
poles of the Bloch solution $\psi(x,E)$, and, $Q$ be the set of
the points where $k'(E)=0$.\\
Recall that the points of $Q$ are (projected) inside open gaps of
the periodic operator (one point per gap), and that the points of
$P$ are (projected) either inside open gaps or at their edges
(also one point per open gap).
\subsubsection{Local construction of the  function $\omega$ and the
differential $\Omega$}
\label{sec:constr-diff-omega}
Let $D\subset\mathcal{S}\setminus \{E_l\}$ be a simply connected
domain. On $D$, fix $k$, an analytic branch of the Bloch
quasi-momentum of $\psi$. Then, the functions $p$ and $\hat p$ are
meromorphic on $D$. We let
\begin{equation}
  \label{eq:18}
  \omega(E)=-\frac{\int_0^1 \hat p(x,\,E)\,
    \frac{\partial p}{\partial E}(x,\,E)dx}
   {\int_0^1 p(x,\, E)\,\hat p(x,\,E)dx},\quad\text{and}\quad
    \Omega(E) = \omega(E)\,dE.
\end{equation}
Note that the function $\omega$ was introduced and analyzed in the
paper~\cite{Fe-Kl:01b}. Using the differential $\Omega$ instead of
this function makes computations more transparent. We have
\begin{Le}
  \label{le:Omega:loc}
  $\Omega$ is a meromorphic differential on $D$. All its poles are
  simple; they are situated at exactly the points of $P\cup Q$. The
  residues are given by the formulae:
  \begin{equation}
    \label{eq:18a}
    \res_p \Omega=1,\quad\forall p\in P\setminus Q,\quad\quad
    \res_q\Omega=-1 / 2,\quad\forall q\in Q\setminus P,\quad\quad
    \res_r\Omega=1 / 2,\quad\forall r\in Q\cap P.
  \end{equation}
\end{Le}
\noindent Lemma~\ref{le:Omega:loc} follows from the analysis made
in~\cite{Fe-Kl:01b} when proving Lemma 3.1. We omit the details.
\subsubsection{Global properties of $\Omega$}
\label{sec:glob-prop-omega}
By means of~\eqref{k->p}, we see that, $\omega$ and $\Omega$ do not
depend on the choice of the branch $k$. Hence, $\omega$ and $\Omega$
are uniquely defined on $\mathcal{S}\setminus\{E_l\}$. One can analyze
$\Omega$ on the whole Riemann surface $\mathcal S$ ($\infty$ was not
``included'' in $\mathcal S$). This gives
\begin{Le}
  \label{le:Omega}
  $\Omega$ is a meromorphic differential on the whole Riemann surface
  $\mathcal{S}$. Its poles and the residues at these poles are
  described in Lemma~\ref{le:Omega:loc}.
\end{Le}
\demo In view of Lemma~\ref{le:Omega:loc}, it suffices to study
$\Omega$ in $V_n$, a sufficiently small neighborhood of $E_n$, an end
of a spectral gap.  Recall that $k'$ has zeros only inside open gaps.
So, as $V_n$ can be taken arbitrarily small, there are two cases to
consider:
\begin{itemize}
\item either $P\cap V_n=\emptyset$,
\item or $E_n\in P$.
\end{itemize}
We have to show that $\Omega$ is holomorphic in $V_n$ with respect to
the local variable $\tau=\sqrt{E-E_n}$.  Consider the first case.
Recall that $k$ is analytic (holomorphic) in $\tau$. So, $p$ and $\hat
p$ are also holomorphic in $\tau$, and we have only to check that the
function $f(\tau)=\int_0^1 p\,\hat p\,dx=\int_0^1 \psi\,\hat \psi\,dx$
does not vanish at $\tau=0$.  At the end of any spectral gap, one has
$\hat \psi=\psi$, and $\psi$ is real. So, $f(0)=\int_0^1 |\psi(x,E_n)|^2
dx>0$. This completes the proof in the first case.\\
In the second case, one has to prove that $\Omega$ has a simple pole
at $\tau=0$, and that $\res_{\tau=0}\Omega=1$. Now, in $V_n$,
$\psi(x,\tau)=\psi_0(x)/\tau+({\rm \ a \ holomorphic \ function})$.
The function $\psi_0$ is a (non-trivial) Bloch solution of~\eqref{PSE}
at $E=E_n$. It is real valued. This and the definitions of $p$ and
$\hat p$ imply that, in $V_n$, one has $\Omega=d\tau/\tau+({\rm \ a \ 
  holomorphic \ differential})$.  This completes the analysis of the
second case and, therefore, the proof of Lemma~\ref{le:Omega}.\qed
\subsubsection{Differential $\Omega$ and analytic Bloch solutions}
\label{sec:diff-omega-analyt}
Let us formulate a very important property of $\Omega$. Consider again
a simply connected domain $D\subset \mathcal{S}$. Pick $E_0\in
D\setminus(P\cup Q)$. In a sufficiently small neighborhood of $E_0$, one can
define the function
\begin{equation}
  \label{40.10}
  \sqrt{k_E'(E)}\,\, e^{\dsize \int_{E_0}^{E}\Omega}\, \psi(x,\,E).
\end{equation}
It is also a Bloch solution of the periodic Schr{\"o}dinger equation.
Lemma~\ref{le:Omega} immediately implies that it can be analytically
continued on the whole domain $D$.
\subsubsection{The function $\omega$ along gaps and bands}
\label{sec:function-omega-along}
In applications, one uses the following observations:
\begin{Le}
  \label{omega:gaps-and-bands}
  Along open gaps, the values of $\omega$ are real. Along bands,
  $\omega(E)$ and $\omega(\hat E)$ only differ by complex conjugation.
\end{Le}
\demo The statements follow from the facts that, along the gaps,
$\psi$ is real and $k$ is purely imaginary modulo $2\pi$, and that
along the bands, $\psi$ and $\hat \psi$ differ by complex conjugation,
and $k$ is real.\qed
%

%
\section{The complex momentum}
\label{S4}
The main analytic object of the complex WKB method is the complex
momentum. We now define and discuss it as well as some related objects
(e.g. the Stokes lines). We complete this section with an example: we
discuss the complex momentum and the Stokes lines for
$W(\zeta)=\alpha\cos\zeta$.
\subsection{Definition and elementary properties}
\label{sec:defin-elem-prop}
\begin{Def}
  For $\zeta\in{\mathcal D}(W)$, the domain of analyticity of the
  function $W$, the complex momentum is defined by
  \begin{equation}
    \label{eq:16}
    \kappa(\zeta)=k(E-W(\zeta))
  \end{equation}
  where $k$ is  the Bloch quasi-momentum of~\eqref{Ho}.
\end{Def}
\noindent Clearly, the complex momentum can also be interpreted
the Bloch quasi-momentum for the periodic Schr{\"o}dinger
equation~\eqref{PSE} regarded as a function of the complex
parameter $\zeta$.
\subsubsection{Branch points}
\label{sec:branch-points}
The relation between $k$ and $\kappa$ shows that the complex momentum
is a multi-valued analytic function, and that its branch points are
related to the branch points of the quasi-momentum by the relations
\begin{equation}
  \label{eq:17}
E_j=E-W\,(\zeta),\quad j=1,2,3,\, \dots
\end{equation}
Note that all of them are situated on $W^{-1}(\R)$, the pre-image
of the real line with respect to $W$.
\smallpagebreak Let $\zeta_0$ be a branch point of $\kappa$. Assume
$W'(\zeta_0)\ne 0$. Then, this branch point is of square root type: in
a neighborhood of $\zeta_0$, $\kappa$ is analytic in
$\sqrt{\zeta-\zeta_0}$, and
\begin{equation}
  \label{k:sqrt}
  \kappa(\zeta)-\kappa(\zeta_0)\sim\kappa_1
  \sqrt{\zeta-\zeta_0},\quad \kappa_1\ne 0.
\end{equation}
\subsubsection{Regular domains and branches of the complex
momentum}\label{kappa:branches}
\begin{Def}
  We say that a set is {\it regular} if it is a simply connected subset
  of the domain of analyticity of $W$ that contains no branch points of
  $\kappa$.
\end{Def}
\noindent  Let $D$ be a regular domain. In $D$, one can fix an
analytic branch of $\kappa$, say $\kappa_0$. By~\eqref{eq:55}, all the
other branches of $\kappa$ analytic on $D$ are described by the
formulas
\begin{equation}\label{allbr}
  \kappa_m^\pm  =\pm  \kappa_0+2\pi m,
\end{equation}
where $\pm $ and $m$ are indexing the branches.
\smallpagebreak Fix a branch point $\zeta_0$ such that $W'(\zeta_0)\ne
0$ and, let $V$ be a neighborhood of $\zeta_0$. Let $c$ be a smooth
curve beginning at $\zeta_0$ and such that $V\setminus c$ is a regular
domain. In $V\setminus c$, fix an analytic branch of the complex
momentum. Then, by~\eqref{k:sqrt}, $\kappa_1(\zeta)$ and
$\kappa_2(\zeta)$, the values of this branch in $V$ on the different
sides of $c$, are related by the formula
\begin{equation}
  \label{two-branches}
  \kappa_1(\zeta)+\kappa_2(\zeta)=2\kappa(\zeta_0),\quad \zeta\in c.
\end{equation}
\subsection{Stokes lines and lines of Stokes type}
\label{sec:stokes-lines-lines}
In the constructions of the complex WKB method, integrals of the form
$\int^{\zeta} \kappa\,d\zeta$ and $\int^{\zeta}(\kappa-\pi)\,d\zeta$
play an important role. Their properties are described in terms of
lines of Stokes type and Stokes lines.
\subsubsection{Lines of Stokes type}
\label{sec:stokes-lines}
Let $D$ be a regular domain. On $D$, fix an analytic branch of the
complex momentum. Pick $\zeta_0\in D$.
\begin{Def}
  \label{def:3}
  The level curves of the harmonic functions $\im
  \int_{\zeta_0}^{\zeta} \kappa\,d\zeta$ and $\im
  \int_{\zeta_0}^{\zeta}(\kappa-\pi)\,d\zeta$ are called {\it lines of
    Stokes type}.
\end{Def}
\noindent Clearly, lines of Stokes type do not depend on
the choice of $\zeta_0$.
\smallpagebreak To analyze the geometry of the lines of Stokes type,
one uses the following lemma (where we identify the complex numbers
with vectors in $\R^2$). One has
\begin{Le}
  \label{le:s-l-1}
  The lines of the family $\im \int^{\zeta} \kappa\,d\zeta=\Const$ are
  tangent to the vector field $\overline{\kappa(\zeta)}$; the lines of
  the family $\im \int^{\zeta} (\kappa-\pi)\,d\zeta=\Const$ are
  tangent to the vector field $\overline{\kappa(\zeta)}-\pi$.
\end{Le}
\noindent This lemma implies that the lines of Stokes type are trajectories
of the differential equations $\D\frac{d\zeta}{dt}=
\overline{\kappa(\zeta)}$ and $\D\frac{d\zeta}{dt}=
\overline{\kappa(\zeta)}-\pi$. So, to study properties of the lines of
Stokes type, one can use standard facts from the theory of
differential equations. In particular, we get
\begin{Cor}
  \label{le:s-l-2}
  The lines of Stokes type of each of the two families fibrate any
  regular domain $D$.
\end{Cor}
\demo For sake of definiteness, consider the lines of the family $\im
\int^{\zeta} \kappa\,d\zeta=\Const$. It suffices to show that the
vector field $\overline{\kappa(\zeta)}$ does not vanish in $D$. But,
we know, that $\kappa$ takes values in $\pi\Z$ only at branch points
of the complex momentum. As $D$ is regular, it does not contain any of
these points. This completes the proof of Corollary~\ref{le:s-l-2}.
\qed
\subsection{Stokes lines}
\label{sec:stokes-lines-2}
\noindent Below, we always work in the domain of analyticity of
$W$. Let $\zeta_0$ be a branch point of the complex momentum. A
{\it Stokes line} beginning at $\zeta_0$ is a curve $\gamma$
defined by the equation $\text{Im}\,\int_{\zeta_0}^\zeta
(\kappa\,(\xi)-\kappa\,(\zeta_0))d\zeta=0$. Here, $\kappa$ is a
branch of the complex momentum continuous on $\gamma$.\\
It follows from~\eqref{allbr} that the Stokes lines starting at
$\zeta_0$ are independent of the choice of the branch of $\kappa$
in the definition of a Stokes line.\\
Assume that $W'(\zeta_0)\ne 0$. Then, in a neighborhood of the
branch point $\zeta_0$, one has~\eqref{k:sqrt}. Hence, there are
three Stokes lines beginning at $\zeta_0$. At the branch point,
the angle between any two of them is equal to $2\pi/3$.\\
%
%
\begin{floatingfigure}{7cm}
  \begin{center}
    \includegraphics[bbllx=71,bblly=577,bburx=360,bbury=721,width=7cm]{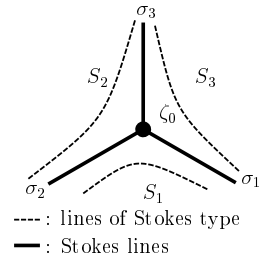}
  \end{center}
    \caption{The Stokes lines in a neighborhood of a branch point}
    \label{stokes:fig:1}
\end{floatingfigure}
%
%
\noindent One can always choose a branch of the complex momentum
(see~\eqref{allbr}) continuous on a given Stokes line $\gamma$ and
such that either $\kappa(\zeta_0)=0$ or $\kappa(\zeta_0)=\pi$. We call
this branch {\it natural}. With respect to the natural
branch, the Stokes lines are lines of Stokes type.\\
Consider $V$, a neighborhood of $\zeta_0$. If $V$ is sufficiently
small, the Stokes lines beginning at $\zeta_0$ divide $V$ into
three domains called {\it sectors}, see Fig.~\ref{stokes:fig:1}.\\
Let $\kappa(\zeta_0)=0$ (resp. $\kappa(\zeta_0)=\pi$). Then, each of
the sectors is fibrated by the lines of Stokes type of the family
$\im\int^{\zeta}\kappa d\zeta=Const$ (resp. $\im\int^{\zeta}
(\kappa-\pi)d\zeta=Const$). In particular, the part of the boundary of
such a sector formed by two Stokes lines can be approximated
arbitrarily well by a line of Stokes type $\im\int^{\zeta}\kappa
d\zeta=Const$ (resp.  $\im\int^{\zeta}(\kappa-\pi) d\zeta=Const$)
intersecting this sector, see Fig.~\ref{stokes:fig:1}.
\subsection{Example: complex momentum and Stokes lines for
  $\mathbf{W(\zeta)=\alpha\cos\zeta}$}
\label{ex:st-ln}
We now discuss the complex momentum and describe the Stokes lines when
$W(\zeta)=\alpha\cos\zeta$. We assume that all the gaps of the
periodic operator~\eqref{Ho} are open, and that the spectral parameter
$E$ satisfies condition (C).
\subsubsection{Complex momentum}
\label{sec:complex-momentum} {\bf 1.} \ The set of branch points
is $2\pi$-periodic and symmetric with respect both to the real line
and to the imaginary axis. For $E$ real, the branch points of the
complex momentum are situated on the lines of the set $W^{-1}(\R)$.
For $W(\zeta)=\alpha\cos\zeta$, this set consists of the real line and
the lines $\re \zeta=\pi l$, $l\in\Z$.
\smallpagebreak Define the half-strip
%
\begin{floatingfigure}{6cm}
  \centering
  \includegraphics[bbllx=71,bblly=583,bburx=247,bbury=721,width=6cm]{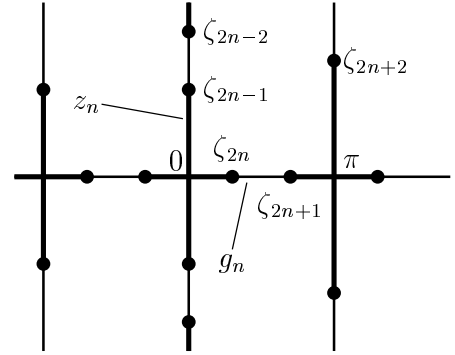}
  \caption{$(z_l)_l$ and $(g_l)_l$}\label{fig:zon-ga}
\end{floatingfigure}
%
\begin{equation}
  \label{Pi}
  \hskip -4cm
  \Pi=\{\zeta\in\C:\ 0< \re\zeta<  \pi,\ \im\zeta>0\}.
\end{equation}
This half-strip is a regular domain. Consider the branch points
situated on $\partial\Pi$, the boundary of $\Pi$. $\partial\Pi$ is
bijectively mapped by $\mathcal{E}:\zeta\to E-W(\zeta)$ onto the real
line. So, for any $j\in \N$, there is exactly one branch point
described by~\eqref{eq:17}. We denote this point by $\zeta_j$.  Under
condition (C), the branch points $\zeta_{2n}$ and $\zeta_{2n+1}$ are
situated on the interval $(0,\pi)$, i.e.
\begin{equation*}
  \hskip -4cm
   0<\zeta_{2n}<\zeta_{2n+1}<\pi;
\end{equation*}
the branch points $\zeta_1$, $\zeta_2$, $\dots$ $\zeta_{2n-1}$ are
situated on the imaginary axis so that
\begin{equation*}
     \hskip -4cm
     0<\im \zeta_{2n-1}<\dots<\im\zeta_{2}<\im \zeta_1;
\end{equation*}
the other branch points are situated on the line $\re\zeta=\pi$ so
that
\begin{equation*}
   0<\im \zeta_{2n+2}<\im \zeta_{2n+3}<\dots
\end{equation*}
In Fig.~\ref{fig:zon-ga}, we show some of the branch points.
\vskip.1cm\noindent {\bf 2.} \ The half-strip $\Pi$ is mapped by
$\mathcal{E}:\zeta\to E-\alpha\cos\zeta$ on the upper half of the
complex plane. So, on $\Pi$, we can define a branch of the complex
momentum by the formula
\begin{equation}
  \label{kappa0k0}
  \hskip-4cm
  \kappa_p(\varphi)=k_p(E-\alpha\cos\varphi),
\end{equation}
$k_p$ being the main branch of the Bloch quasi-momentum for the
periodic operator~\eqref{Ho}. We call $\kappa_p$  {\it the main
branch} of the complex momentum.
\smallpagebreak The main branch of the Bloch quasi-momentum was
discussed in details in section~\ref{SS3.2}. The properties of $k_p$
are ``translated'' into properties of $\kappa_p$ using
formula~\eqref{kappa0k0}. In particular, $\kappa_p$ conformally maps
$\Pi$ into $\C_+$. Fix $l$, a positive integer.  The closed segment
$z_l:=[\zeta_{2l-1},\zeta_{2l}]\subset\partial \Pi$ is bijectively
mapped on the interval $[\pi(l-1),\pi l]$, and, on the open segment
$g_l:=(\zeta_{2l}, \zeta_{2l+1})\subset\partial\Pi$, the real part of
$\kappa$ equals to $\pi l$, and its imaginary part is positive. The
intervals $(z_l)_l$ and $(g_l)_l$ are shown in Fig.~\ref{fig:zon-ga}.
\subsubsection{Stokes lines}
\label{sec:stokes-lines-1}
Let us discuss the set of Stokes lines for $W(\zeta)=\alpha\cos\zeta$.
Due to the symmetry properties of $\mathcal E$, the set of the Stokes
lines is $2\pi$-periodic and symmetric with respect to both the real
and imaginary axes.
\smallpagebreak In Fig.~\ref{bem:St-lines}, we have represented Stokes
lines in $\Pi$ by dashed lines. Consider the Stokes lines beginning at
the branch points $\zeta_l$ with $l\geq 2n$. The other Stokes lines
beginning at points of $\partial\Pi$ are analyzed similarly.  We begin
with properties following immediately from the definition of Stokes
lines.
\smallpagebreak{\it Elementary properties of Stokes lines.\/}
Consider the Stokes lines beginning at $\zeta_{2n+1}$. The
interval $[\zeta_{2n+1},\pi]$ is a part of $z_{n+1}$. So,
$\kappa_p$ is real on this interval, and, therefore, this interval
is a part of a Stokes line beginning at $\zeta_{2n+1}$.  The two
other Stokes lines beginning at $\zeta_{2n+1}$ are symmetric with
respect to the real line, see Fig.~\ref{bem:St-lines}. We denote
by ``a'' the Stokes line going upward from $\zeta_{2n+1}$.
\smallpagebreak Consider the Stokes lines beginning at $\zeta_{2n+2}$.
As $\kappa_p(\zeta_{2n+2})=\pi (n+1)$, they satisfy
$\im\int_{\zeta_{2n+2}}^\zeta(\kappa_p(\zeta)-\pi(n+1))d\zeta=0$.
Recall that, along the segment $g_{n+1}=(\zeta_{2n+2},\zeta_{2n+3})$
of the line $\re \zeta=\pi$, one has $\re \kappa_p=\pi (n+1)$. So,
this segment is a Stokes line beginning at $\zeta_{2n+2}$. The two
other Stokes lines beginning at $\zeta_{2n+2}$ are symmetric with
respect to the line $\re\zeta=\pi$, see Fig.~\ref{bem:St-lines}. We
denote by ``b'' the Stokes line going to the left from $\zeta_{2n+2}$.
\vskip.1cm\noindent The Stokes lines beginning at other branch points
situated on the right part of $\partial \Pi$ are analyzed similarly to
the ones beginning at $\zeta_{2n+2}$.
\smallpagebreak {\it Global properties of ``a'', ..., ``d'' and
  ``e''.} These Stokes lines shown in Fig.~\ref{bem:St-lines} are
described by
\begin{Le}
  \label{BEM:le:Sl}
  The Stokes lines ``a'',..., ``d'' and ``e'' have the following
  properties:
%
\begin{floatingfigure}{6cm}
  \centering
  \includegraphics[bbllx=59,bblly=563,bburx=247,bbury=721,width=6cm]{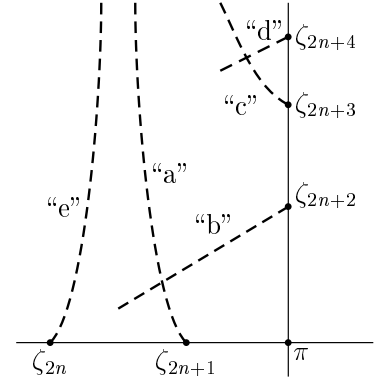}
  \caption{The Stokes lines}\label{bem:St-lines}
\end{floatingfigure}
                                %
  \begin{itemize}
  \item the Stokes lines ``a'' and ``e'' stay inside $\Pi$, are
    vertical and do not intersect one another;
  \item the Stokes line ``c'' stays between ``a'' and the line
    $\re\zeta=\pi$ (without intersecting them) and is vertical;
  \item before leaving $\Pi$, the Stokes lines ``b'' stays vertical,
    and it intersects ``a'' first at a point with positive imaginary
    part;
  \item before leaving $\Pi$, the Stokes lines ``d'' stays vertical
    and intersects ``c'' first above $\zeta_{2n+3}$, the beginning of
    ``c''.
  \end{itemize}
\end{Le}
\demo The main tool in the proof is Lemma~\ref{le:s-l-1}. Below,
we use it without referring to it.\\
First, we note that, as $\im\kappa\ne 0$ in $\Pi$, the Stokes lines
``a'', ``b'',..., ``e'' stay vertical as long as they stay in $\Pi$.\\
Second, one checks that the Stokes lines ``a'',..., ``d'' cannot leave
$\Pi$ by intersecting the line $\re\zeta=\pi$ (the right boundary of
$\Pi$), and that ``e'' cannot leave $\Pi$ by intersecting the
imaginary axis (the left boundary of $\Pi$). We check this property
for ``a'' only; the analysis of the other lines is similar. Note that
``a'' is tangent to the vector field $\overline{\kappa(\zeta)}-\pi n$.
Consider this vector field in a sufficiently small neighborhood of the
line $\re\zeta=\pi$ (in $\Pi$).  There, we have $\re\kappa>\pi n$ and
$\im\kappa>0$. Therefore, ``a'' can intersect the line $\re\zeta=\pi$
only when coming from above to the right. But, this is impossible as
``a'' begins at $\zeta_{2n+1}$ and stays vertical while in $\Pi$.\\
To prove the first point of Lemma~\ref{BEM:le:Sl}, it suffices to
check that ``a'' and ``e'' do not intersect one another while in
$\Pi$. Therefore, we note that both lines belong to the family
$\im\int^\zeta(\kappa-\pi n)d\zeta=\Const$. Therefore, by
Lemma~\ref{le:s-l-2}, while in $\Pi$, ``a'' and ``e'' either stay
disjoint or coincide. The second is impossible as they begin at
distinct points of the real line, and, inside $\Pi$, each of them
is smooth and vertical.\\
To prove the second point of Lemma~\ref{BEM:le:Sl}, it suffices to
check that ``a'' and ``c'' do not intersect one another while in
$\Pi$. Therefore, we note that ``a'' is tangent to the vector
field $v_1(\zeta)=\overline{\kappa(\zeta)}-\pi n$, and that ``c''
is tangent to the vector field
$v_2(\zeta)=\overline{\kappa(\zeta)}-\pi (n+1)$.  Pick
$\zeta_0\in\Pi$. As $\im\kappa(\zeta_0)>0$, both vectors
$v_1(\zeta_0)$ and $v_2(\zeta_0)$ are oriented downward, and $v_1$
is oriented to the right of $v_2$. So, to intersect ``a'', the
line ``c'' has to approach it going from left to right. But, this
is impossible as ``c'' begins to the right of ``a''.\\
To prove the third point of Lemma~\ref{BEM:le:Sl}, it suffices to
check that ``b'' can not leave $\Pi$ intersecting the segment
$[\zeta_{2n+1},\pi]$ of the real line. Therefore, we note that both
this segment and ``b'' belong to the family of lines
$\im\int^\zeta(\kappa-\pi (n+1))d\zeta=\Const$. So, by
Lemma~\ref{le:s-l-2}, ``b'' cannot intersect the segment
$(\zeta_{2n+1},\pi]$. Finally, a local analysis using the Implicit
Function Theorem shows that ``b'' can not contain the point
$\zeta_{2n+1}$.\\
The last point of Lemma~\ref{BEM:le:Sl} follows from the second one as
we have seen that, in $\Pi$, ``d'' goes downwards from $\zeta_{2n+3}$
and stays vertical; moreover, it cannot leave $\Pi$ intersecting
$\Pi$'s right boundary.\\
This completes the proof of Lemma~\ref{BEM:le:Sl} \qed
\smallpagebreak The analysis of the other Stokes lines situated
inside $\Pi$ is analogous to the one made in the proof of
Lemma~\ref{BEM:le:Sl}.
\section{Standard behavior of solutions}
\label{standard-behavior}
Here, we introduce the concept of the standard behavior of
solutions of~\eqref{family} studied in the framework of the
complex WKB method. Then, we consider the canonical domains, an
important example of domains on the complex plane of $\zeta$ where
one can construct solutions having standard behavior.
\subsection{Canonical Bloch solutions} \label{CBS}
To describe the asymptotic formulae of the complex WKB method, one
needs Bloch solutions of equation~\eqref{PSE} analytic in $\zeta$
on a given regular domain. We build them using the 1-form
$\Omega=\omega\,d\mathcal{E}$ introduced in
section~\ref{sec:Omega}.
\smallpagebreak Pick $\zeta_0$ a regular point. Let
$\mathcal{E}_0=\mathcal{E}(\zeta_0)$. Assume that $\mathcal{E}_0
\not\in P\cup Q\cup$. Let $U_0$ be small enough neighborhood of
${\mathcal E}_0$ and let $V_0$ be a neighborhood of $\zeta_0$ such
that $\mathcal {E}(V_0)\subset U_0$. In $U_0$, we fix a branch of the
function $\sqrt{k'(\mathcal{E})}$ and consider
$\psi_\pm(x,\mathcal{E})$, two branches of the Bloch solution
$\psi(x,\mathcal{E})$, and $\Omega_\pm$, two corresponding branches of
$\Omega$. For $\zeta\in V_0$, put
\begin{equation}
  \label{canonicalBS}
  \Psi_\pm (x,\zeta)=
  q(\mathcal{E})\,e^{\int_{\mathcal{E}_0}^\mathcal{E} \Omega_\pm}
  \psi_\pm (x,\mathcal{E}),\quad q(\mathcal{E})=\sqrt{k'(\mathcal{E})},\quad
  \mathcal{E}=\mathcal{E}(\zeta).
\end{equation}
The functions $\Psi_\pm$ are called the {\it canonical Bloch
solutions
  normalized at the point $\zeta_0$}.
\smallpagebreak The properties of the differential $\Omega$ imply that
the solutions $\Psi_\pm$ can be analytically continued from
$V_0$ to any regular domain $D$ containing $\zeta_0$. \\
One has
\begin{equation}
  \label{Wcanonical}
  w(\Psi_+(\cdot ,\zeta),\Psi_-(\cdot ,\zeta))=
  w(\Psi_+(\cdot ,\zeta_0),\Psi_-(\cdot ,\zeta_0))=
  k'(\mathcal{E}_0)w(\psi_+(x,\mathcal{E}_0),\psi_-(x,\mathcal{E}_0))
\end{equation}
This formula is proved in~\cite{Fe-Kl:02}. It shows that the Wronskian
is independent of $\zeta$ and depends only on the normalization point
$\zeta_0$ and the spectral parameter. As $\mathcal{E}_0\not \in
Q\cup\{E_l\}$, the Wronskian $w(\Psi_+(\cdot ,\zeta),\Psi_-(\cdot ,\zeta))$ is
non-zero.
\subsection{Solutions having standard asymptotic behavior}
\label{sec:standard-asymptotics}
Here, we discuss behavior of solutions to~\eqref{family}
satisfying~\eqref{consistency}. Speaking about a solution having {\it
  standard behavior}, first of all, we mean that this solution has the
asymptotics
\begin{equation}
  \label{stand:as}
  f=e^{\dsize\sigma\,\frac{i}{\varepsilon} \int^{\zeta}
    \kappa\,d\zeta}\, (\Psi_\sigma+o\,(1)),\quad \text{as}\quad
  \varepsilon\to 0,
\end{equation}
where $\sigma$ is either the sign ``$+$'' or ``$-$''. The solutions
constructed by the complex WKB method also have other important
properties. When speaking of standard behavior, we mean all these
properties. Let us formulate the precise definition.
\smallpagebreak Fix $E=E_0$. Let $D$ be a regular domain. Fix
$\zeta_0\in D$ so that $\mathcal{E}(\zeta_0)\not\in P\cup Q$. Let
$\kappa$ be a branch of the complex momentum continuous in $D$, and
let $\Psi_\pm$ be the canonical Bloch solutions defined on $D$,
normalized at $\zeta_0$ and indexed so that $\kappa$ be the
quasi-momentum for $\Psi_+$.
\begin{Def}
  \label{def:1}
  We say that, in $D$, a solution $f$ has standard behavior (or
  standard asymptotics) $f\sim\exp(\sigma\frac{i}{\varepsilon}
  \int^{\zeta} \kappa\,d\zeta)\cdot\Psi_\sigma$ if
  \begin{itemize}
  \item there exists $V_0$, a complex neighborhood of $E_0$, and $X>0$
    such that $f$ is defined and satisfies~\eqref{family}
    and~\eqref{consistency} for any $(x,\zeta,E)\in [-X,X]\times
    D\times V_0$;
  \item $f$ is analytic in $\zeta\in D$ and in $E\in V_0$;
  \item for any $K$, a compact subset of $D$, there is $V\subset V_0$,
    a neighborhood of $E_0$, such that, for $(x,\zeta,E)\in
    [-X,X]\times K\times V$, $f$ has the uniform
    asymptotic~\eqref{stand:as};
  \item this asymptotic can be differentiated once in $x$ retaining
    its uniformity properties.
\end{itemize}
\end{Def}
\subsection{Canonical domains}
\label{sec:Canonical-domains}
An important example of a domain where one can construct a solution
with standard asymptotic behavior is a canonical domain. Let us define
canonical domains and formulate one of the basic results of the
complex WKB method.
\subsubsection{Canonical lines}
\label{sec:canonical-lines}
We say that a piecewise $C^1$-curve $\gamma$ is {\it vertical} if
it intersects the lines $\{\im \zeta=\Const\}$ at non-zero angles
$\theta$, \ $0<\theta<\pi$. Vertical lines are naturally
parameterized by $\im\zeta$.
\smallpagebreak Let $\gamma$ be a $C^1$ regular vertical curve. On
$\gamma$, fix $\kappa$, a continuous branch of the complex
momentum.
\begin{Def}
  \label{def:2}
  The curve $\gamma$ is {\it canonical} if, along $\gamma$,
  \begin{enumerate}
  \item $ \im\int^{\zeta}\kappa d\zeta$ is strictly monotonously
    increasing with $\im\zeta$,
  \item $\im\int^{\zeta} (\kappa-\pi)d\zeta$ is strictly monotonously
    decreasing with $\im\zeta$.
  \end{enumerate}
\end{Def}
\noindent Note that canonical lines are stable under small
$\mathcal{C}^{1}$-perturbations.
\subsubsection{Canonical domains}
\label{sec:canonical-domains}
Let $K$ be a regular domain. On $K$, fix a continuous branch of
the complex momentum, say $\kappa$.
\begin{Def}
  \label{def:4}
  The domain $K$ is called {\it canonical} if it is the union of
  curves that are connecting two points $\zeta_1$ and $\zeta_2$
  located on $\partial K$ and that are canonical with respect to
  $\kappa$.
\end{Def}
\noindent One has
\begin{Th}[\cite{Fe-Kl:01b, Fe-Kl:02}]
  \label{T5.1}
  Let $K$ be a bounded domain canonical with respect to $\kappa$.  For
  sufficiently small positive $\varepsilon$, there exists $(f_\pm )$,
  two solutions of \eqref{family}, having the standard behavior in $K$
  so that
  \begin{equation*}
    f_\pm \sim \exp\left(\pm \frac{i}{\varepsilon}
      \int_{\zeta_{0}}^{\zeta} \kappa d\zeta\right)\Psi_\pm.
  \end{equation*}
  For any fixed $x\in\R$, the functions $f_\pm (x,\zeta)$ are analytic
  in $\zeta$ in $S(K):=\{Y_1<\im\zeta<Y_2\}$, the smallest strip
  containing $K$.
\end{Th}
\noindent In~\cite{Fe-Kl:01b}, we haven't discussed the dependence of
$f_\pm$ on $E$: we have proved Theorem~\ref{T5.1} without requiring
all the properties in the definition of the standard behavior; in
particular, we did not impose requirements in the behavior in $E$.
In~\cite{Fe-Kl:02}, we have formulated Definition~\ref{def:1} and
observed that $f_\pm$ (constructed in Theorem 5.1 of~\cite{Fe-Kl:01b})
have the standard behavior on $K$.
\noindent One easily calculates the Wronskian of the solutions
$f_\pm (x,\zeta)$ to get
\begin{equation}
  \label{W_of_f_pm}
  w(f_+,f_-)=w(\Psi_+,\Psi_-)+o(1).
\end{equation}
By~\eqref{Wcanonical}, for $\zeta$ in any fixed compact subset of
$K$ and $\varepsilon$ sufficiently small, the solutions $f_\pm $
are linearly independent.
\section{Local canonical domains}
\label{local-constructions}
In this section, following~\cite{Fe-Kl:02}, we present a simple
approach to find ``local'' canonical domains. We then give an example
of a local canonical domain for the case of
$W(\zeta)=\alpha\cos\zeta$.
\smallpagebreak Below, we assume that $D$ is a regular domain, and
that $\kappa$ is a branch of the complex momentum analytic in $D$. A
{\it segment} of a curve is a connected, compact subset of that curve.
\subsection{General constructions}
\label{lCD}
\subsubsection{Definition}
\label{sec:definition}
Let $\gamma\subset D$ be a line canonical with respect to $\kappa$.
Denote its ends by $\zeta_{1}$ and $\zeta_{2}$.  Let a domain
$K\subset D$ be a canonical domain corresponding to the triple
$\kappa$, $\zeta_{1}$ and $\zeta_{2}$. If $\gamma\subset K$,
then, $K$ is called a canonical domain {\it enclosing} $\gamma$.\\
As any line close enough in $C^{1}$-norm to a canonical line is
canonical, one has
\begin{Le}[\cite{Fe-Kl:02}]
  \label{LCD}
  One can always construct a canonical domain enclosing any given
  canonical curve.
\end{Le}
\noindent Canonical domains, whose existence is established using
this lemma are called {\it local}.
\subsubsection{Constructing canonical curves}
\label{pre-cl}
To construct a local canonical domain we need a canonical line to
start with. To construct such a line, we first build
pre-canonical lines made of some ``elementary'' curves.\\
Let $\gamma\subset D$ be a vertical curve.  We call $\gamma$ {\it
  pre-canonical} if it is a finite union of bounded segments of
canonical lines and/or lines of Stokes type. In section~\ref{ex:lcd},
we shall see that, in practice, pre-canonical lines are easy to find.
One has
\begin{Pro}[\cite{Fe-Kl:02}]
  \label{pro:pcl:1}
  Let $\gamma$ be a pre-canonical curve. Denote the ends of $\gamma$
  by $\zeta_a$ and $\zeta_b$.\\
  Fix $V\subset D$, a neighborhood of $\gamma$ and $V_a\subset D$, a
  neighborhood of $\zeta_a$. Then, there exists a canonical line
  $\tilde\gamma\subset V$ connecting the point $\zeta_b$ to some point
  in $V_a$.
\end{Pro}
\noindent Proposition~\ref{pro:pcl:1} tells us that, arbitrarily close
to any pre-canonical line, one can construct a canonical line.
\subsection{Example: constructing a canonical line for
  $\mathbf{W(\zeta)=\alpha\cos\zeta}$}
\label{ex:lcd}
We again turn to the case $W(\zeta)=\alpha\cos\zeta$ and assume
that all the gaps of the periodic operator~\eqref{Ho} are open,
and that $E$ satisfies condition (C). Recall that, in this case,
the branch points and the Stokes lines were studied in
section~\ref{ex:st-ln} (see Fig.~\ref{bem:St-lines}). In the
sequel, we assume that $n$ in (C) is even. The case $n$ odd is
treated similarly; only some details differ.
\smallpagebreak Let $Y>0$. We now explain the details of the
construction of a canonical line going from $\{\im\zeta=Y\}$ to
$\{\im\zeta=-Y\}$ (where $Y$ can be taken arbitrarily large).
%
%
\begin{floatingfigure}{9cm}
  \centering
  \includegraphics[bbllx=71,bblly=545,bburx=375,bbury=721,width=9cm]{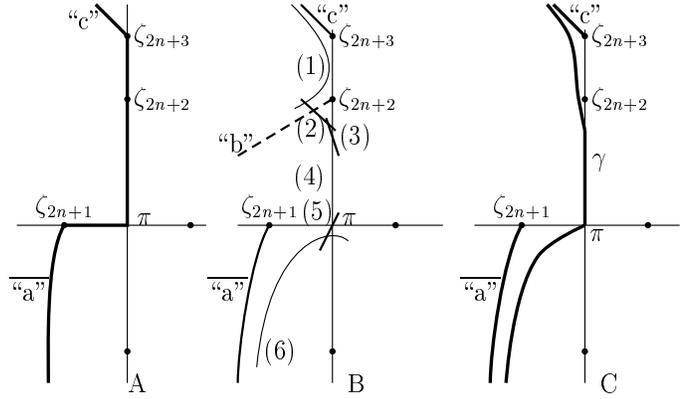}
  \caption{The construction of the canonical line}\label{BEM:fig:pcl}
\end{floatingfigure}
%
\noindent By Lemma~\ref{LCD} and Theorem~\ref{T5.1}, this
canonical line enables us to construct a solution of~\eqref{family},
analytic in $S_Y=\{|\im\zeta|< Y\}$. In later sections, we shall study
the global asymptotics of this solution.
\smallpagebreak \noindent To find a canonical line, we first find a
pre-canonical line. Consider the curve $\beta$ which is the union of
the Stokes line $\overline{\text{``a''}}$ (symmetric to ``a'' with
respect to $\R$), the segment $[\zeta_{2n+1},\pi]$ of the real line,
the segment $[\pi,\zeta_{2n+3}]$ of the line $\re\zeta=\pi$ and the
Stokes line ``c'', see Fig.~\ref{BEM:fig:pcl}, part A. We now
construct $\alpha$, a pre-canonical line close to the line $\beta$.\\
When speaking of $\kappa_p$ along $\alpha$, we mean the branch of the
complex momentum obtained from $\kappa_p$ by analytic continuation
along $\alpha$ (the analytic continuation can be done by means of
formula~\eqref{eq:14} relating the complex momentum to the Bloch
quasi-momentum).
\vskip.1cm\noindent Actually, the line $\alpha$ is pre-canonical with
respect to the branch of the complex momentum related to $\kappa_p$ by
the formula:
\begin{equation}
  \label{kappaPkappa}
  \kappa= \kappa_p-\pi n.
\end{equation}
Note that, as $n$ is even, $\pi n\in 2\pi\Z$, and $\kappa$ indeed is a
branch of the complex momentum; it is the natural branch for the
points $\zeta_{2n}$, $\zeta_{2n+1}$, $\zeta_{2n+2}$ and
$\zeta_{2n+3}$. We prove
\begin{Pro}
  \label{pro:2}
  Fix $\delta>0$ and $Y> \zeta_{2n+3}$. In the $\delta$-neighborhood
  of $\beta$, there exists $\alpha$, a line pre-canonical with respect
  to the branch $\kappa$ having the following properties:
  \begin{itemize}
  \item at its upper end, one has $\im\zeta>Y$;
  \item at its lower end, one has $\im\zeta<-Y$;
  \item it goes around the branch points of the complex momentum as
   the curve $\gamma$
    shown in Fig.~\ref{BEM:fig:pcl}, part C;
  \item it contains a canonical line which stays in $\Pi$, goes from a
    point in $\Pi$ to the line $\re\zeta=\pi$ and, then, continues
    along this line until it intersects the real line.
  \end{itemize}
\end{Pro}
\demo In Fig.~\ref{BEM:fig:pcl}, part B, we illustrate the
construction of $\alpha$.  In this figure, we show the ``elementary''
segments (1), (2),..., (6) we use to build $\alpha$. Let us describe
these segments in details. Below, we denote by $V_\delta$ the left
hand side of the $\delta$-neighborhood of the line $\beta$.
\\
{\bf The segment (1).\/} It is a segment of $l_1$, a line of Stokes
type $\im\int^\zeta(\kappa-\pi)d\zeta=\Const$. Note that the Stokes
lines ``b'', $[\zeta_{2n+2},\zeta_{2n+3}]$ and ``c'' are also level
curves of the harmonic function $\im\int^\zeta(\kappa-\pi)d\zeta$. So,
we choose $l_1$ so that it go to the left of these three Stokes lines
as close to them as needed (inside a given compact set).\\
Note that the part of $l_1$ situated in $\Pi$ is vertical.\\
We choose $l_1$ and $a_1$ and $a_2$, the upper and the lower ends of
the segment (1) so that
\begin{itemize}
\item the segment (1) be situated in $V_\delta$;
\item the segment (1) be situated in $\Pi$ and, thus, be vertical;
\item $\im a_1>Y$, and $\im a_2<\im\zeta_{2n+2}$.
\end{itemize}
The precise choice of $a_2$ will be described later.\\
{\bf The segment (2).\/} It is a segment of $l_2$, a line of Stokes type
$\im\int^\zeta\kappa d\zeta=\Const$ which contains $a$, a point of
the line $\re\zeta=\pi$ such that $0<\im a<\im\zeta_{2n+2}$.\\
Let us show that 
\begin{description}
\item[(a)] the line $l_2$ is horizontal (i.e. parallel to the real
  line) at the point $a$;
\item[(b)] having entered in $\Pi$ at $a$, the line $l_2$ becomes
  vertical and goes upward;
\item[(c)] the line $l_2$ stays vertical in $\Pi$;
\item[(d)] above $a$, it intersects the Stokes line ``b'' staying
  inside $\Pi$.
\end{description}
Therefore, note that the line $l_2$ is tangent to the vector field
$\overline{\kappa(\zeta)}$.  As $a\in z_{n+1}$, one has
$\im\kappa(a)=0$ which implies (a). In $\Pi$, near the line
$\re\zeta=\pi$ and above $a$, one has $\im\kappa>0$ and $0<\re\kappa$.
This implies (b). As $\im\kappa\ne 0$ inside $\Pi$, we get (c).
Finally, if $l_2$ does not intersect ``b'', it has to come back to the
line $\re\zeta=\pi$. It can come to this line going downwards to the
right. This is impossible in view of (c).\\
We assume that $l_1$ is chosen close enough to the Stokes lines ``c'',
$[\zeta_{2n+2},\zeta_{2n+3}]$ and ``b'' so that $l_2$ intersects $l_1$
(after having intersected ``b'').\\
As $a_2$, the lower end of the segment (1) and the upper end of the
segment (2), we choose  the intersection point.\\
We choose $a_3$, the lower end of the segment (2), between $a_2$ and
$a$. Then,
\begin{itemize}
\item the segment (2) stays inside $\Pi$, and, thus, is vertical.
\end{itemize}
We choose $a$ so to close $\zeta_{2n+2}$ that
\begin{itemize}
\item the segment (2) is inside $V_\delta$.
\end{itemize}
We describe the precise choice of $a_3$ later.\\
{\bf The segment (3), (4) and (5).\/} They form a canonical line. To
describe them, consider $l_3$ an internal subsegment of the segment
$(\overline{\zeta_{2n+2}},\zeta_{2n+2})\subset\{\re\zeta=\pi\}$. We
assume that $l_3$ begins above the point $a$ and ends below the real
line.\\
The segment $l_3$ is a canonical line with respect to $\kappa$.
Indeed, $(\overline{\zeta_{2n+2}},\zeta_{2n+2})$ is a part of a
connected component of the pre-image (with respect to $\mathcal E$) of
the $(n+1)$-st spectral band of the periodic operator~\eqref{Ho}. So,
along $l_3$, one has $0<\kappa<\pi$. This implies that $l_3$ is a
canonical line. \\
Recall that any line $C^1$-close enough to a canonical line is
canonical. This enables us to choose the ``elementary segments (3),
(4) and (5) so that
\begin{itemize}
\item they form a canonical line;
\item the segment (3) connects in $\Pi\cap V_\delta$ the point $a_3$,
  an internal point of $l_2$, to $a_4$, a point of $l_3$ such that
  $\im a_4>0$;
\item the segment (4) goes along $l_3$ from $a_4$ to $\pi$;
\item the segment (5) connects in $\overline\Pi$ (the domain symmetric
  to $\Pi$ with respect to $\R$) the point $\pi$ to $a_6$, a point of
  $\overline\Pi$.
\end{itemize}
We describe the precise choice of $a_6$ later.\\
{\bf The segment (6).\/} It is a segment of $l_4$, a line of Stokes
type $\im\int^\zeta\kappa d\zeta=\Const$ containing  the point $a_6$.\\
To describe $l_4$ more precisely, consider the Stokes line
$[\zeta_{2n+1},\pi]$ and the Stokes line $\overline{\text{``a''}}$
symmetric to ``a'' with respect to the real line. As $l_4$, they are
level curves of the function $\im\int^\zeta\kappa d\zeta$.  So, we can
and do construct $l_4$ and (6) so that $l_4$ go below
$[\zeta_{2n+1},\pi]$ and to the left of $\overline{\text{``a'' }}$ as
as close to these lines as needed (inside any given compact set).\\
We choose $a_6$ and $a_7$ the upper and the lower ends of (6) so
that
\begin{itemize}
\item the segment (6) be in the $\delta$-neighborhood of $\beta$;
\item the segment (6) be inside $\overline\Pi$ (and, so, be vertical);
\item $a_7$ be below the line $\im\zeta=-Y$.
\end{itemize}
{\bf The curve $\alpha$.\/} It is made of the ``elementary'' segments
(1) --- (6); it is vertical and, by construction, consists of segments
of lines of Stokes type and a line canonical with respect to $\kappa$.
So, it is pre-canonical with respect to $\kappa$.  By construction, it
has all the properties described in Proposition~\ref{pro:2}.\qed
\smallpagebreak{\it Remark to the proof of Proposition~\ref{pro:2}.\/}
The lines $l_2$ and $l_4$ are not vertical at the points of
intersection with the line $\re\zeta=\pi$ (as there
$\im\kappa=\im\kappa_0=0$). The ``elementary'' segments (3) and (5)
were included into $\alpha$ to make it vertical.
\smallpagebreak Now, construct a canonical line close to $\alpha$.
One obtains:
\begin{Pro}
  \label{le:ex-c-l}
  In arbitrarily small neighborhood of the pre-canonical line
  $\alpha$, there exists a canonical line $\gamma$ which has all the
  properties of the line $\alpha$ listed in Proposition~\ref{pro:2}.
\end{Pro}
\demo Denote by $\gamma_0$ the canonical line mentioned in the fourth
point of Proposition~\ref{pro:2}. The proof of
Proposition~\ref{le:ex-c-l} consists of two steps. Fix $V$, a
neighborhood of $\alpha$. First, using Proposition~\ref{pro:pcl:1}, we
construct $\gamma_a$ and $\gamma_b$, two canonical lines situated in
$V$ and such that:
\begin{enumerate}
\item $\gamma_a$ connects the upper end of $\gamma_0$ to a point
  situated above the line $\im\zeta=Y$;
\item $\gamma_b$ connects the lower end of $\gamma_0$ to a point
  situated below the line $\im\zeta=-Y$.
\end{enumerate}
In the second step, one considers the line $\tilde\gamma=\gamma_a
\cup\gamma_0\cup\gamma_b$. It is vertical and consists of three
canonical lines. To get the desired canonical line, one smoothes
$\tilde \gamma$ out near the ends of $\gamma_0$.  \qed
%


%
\section{The main continuation principles}
\label{sec:m-t}
\noindent This section is devoted to the main continuation
principles, namely, the Rectangle Lemma, the Adjacent Canonical Domain Principle
and the Stokes Lemma. In section~\ref{ex:cont-diag}, we give a
detailed example explaining how to use them.
\smallpagebreak In the sequel, a set is called {\it constant} if it is
independent of $\varepsilon$.
\subsection{The Rectangle Lemma: asymptotics of increasing solutions}
\label{sec:rectangle-lemma}
Fix $\eta_m<\eta_M$. Define the strip $S=\{\zeta\in\C:\
\eta_m\leq\im\zeta\leq\eta_M\}$. Let $\gamma_1$ and $\gamma_2$ be two
vertical lines such that $\gamma_1\cap \gamma_2=\emptyset$. Assume
that both lines intersect the strip $S$ at the lines $\im\zeta=\eta_m$
and $\im\zeta=\eta_M$, and that $\gamma_1$ is situated to the left of
$\gamma_2$.\\
Consider $R$, the compact set bounded by $\gamma_1$, $\gamma_2$ and
the boundaries of $S$. Let $D$=$R\setminus(\gamma_1\cup\gamma_2)$.
\smallpagebreak One has
\begin{Le}[The Rectangle Lemma~\cite{Fe-Kl:01b}]
  \label{Rectangle}
  Fix $E=E_0$. Assume that the ``rectangle'' $R$ is
  regular. Let $f$ be a solution of~\eqref{family}
  satisfying~\eqref{consistency}.  Then, for sufficiently small
  $\varepsilon$, one has
  \begin{description}
  \item[1] If $\im\kappa<0$ in $D$, and if, in a neighborhood of
    $\gamma_1$, $f$ has standard behavior $f\sim
    \exp(\frac{i}{\varepsilon} \int^{\zeta}\kappa d\zeta)
    \cdot \Psi_+$, then, it has standard behavior in a constant domain
    containing the ``rectangle'' $R$.
  \item[2] If $\im\kappa>0$ in $D$, and if, in a neighborhood of
    $\gamma_2$, $f$ has standard behavior $f\sim
    \exp(\frac{i}{\varepsilon} \int^{\zeta}\kappa d\zeta)
    \cdot \Psi_+$, then, it has the standard behavior in a constant domain
    containing the ``rectangle'' $R$.
  \end{description}
\end{Le}
\noindent Lemma~\ref{Rectangle} was proved in~\cite{Fe-Kl:01b} where
one can find more details and references.
\subsection{Estimates of ``decreasing'' solutions}
\label{sec:lempro}
The Rectangle Lemma allows us to ``continue'' standard behavior as
long as the leading term increases along a horizontal line. If the
leading term decreases, then, in general, we can only estimate the
solution, but not get an asymptotic behavior.
\begin{Le}[\cite{Fe-Kl:01b}]
  \label{lempro}
  Fix $E=E_0$.  Let $\zeta_1$ and $\zeta_2$ be fixed points such that
  \begin{enumerate}
  \item $\im\zeta_1=\im\zeta_2$;
  \item $\re\zeta_{1}<\re\zeta_{2}$;
  \item the segment $[\zeta_1,\zeta_2]$ of the line
    $\im\zeta=\im\zeta_{1}$ is regular.
  \end{enumerate}
  Fix a continuous branch of $\kappa$ on $[\zeta_1,\zeta_2]$.  Assume
  that $\im\kappa(\zeta)>0$ on the segment $[\zeta_1,\zeta_2]$.  Let
  $\psi$ be a solution having standard behavior $\psi\sim
  e^{\frac{i}{\varepsilon}\int_{\zeta_1}^{\zeta}\kappa d\zeta}
  \Psi_+$ in a neighborhood of $\zeta_1$.\\
  Then, there exists $C>0$ such that, for $\varepsilon$ sufficiently
  small, one has
  \begin{equation}
    \label{eq:8} \left\vert\frac{d\psi}{dx}(x,\zeta)\right\vert+\vert
    \psi(x,\zeta)\vert\leq
    Ce^{\dsize\frac1{\varepsilon}\int_{\zeta_1}^{\zeta}
      \vert\im\kappa\vert d\zeta}, \quad \zeta\in [\zeta_1,\zeta_2].
  \end{equation}
  uniformly in $E$ in a constant neighborhood of $E_{0}$.
\end{Le}
\noindent One also has the ``symmetric'' statement when $\im\kappa<0$
and $f$ has standard behavior $f\sim e^{\frac{i}{\varepsilon}
  \int_{\zeta_2} ^{\zeta}\kappa d\zeta}\Psi_+$ in a neighborhood of
$\zeta_{2}$.
\subsection{The Adjacent Canonical Domain Principle}
\label{sec:addCD}
The estimate we obtained in Lemma~\ref{lempro} can be far from
optimal: the estimate only says that the solution $\psi$ cannot
increase faster than $\dsize\exp\left(\frac1{\varepsilon}
  \int_{\zeta_1}^{\zeta}\vert\text{Im}\kappa\vert d\zeta\right)$
whereas it can, in fact, decrease along $[\zeta_{1},\zeta_{2}]$. The
Adjacent Canonical Domain Principle enables us to justify the asymptotics of
decreasing solution.
\subsubsection{The statement}
\label{sec:adjac-doma-princ}
Let $\gamma$ be a segment of a vertical curve.  Let $S$ be the
smallest strip of the form $\{C_1\le \im\zeta\le C_2\}$ containing
$\gamma$.
\begin{Def}
  \label{def:7}
  Let $U\subset S$ be a regular domain. We say that $U$ is adjacent to
  $\gamma$ if $\gamma\subset \partial U$.
\end{Def}
\noindent We have proved
\begin{Pro}[The Adjacent Canonical Domain Principle~\cite{Fe-Kl:01b}]
  \label{AddCD}
  Let $\gamma$ be a segment of a canonical line. Assume that a
  solution $f$ has standard behavior in a domain adjacent to $\gamma$.
  Then, $f$ has the standard behavior in any bounded canonical domain
  enclosing $\gamma$.
\end{Pro}
\noindent To apply the Adjacent Canonical Domain Principle, one needs to
describe canonical domains enclosing a given canonical line.
Therefore, we now discuss such domains.
\subsubsection{General description of enclosing canonical domains}
\label{sec:gener-descr-encl}
We work in a regular domain $D$. We assume that $\kappa$ is a branch
of the complex momentum analytic in $D$. We discuss only lines
pre-canonical (e.g. canonical lines or lines of Stokes type) with
respect to $\kappa$.\\
The general tool for constructing the enclosing canonical domains is
\begin{Pro}[\cite{Fe-Kl:01b}]
  \label{pro:pcl:2}
  Let $\gamma$ be a segment of a canonical line.  Assume that
  $K\subset D$ is a simply connected domain containing $\gamma$
  (without its ends). The domain $K$ is a canonical domain enclosing
  $\gamma$ if and only if it is the union of pre-canonical lines
  obtained from $\gamma$ by replacing some of $\gamma$'s internal
  segments by pre-canonical lines.
\end{Pro}
\subsubsection{Adjacent canonical domains}
\label{sec:adjac-canon-doma}
It can be quite difficult to find the ``maximal'' canonical domain
enclosing a given canonical line. In practice, it is much more
convenient to use ``simple'' canonical domains obtained with
Lemma~\ref{parallelogram-le}. To make the formulation of this result
more transparent, we first list elementary properties of canonical
lines and lines of Stokes type.
\smallpagebreak The following lemma is a simple corollary of
Lemma~\ref{le:s-l-1} and of the definition of canonical lines:
\begin{Le}
  \label{s-l:prop}
  One has
  \begin{itemize}
  \item If $\im\kappa\ne 0$ in a regular domain $U$, then, all the
    lines of Stokes type inside $U$ are vertical.
  \item Let $\gamma$ be a canonical curve. Then, any line of Stokes
    type intersecting $\gamma$ intersects it transversally.
  \item Let $\gamma$ be a canonical curve. Any of its internal segment
    is a canonical curve. Moreover, $\gamma$ is an internal segment of
    another canonical curve.
  \item Let $\gamma$ be a canonical curve.  Let $U$ be a domain
    adjacent to $\gamma$. Assume that $\im\kappa\ne 0$ in $U$.
    Consider two lines of Stokes type (from the two different
    families) containing $\zeta_0$, an internal point of $\gamma$. In
    $U$, one of these lines goes upward from $\zeta_0$, and the second
    one is going downward from $\zeta_0$.
  \end{itemize}
\end{Le}
\noindent Now, we can formulate the statement about ``simple''
canonical domains.
\begin{Le}[The Trapezium Lemma]
  \label{parallelogram-le}
  Let $\gamma_0$ be a segment of a canonical line. Let $U$ be a domain
  adjacent to $\gamma$, a canonical line containing $\gamma_0$ as an
  internal segment. Assume that $\im\kappa\ne 0$ in $U$. Denote by
  $\sigma_u\subset U$ (resp. $\sigma_d\subset U$), the line of Stokes
  type starting from the upper (resp. lower) end of $\gamma_0$ and
  going downwards (resp. upwards). One has:
  \begin{itemize}
  \item Pick $\tilde\gamma$, one more canonical line not intersecting
    $\gamma_0$. If $T\subset U$ is the simply connected domain bounded
    by $\sigma_u$, $\sigma_d$, $\gamma_0$ and $\tilde\gamma$, then,
    $T$ is a part of a canonical domain enclosing $\gamma_0$.
  \item Assume that $\sigma_u$ intersects $\sigma_d$. Let $T\subset U$
    be the simply connected domain bounded by $\sigma_u$, $\sigma_d$
    and $\gamma_0$. Then, $T$ is a part of canonical domain enclosing
    $\gamma_0$.
  \end{itemize}
\end{Le}
\noindent We prove this lemma in section~\ref{proof:T-l}.
\smallpagebreak To use the second part of the Trapezium Lemma, one has
to check that $\sigma_d$ and $\sigma_u$ intersect. Therefore, in
practice, one uses
\begin{Le}
  \label{le:intersections}
  Inside any regular domain, a canonical line and a line of Stokes
  type can intersect at most once. Two line of Stokes type from the
  different families also can intersect at most once. Two lines of
  Stokes type from the same family either are disjoint or they
  coincide.
\end{Le}
\noindent The first two statements of this lemma easily follow from
the definitions. The last one follows from Lemma~\ref{le:s-l-2}. We
omit the elementary details.
\subsection{The Stokes Lemma}
\label{sec:St-Lemme}
\smallpagebreak {\it Notations and assumptions.\/} Assume that
$\zeta_0$ is a branch point of the complex momentum such that
$W'(\zeta_0)\ne 0$.\\
There are  three Stokes lines beginning at $\zeta_0$. The angles
between them at $\zeta_0$ are equal to $2\pi/3$. We denote these
lines by $\sigma_1$, $\sigma_2$ and $\sigma_3$ so that $\sigma_1$
is vertical at $\zeta_0$ (see Fig.~\ref{stokes:fig:1}).\\
Let $\tilde\sigma_1$ be a (compact) segment of $\sigma_1$ which
begins at $\zeta_0$, is vertical and contains only one branch
point, i.e. $\zeta_0$.\\
Let $V$ be a neighborhood of $\tilde\sigma_1$. Assume that $V$ is
so small that the Stokes lines $\sigma_1$, $\sigma_2$ and
$\sigma_3$ divide it into three sectors. We denote them by $S_1$,
$S_2$ and $S_3$ so that $S_1$ be situated between $\sigma_1$ and
$\sigma_2$, and the sector $S_2$ be between $\sigma_2$ and
$\sigma_3$ (see Fig.~\ref{stokes:fig:1}).
\smallpagebreak{\it The statement.\/} We prove
\begin{Le}[The Stokes Lemma]
  \label{st-lm}
  Let $V$ be sufficiently small. Let $f$ be a solution that has
  standard behavior $f\sim e^{\frac i\varepsilon\int^\zeta\kappa
    d\zeta}\Psi_+$ inside the sector $S_1\cup \sigma_2\cup S_2$ of
  $V$. Moreover, assume that, in $S_1$ near $\sigma_1$, one has
  $\im\kappa(\zeta)>0$ if $S_1$ is to the left of $\sigma_1$ and
  $\im\kappa(\zeta)<0$ otherwise.
  Then, $f$ has standard behavior inside $V\setminus\sigma_1$, the
  leading term of the asymptotics being obtained by analytic
  continuation from $S_1\cup\sigma_2\cup S_2$ to $V\setminus\sigma_1$.
\end{Le}
\noindent We prove the Stokes Lemma in section~\ref{Principles-proof}.

%

%
\section{Computing  a continuation diagram: an example}
\label{ex:cont-diag}
We again consider the case of $W(\zeta)=\alpha\cos\zeta$, assuming
that all the gaps of the periodic operator are open and that $E$
satisfies hypothesis (C). For sake of definiteness, we assume
additionally that $n$ in (C) is even. In the case of $n$ odd, one
obtains similar results. In section~\ref{ex:lcd}, we have constructed
a canonical line $\gamma$ going around the branch points of the
complex momentum as in Fig.~\ref{BEM:fig:pcl}, part C. Its properties
are described by Proposition~\ref{le:ex-c-l}. By means of
Theorem~\ref{T5.1}, we construct $f$, a solution having the standard
behavior $f\sim \exp\left(\frac i\varepsilon\int_{\pi}^\zeta \kappa
  d\zeta\right)\cdot \Psi_+$ on $K$, a local canonical domain enclosing
$\gamma$. Here, $\kappa$ is the branch of the complex momentum defined
by~\eqref{kappaPkappa}.  The solution $f$ is analytic in
$S(K):=\{Y_1<\im\zeta<Y_2\}$, the smallest strip containing $K$.  In
this section, using our continuation tools, we study the asymptotic
behavior of $f$ in $S=S(K)$ outside $K$.
%
%
\begin{floatingfigure}{4.5cm}
  \centering
  \includegraphics[bbllx=70,bblly=600,bburx=176,bbury=721,width=4.5cm]{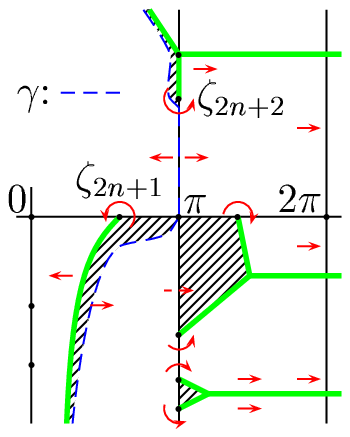}
  \caption{The continuation diagram}\label{BEM:fig:dp:1}
\end{floatingfigure}
%
\smallpagebreak Let $D=\{|\im\zeta|\le Y,\,\,\,0<\re \zeta<2\pi\}$
($Y$ is as in Proposition~\ref{pro:2}). Consider also $D'$, the domain
obtained from $D$ by cutting it along segments of Stokes lines and
along lines $\re \zeta=\Const$ as shown in Fig.~\ref{BEM:fig:dp:1}.
Note that, we have cut away (i.e. $D'$ does not contain) the part of
$D$ situated to right of the Stokes line ``c'' (see
Fig.~\ref{bem:St-lines}). The domain $D'$ is simply connected; thus,
both the branch $\kappa$ and the leading term of the standard
asymptotics of $f$ can be analytically continued on $D'$ in a unique
way. Using the continuation principles, we prove
\begin{Pro}
  \label{f:global}
  If $\delta$ (from Proposition~\ref{pro:2}) is chosen sufficiently\\
  small, then, inside $D'$, the solution $f$ has the standard behavior
  \begin{equation*}
    \hskip-4cm
    f\sim \exp\left(\frac i\varepsilon\int_{\pi}^\zeta \kappa
      d\zeta\right)\,\Psi_+.
  \end{equation*}
\end{Pro}
\noindent The rest of this section devoted to the proof of this
proposition. The proof is naturally divided into ``elementary'' steps.
In each step, applying just one of the three continuation tools (i.e.
the Rectangle Lemma, the Adjacent Canonical Domain Principle and the
Stokes Lemma), we extend the continuation diagram, justifying the
standard behavior of $f$ on a larger subdomain of $D'$.
Fig.~\ref{BEM:fig:dp:1} shows where we use each of the continuation
principles. The full straight arrows indicate the use of the Rectangle
Lemma, the circular arrows, the use of the Stokes Lemma, and, the
dashed arrows and the hatched zones, the use of the Adjacent Canonical
Domain Principle. When proving Proposition~\ref{f:global}, one repeats
the same arguments quite often. So, we explain in details only the
first few steps of the proof to show how to use each of the
continuation tools.
\subsection{Behavior of $\mathbf{f}$ between the lines $\mathbf{\gamma}$
  and $\mathbf{\beta}$: applying the Adjacent Canonical Domain Principle}
\label{sec:behavior-f-between}
Recall that $\gamma$ first goes downwards staying to the left of
$\beta$, and, then, $\gamma$ and $\beta$ meet at a point $a$,
$0<\im a<\im\zeta_{2n+2}$. They coincide up to a point $b$, $\im
b<0$. Here, by means of the Adjacent Canonical Domain Principle, we prove
that $f$ has the standard behavior inside a subdomain of $D$
situated above $a$ between $\gamma$ and $\beta$. Our strategy is
the following. First, we use the Trapezium Lemma,
Lemma~\ref{parallelogram-le}, to describe a part of a canonical
domain enclosing to the upper part of $\gamma$, and, then, we use
the Adjacent Canonical Domain Principle.
\subsubsection{Describing $U$, $\gamma_0$, $\sigma_u$ and $\sigma_d$}
\label{sec:describing-u-gamma_0}
Let us describe the domain $U$ and the curves $\gamma_0$, $\sigma_d$
and $\sigma_u$ needed to apply Lemma~\ref{parallelogram-le}.
\smallpagebreak {\it The domain $U$.\/} It is the domain bounded by
$\beta$, $\gamma$ and the line $\im\zeta=\Const$ containing the upper
end of $\gamma$. Inside $U$, one has $\im\kappa>0$.
\smallpagebreak {\it The line $\sigma_u$.\/} As $\sigma_u$, we take
the line which intersects $\beta$ at $\tilde \zeta_u$, the point with
the imaginary part equal to $Y$, and belongs to the family
$\im\int^{\zeta}\kappa d\zeta=\Const$.  Recall that $\gamma$ is
constructed in the $\delta$-neighborhood of $\beta$ where $\delta$ can
be fixed arbitrarily small. One has
\begin{Le}
  \label{le:2}
  The line $\sigma_u$ enters $U$ at the point $\tilde\zeta_u$ and goes
  upwards. If $\delta$ is sufficiently small, then, $\sigma_u$
  intersects $\gamma$ at $\zeta_u$, an internal point of $\gamma$.
\end{Le}
\demo Recall that $Y>\im\zeta_{2n+3}$, and that, above $\zeta_{2n+3}$,
$\beta$ coincides with the Stokes line ``c'' tangent to the vector
field $\overline{\kappa(\zeta)-\pi}$. The line $\sigma_u$ is tangent
to the vector field $\overline{\kappa(\zeta)}$. One has
$\im\kappa(\tilde \zeta_u)>0$. Therefore, at $\tilde\zeta_u$, the
tangent vector to $\beta$ (oriented upwards) is directed to the left
with respect to the tangent vector to $\sigma_u$ (oriented upwards).
So, $\sigma_u$ enters $U$ at $\tilde\zeta_u$ going upwards. As
$\im\kappa\ne 0$ in $U$, $\sigma_u$ stays vertical (in $U$). As
$\sigma_u$ is independent of $\delta$, if $\delta$ is sufficiently
small, $\sigma_u$ intersects $\gamma$. This completes the proof of
Lemma~\ref{le:2}.\qed
\smallpagebreak {\it The line $\sigma_d$.\/} It is the line which
intersects $\beta$ at $\tilde \zeta_d$, a point such that $\im
a<\tilde\zeta_d<\im\zeta_{2n+2}$, and belongs to the family
$\im\int^{\zeta}(\kappa-\pi) d\zeta=\Const$. One has
\begin{Le}
  \label{le:3}
  The line $\sigma_d$ enters $U$ at $\tilde\zeta_d$, goes downwards
  and then, staying in $U$, it intersects $\gamma$ at a point
  $\zeta_d$. This point can be made arbitrarily close to $a$ by
  choosing $\tilde\zeta_d$ sufficiently close to $a$.
\end{Le}
\demo Recall that the segment $[\pi,\zeta_{2n+2}]$ of the line
$\re\zeta=\pi$ belongs to the pre-image (by $\mathcal E$) of the
$(n+1)$-st spectral band of the periodic operator. So,
$\im\kappa=0$ and $0<\re\kappa<\pi$ on $[\pi,\zeta_{2n+2}]$.
Moreover, in $U$, one has $\im\kappa>0$. As $\sigma_d$ is tangent
to the vector field $\overline \kappa-\pi$, arguing as usual, we
deduce from these properties of $\kappa$  that 
\begin{enumerate}
\item $\sigma_d$ is orthogonal to $\beta$ at $\tilde\zeta_d$, enters
  $U$ at this point;
\item having entered $U$, it goes downwards and stays vertical while
  in $U$;
\item it leaves $U$ intersecting $\gamma$.
\end{enumerate}
Being an integral curve of a smooth vector field, $\sigma_d$
intersects $\gamma$ as close to $a$ as desired provided that
$\tilde\zeta_d$ is sufficiently close to $a$. This completes the proof
of Lemma~\ref{le:3}.\qed
\smallpagebreak {\it The line $\gamma_0$.\/} We choose $\delta$ so
that $\sigma_u$ intersect $\gamma$. Then, $\gamma_0$ is the segment of
$\gamma$ between its intersections with $\sigma_d$ and $\sigma_u$.
\subsubsection{Describing the curve $\tilde\gamma$}
\label{sec:descr-curve-tild}
We shall use the first variant of the Trapezium Lemma (i.e. the first
point of Lemma~\ref{parallelogram-le}). Let us describe the canonical
line $\tilde\gamma$ needed to apply it. In
Proposition~\ref{le:ex-c-l}, we have constructed $\gamma$ by means of
Proposition~\ref{pro:2}. In the same way, we can construct another
canonical line situated arbitrarily close to $\beta$. So, we can
assume that it is strictly between $\gamma_0$ and $\beta$. This
canonical line is the one we use as $\tilde\gamma$.\\
As $\sigma_u$ and $\sigma_d$ intersect $\gamma$ and $\beta$, they
also intersect $\tilde \gamma$.
\subsubsection{Completing the analysis}
\label{sec:adp:1}
By the Trapezium Lemma, the domain bounded by $\gamma_0$, $\sigma_u$,
$\sigma_d$ and $\tilde\gamma$ is a part of a canonical domain
enclosing $\gamma_0$. So, by the Adjacent Domain Canonical Principle,
$f$ has the standard behavior here.\\
As $\zeta_d$ can be chosen arbitrarily close to $a$ and $\tilde
\gamma$ can be constructed arbitrarily close to $\beta$, we conclude
that $f$ has the standard behavior in the domain bounded by $\beta$,
$\gamma$ and the line $\im\zeta=Y$.
\subsection{``Crossing'' the segment $\mathbf{[0,\zeta_{2n+2}]
    \subset\beta}$: another example of how to use the Adjacent
  Canonical Domain Principle}
\label{sec:crossing-segment-0}
Pick $c$ so that $0\leq c<\im\zeta_{2n+2}$. Let $s_c$ be the segment
$[0,\zeta_{2n+2}-c]$ of the line $\beta$ (i.e of the line
$\re\zeta=\pi$). We shall check
\begin{Le}
  \label{sc}
  For $c>0$, \ $s_c$  is a canonical line.
\end{Le}
\noindent This and the Adjacent Canonical Domain Principle will imply
\begin{Le}
  \label{ACD:2}
  The solution $f$ has standard behavior in a neighborhood of any
  internal point of $s_0$ (i.e. $s_c$ with $c=0$).
\end{Le}
\demo Indeed, let $c>0$. As, $I_c$ is canonical, by Lemma~\ref{LCD},
there is $K_c$, a canonical domain enclosing $s_c$.  Moreover, by the
previous step, see section~\ref{sec:adp:1}, $f$ has the standard
behavior to the left of $s_c$. Applying the Adjacent Canonical Domain
Principle, we prove that $f$ has standard behavior in $K_c$. As $c$
can be taken arbitrarily small, we obtain Lemma~\ref{ACD:2}. \qed
\smallpagebreak Before proving Lemma~\ref{sc}, note that $s_0$
contains the branch point $\zeta_{2n+2}$. So, $s_0$ itself cannot be a
canonical line.
\smallpagebreak{\it Proof of Lemma~\ref{sc}}. Note that $s_c\subset
z_{n+1}$, i.e. $s_c$ is a part of a connected component of the
pre-image of the $(n+1)$-st spectral band of the periodic
operator~\eqref{Ho} with respect to the mapping
$\mathcal{E}:\,\zeta\to E-W(\zeta)$. For $c>0$, $\mathcal{E}$ maps
$s_c$ strictly into the $(n+1)$-st spectral band. This implies that,
along $s_c$, one has $0<k(\zeta)<\pi$. Now, Lemma~\ref{sc} follows
from the definition of canonical lines. \qed
\subsection{Behavior of $\mathbf{f}$ to the right of $\mathbf{s_0}$: using
  the Rectangle Lemma}
\label{sec:Rl:appl}
Let $R_0$ be the rectangle bounded by the real line, the segment
$s_0$, the line $\im\zeta=\im\zeta_{2n+2}$ and the line
$\re\zeta=2\pi$. By means of the Rectangle Lemma, we prove
\begin{Le}
  \label{le:R:1}
  Inside $R_0$, the solution $f$ has the standard behavior.
\end{Le}
\demo First, we note that, in the interior of $R_0$, one has
$\im\kappa<0$. Indeed, $\im\kappa$ vanishes only at points of the
pre-image of the set of spectral bands of the periodic operator with
respect to $\mathcal E$. Therefore, in the interior of $R_0$, one has
$\im\kappa\ne0$. Furthermore, in $\Pi$, the imaginary part of $\kappa$
is positive, and to go from $\Pi$ to $R_0$ (while staying inside
$D'$), one has to intersect $s_0$, i.e. a connected component of the
pre-image of the $(n+1)$-st spectral band. So, in the interior of
$R_0$, the imaginary part of $\kappa$ is negative.\\
Now, fix $c$, a sufficiently small positive constant. Consider the
closed ``rectangle'' $R_c\subset R_0$ delimited by the lines
$\re\zeta=\pi$, $\im\zeta=c$, the line $\re\zeta=2\pi-c$ and the line
$\im\zeta=\im\zeta_{2n+2}-c$. As $R_c\subset R_0$, the imaginary part
of $\kappa$ is negative in $R_c$. Moreover, by Lemma~\ref{ACD:2}, the
solution $f$ has the standard behavior $f\sim e^{\frac i\varepsilon
  \int_\pi^\zeta \kappa d\zeta} \Psi_+$ in a neighborhood of the left
boundary of $R_c$. So, the rectangle $R_c$ satisfies the assumptions
of the Rectangle Lemma, and, therefore, $f$ has the standard behavior
inside $R_c$. As $c$ can be taken arbitrarily small, this implies that
$f$ has standard asymptotics inside the whole rectangle $R_0$. \qed

\subsection{Applying the Stokes Lemma}
\label{sec:apply-stok-lemma}
Recall that the segment $\sigma=[\zeta_{2n+2},\zeta_{2n+3}]$ of
the line $\re \zeta=\pi$ is a Stokes line. By the previous steps,
we know that, at least near $\sigma$, the solution $f$ has the
standard behavior to the left of and below $\sigma$. To justify
the standard behavior of $f$ to the right of $\sigma$, one uses
the Stokes Lemma.\\
Let $V$ be a neighborhood of $\sigma$. Pick $c$ so that
$0<c<\im(\zeta_{2n+3}-\zeta_{2n+2})$. Let $V_c=\{\zeta\in
V,\,\,\im\zeta<\im\zeta_{2n+3}-c\}$.
We prove
\begin{Le}
  \label{le:Sl:appl}
  If $V$ is sufficiently small, $f$ has the standard behavior in
  $V_c\setminus\sigma$.
\end{Le}
\demo There are three Stokes lines beginning at $\zeta_{2n+2}$. These
are the lines $\sigma$, ``b'' and the line ``$\tilde {\rm b}$''
symmetric to ``b'' with respect to the line $\re\zeta=\pi$. Suppose
that $V$ is chosen sufficiently small. Then,
\begin{enumerate}
\item the three Stokes lines divide $V_c$ into three sectors;
\item by the first three steps of the continuation process, we know
  that $f$ has the standard behavior outside the sector bounded by
  $\sigma$ and ``$\tilde{\rm b}$'';
\item in $V_c$, to the left of $\sigma$, \ $\im\kappa>0$.
\end{enumerate}
So, the conditions of the Stokes Lemma are satisfied, and,
therefore, $f$ has the standard behavior in $V_c\setminus\sigma$.
This completes the proof of Lemma~\ref{le:Sl:appl}.\qed
\subsection{Completing the analysis of $\mathbf{f}$ in $\mathbf{D'}$}
\label{sec:compl-analys-f}
One completes the analysis of $f$ using our continuation tools as
indicated in Fig.~\ref{BEM:fig:dp:1}. Applying each of the
continuation principles, one argues essentially as in the previous
steps. Let us outline the analysis concentrating only on the new
elements.
\subsubsection{The solution $f$ in $D'\cap\C_+$}
\label{sec:solution-f-dcapc_+}
By means of the Rectangle Lemma, one justifies the standard
behavior of $f$  first to the left of $\gamma$ and, second, to the
right of the line $\re\zeta=\pi$.
\subsubsection{Beginning the analysis of $f$ in $D'\cap \C_-$: standard steps}
\label{sec:beginning-analysis-f}
{\bf 1.} \ One begins with justifying the standard behavior between
the lines $\beta$ and $\gamma$ below the real line. Therefore, one
uses the Adjacent Canonical Domain Principle.\\
{\bf 2.} \ Then, one ``continues the asymptotics'' of $f$ to the right
of $\gamma$. First, one tries to use the Rectangle Lemma. However, on
the line $\re\zeta=\pi$, one meets a problem: $\im\kappa=0$ on the
segments $s_{n+1}=[\pi,\bar\zeta_{2(n+1)}]$ and
$s_j=[\bar\zeta_{2j-1},\bar\zeta_{2j}]$ for $j=n+2,\,n+3\dots$.
Indeed, $s_j$ is a connected component of the pre-image of the $j$-th
spectral band of the periodic operator.\\
In result, one obtains standard behavior by means of the Rectangle
Lemma only outside the domains
\begin{equation*}
  d_j=\{\zeta=s+t,\,s\in s_j,\,\,\pi\le t<2\pi\},\quad
  j=n+1,n+2\dots.
\end{equation*}
{\bf 3.} \ Consider the hatched domains in Fig.~\ref{BEM:fig:dp:1}.
Each of them is adjacent to one of the segments $s_j$ and bounded by
Stokes lines. Denote by $T_j$ the hatched domain adjacent to $s_j$.
One justifies the standard behavior in $T_j$ by means of the Adjacent
Domain Principle and the second variant of the Trapezium Lemma (second
point of Lemma~\ref{parallelogram-le}). Let us describe the domain $U$
and the lines $\gamma_0$, $\sigma_u$ and $\sigma_d$ needed to apply
the Trapezium Lemma to study $f$ in $T_j$.\\
{\it The line $\gamma_0$.\/} \ Let $\zeta_u$ and $\zeta_d$ be two
internal points of $s_j$ such that $\im\zeta_d<\im\zeta_u$. The line
$\gamma_0$ is the segment $[\zeta_d,\zeta_u]$ of $s_j$. We define the
branch of the complex momentum with respect to which $\gamma_0$ is a
canonical line. Therefore, we note that $ \pi(j-1)<\kappa+\pi n<\pi j$
as $\zeta$ is inside $s_j$ and set
\begin{equation}
  \label{eq:19}
  \kappa_j=\left\{\begin{array}{c}\kappa+\pi n-\pi(j-1),
      \quad\text{if \ } j\text{\ is \ odd},\\ \pi j-\pi n-
      \kappa,\quad \text{otherwise}.\end{array}\right.
\end{equation}
As seen from the section~\ref{kappa:branches}, the function $\kappa_j$
is a branch of the complex momentum. Along $s_j$, one has
$0<\kappa_j<\pi$.  This implies that $\gamma_0$ is a canonical
line with respect to $\kappa_j$.\\
For sake of definiteness, below, we assume that $j$ is odd. The
case $j$ even is treated similarly.\\
{\it The domain $U$\/}. It is a subdomain of $T_j$. In $T_j$, one has
$\im\kappa_j>0$. Indeed, to go from $\Pi$ to $T_j$, one has to twice
intersect connected components of the pre-image (with respect to
$\mathcal E$) of the set of the spectral bands. So, in $T_j$, one has
$\im\kappa>0$. As $j$ is odd,~\eqref{eq:19} implies
that $\im\kappa_j>0$ in $T_j$.\\
{\it The lines $\sigma_u$ and $\sigma_d$\/}. They are respectively
defined by the relations $\im\int_{\zeta_u}^{\zeta} \kappa_j d\zeta=0$
and $\im\int_{\zeta_d}^{\zeta}(\kappa_j-\pi) d\zeta=0$.  Note that,
$\sigma_d$ contains $\zeta_d$, and $\sigma_u$ contains $\zeta_u$. So,
if $\zeta_d$ and $\zeta_u$ would be respectively the lower and the
upper end of $s_j$, then, the lines $\sigma_u$ and $\sigma_d$ are the
lines of Stokes type bounding $T_j$.\\
By means of Lemma~\ref{le:s-l-1}, one proves that, in $T_j$, the lines
$\sigma_u$ and $\sigma_d$ are vertical, $\sigma_u$ is going downward
from $\zeta_u$, and $\sigma_d$ is going upward from $\zeta_d$.\\
Finally, one checks that, having entered in $T_j$, the lines
$\sigma_u$ and $\sigma_d$ intersect one another before leaving $T_j$.
Indeed, Lemma~\ref{le:intersections} implies that the line $\sigma_d$
(resp. $\sigma_u$) can leave $T_j$ only intersecting its upper
(resp. lower) boundary.\\
{\it Completing the analysis.\/} The Trapezium Lemma implies that the
domain bounded by $\gamma_0$, $\sigma_d$ and $\sigma_u$ is a part of a
canonical domain enclosing $\gamma_0$. Therefore, by Adjacent
Canonical Domain Principle, $f$ has the standard behavior in this
domain. Note that, as $\zeta_u$ and $\zeta_d$ approach the upper and
lower ends of $s_j$, the curves $\sigma_u$ and $\sigma_d$ approach the
upper and lower boundary of $T_j$. This implies that, in fact, $f$ has
the standard behavior inside the whole domain $T_j$.\\
{\bf 4.} \ One justifies the standard behavior of $f$ to the left of
the hatched domains using the Stokes Lemma and the Rectangle Lemma
(see Fig.~\ref{BEM:fig:dp:1}). We omit the details and note only that,
to do this to the right of $T_{n+1}$, one first has to check that $f$
has the standard behavior along the interval $(\zeta_{2n+1},
2\pi-\zeta_{2n+1})$ of the real line (this was not done before!). We
do this in the next subsection.
\subsubsection{The analysis of $f$ in $D'$ and along the interval
$(\zeta_{2n+1},2\pi-\zeta_{2n+1})$ of the real line}
\label{sec:analysis-f-d}
First, as $f$ has the standard behavior in a neighborhood of $\gamma$,
it has the standard behavior in a neighborhood of $\zeta=\pi$, the
point of intersection of $\gamma$ and the real line.  Hence, there
exists a point $a$ such that $\zeta_{2n+1}\le a<\pi$ such that $f$ has
the standard behavior in a neighborhood of any point situated between
$\pi$ and $a$, but not at $a$. Assume that $a>\zeta_{2n+1}$. Let
$\alpha$ be the segment of the line $\re\zeta=a$ connecting a point
$a_1\in\C_-$ to a point $a_2\in\C_+$. One has $0<\kappa(a)<\pi$. So,
if $\alpha$ is sufficiently small, it is canonical. The solution $f$
has the standard behavior to the right of $\alpha$ (this follows from
the definition of $a$ and the previous analysis). So, we are in the
case of the Adjacent Canonical Domain Principle; it implies that $f$
has the standard behavior in a local canonical domain enclosing
$\alpha$.  Therefore, $f$ has the standard behavior in a constant
neighborhood of $a$. So, we obtain a contradiction, and, thus
$a=\zeta_{2n+1}$. This completes the analysis of $f$ along the
interval $(\zeta_{2n+1},\pi)$.  Similarly one studies $f$ along
$(\pi,2\pi-\zeta_{2n+1})$.
\subsubsection{Completing the proof}
\label{sec:completing-proof}
We still have to check that $f$ has the standard behavior to the left
of the Stokes line $\overline{\text{``a''}}$ symmetric to ``a'' with
respect to the real line. Therefore, one first uses the Stokes Lemma
to justify the standard behavior in the left hand side of a small
neighborhood of $\overline{\text{``a''}}$, and, then, one uses The
Rectangle Lemma to justify the standard behavior in the rest of the
part of $D'$ situated to the left of
$\overline{\text{``a''}}$.\\
This completes the analysis of the behavior of $f$ in the domain
$D'$.\qed


%
\section{Behavior of solutions outside the continuation diagrams}
\label{sec:2waves}
\noindent In this section, we formulate and prove the Two-Waves
Principle.
\subsection{Formulation of the problem}
\label{sec:formulation-problem}
\subsubsection{Geometry of the problem}
\label{tw:geom}
\begin{figure}[h]
  \includegraphics[bbllx=71,bblly=618,bburx=474,bbury=721,width=14cm]{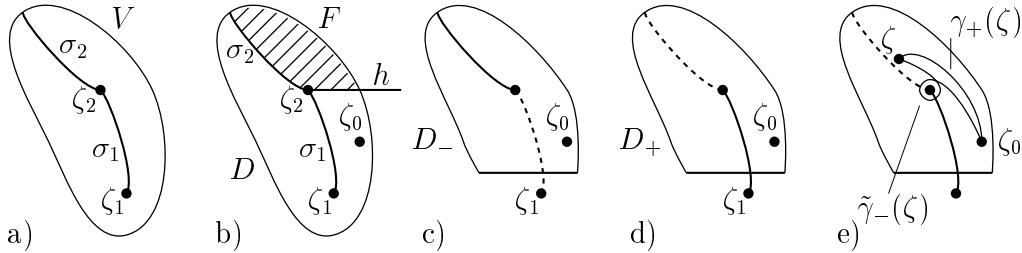}
  \caption{The Two-Waves Principle}\label{FDP:1}
\end{figure}

Assume that for $E=E_0$, one has the geometrical situation shown
in part a) of Fig.~\ref{FDP:1}. There, $\zeta_1$ and $\zeta_2$ are
two branch points of the complex momentum such that $W'(\zeta_1)$
and $W'(\zeta_2)$ are non zero. The line $\sigma_1$ is
simultaneously a Stokes line beginning at $\zeta_1$ and at
$\zeta_2$.  The line $\sigma_2$ is a segment of a Stokes line
beginning at $\zeta_2$. We assume that both $\sigma_1$ and
$\sigma_2$ are vertical.
%
%
\smallpagebreak Let $V$ be a neighborhood of $\sigma_1\cup\sigma_2$
containing only two branch points, precisely $\zeta_1$ and $\zeta_2$.
Let $h=\{\zeta:\,\, \im\zeta=\zeta_2,\,\,\re\zeta>\re\zeta_2\}$. Also,
denote by $F$ the part of $V$ situated above $h$ and to the right of
$\sigma_2$ , see Fig.~\ref{FDP:1}, b).
\subsubsection{Formulation of the problem}
\label{sec:prel-descr-probl}
Pick $\zeta_0\in V$ so that $\mathcal{E}(\zeta_0)\not\in P\cup Q$.
Assume that a solution $f$ has the standard behavior $f\sim\exp\left
  (\frac i\varepsilon\int_{\zeta_0}^\zeta\kappa\,d\zeta\right)
\cdot \Psi_+$ in the domain $D=V\setminus (F\cup \sigma_1)$. Assume,
moreover, that the imaginary part of $\kappa$ is positive in $D$ to
the left of $\sigma_1\cup\sigma_2$. Our aim is then to describe $f$ in
the domain $F$.
\smallpagebreak The problem described above comes about in the case
studied in section~\ref{ex:cont-diag}. The lines $\sigma_1$ and
$\sigma_2$ are respectively the Stokes lines $[\zeta_{2n+2},
\zeta_{2n+3}]$ and ``c'', and the domain $F$ is situated to the right
of ``c'' above the line $\im\zeta=\im\zeta_{2n+3}$, see
Fig.~\ref{BEM:fig:dp:1}.
\subsection{Two-Waves Principle}
\label{sec:two-waves-principle}
The natural idea is to try to represent $f$ as a linear combination of
solutions having standard behavior in $F$. This leads to the following
construction.
\smallpagebreak Consider $D_\pm$, the subdomains of $V$ shown in
Fig.~\ref{FDP:1}, parts c) and d). On each of them, fix the branch
of the complex momentum so that, in some neighborhood of
$\zeta_0$, it coincide with the branch from the asymptotics of
$f$. It will be convenient to assume that $\zeta_0$ is to the
right of $\sigma_1\cup\sigma_2$. One has
\begin{Le}[Two-Waves Principle]
  \label{tw:1}

  Assume that there are two solutions $h_\pm$ having the standard
  behavior $h_\pm\sim\exp(\pm\frac i\varepsilon\int_{\zeta_0}^{\zeta}
  \kappa\,d\zeta)\cdot \Psi_\pm$ in $D_\pm$. Then,
  \begin{equation}\label{f:hpm}
    f(\zeta)=g(\zeta)\,h_+(\zeta)+ G(\zeta)\, h_-(\zeta),\quad \zeta\in F,
  \end{equation}
  where $\zeta\mapsto G(\zeta)$ and $\zeta\mapsto g(\zeta)$ are two
  $\varepsilon$-periodic functions. In $F$, these functions admit the
  asymptotic representations
                                %
                                %
  \begin{floatingfigure}{6cm}
    \begin{center}
\includegraphics[bbllx=71,bblly=620,bburx=247,bbury=721,width=6cm]{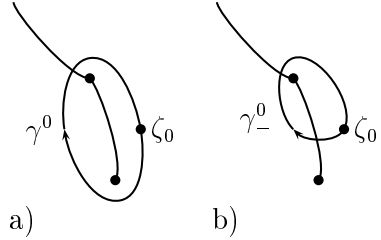}
    \end{center}
    \caption{The curves $\gamma^0$ and $\gamma_-^0$}\label{FDP:2}
  \end{floatingfigure}
                                %
                                %
  \begin{equation}
    \label{gG}
    \begin{split}
      \hskip-5cm
      G&=e^{\dsize\frac{2i\kappa(\zeta_2)}\varepsilon\,(\zeta-\zeta_2)}\,\,
      \frac{(A -1+ o(1))}B\,\,(1+o(1)),\\
      \hskip-5cm g&=1+o(1),
    \end{split}
  \end{equation}
  where $A$ and $B$ are constants given by the formulae
  \begin{equation}
    \label{tw:AB}
    \begin{split}
      \hskip-5cm A&=\exp\left(\frac i\varepsilon\,\oint_{\gamma_0}
        \kappa d\zeta+\oint_{\gamma_0}\Omega_++\ind(\gamma_0)\right),\\
      \hskip-5cm B&=\exp\left(-\frac i\varepsilon \oint_{\gamma_-^0}
        \kappa d\zeta+\oint_{\gamma_-^0}\Omega_-+\ind(\gamma_-^0)\right).
    \end{split}
  \end{equation}
  Here, $\gamma_0$ and $\gamma_-^0$ are loops going around the branch
  point as shown in Fig.~\ref{FDP:2} and do not containing any points
  of ${\mathcal E}(P\cup Q)$; \ $\ind(\alpha)$ denotes the increment
  of $\arg\left(\sqrt{k'({\mathcal E}(\zeta))}\right)$ along a closed
  curve $\alpha$. The representations~\eqref{gG} are uniform in
  $\zeta$ and $E$ provided that $\zeta$ is in a compact subset of $F$
  and $E$ is in a sufficiently small neighborhood of $E_0$.
\end{Le}
\subsection{Comments and remarks}
\label{tw:comments}
Let us comment on the Two-Waves Principle.
\subsubsection{Solutions $h_\pm$}
\label{sec:solutions-h_pm}
Recall that $V$ is a neighborhood of $\sigma_1\cup\sigma_2$. If $V$ is
sufficiently small (and, thus, ``thin'' and ``stretched'' along
$\sigma_1$ and $\sigma_2$), the solutions $h_\pm$ can be easily
constructed using our standard techniques. However, in practice, one
does not use these local constructions. Instead, one tries to
construct $h_\pm$ so that they have the standard behavior on domains
as large as possible. Thus, their construction is determined by the
concrete geometry of the problem.  Detailed examples can be found in
section~\ref{ex:tw}.
\subsubsection{A convenient representation for $f$}
\label{new-hm}
We have formulated the Two-Waves Principle in terms of the solutions
$h_\pm$ to simplify the exposition. However, to makes the results more
transparent, let us change the normalization of $h_-$. Let
\begin{equation*}
  h_{-}^o=
  e^{\frac{2i\kappa(\zeta_2)}\varepsilon\,(\zeta-\zeta_2)}\,B^{-1}\,h_-.
\end{equation*}
It will follow from the proof of Lemma~\ref{tw:1} that the
solution $h_{-}^o$ has the standard behavior
\begin{equation}
  \label{new-hm:as}
  h_-^o\sim e^{\frac i\varepsilon
    \int_{\tilde\gamma_-(\zeta)}\kappa\,d\zeta}\Psi_+,\quad\quad
    \zeta\in F,
\end{equation}
where the curve $\tilde\gamma_-$ is shown in Fig.~\ref{FDP:1}, part
e), and $\kappa$ and $\Psi_+$ are obtained by the analytic
continuation from $\zeta_0$ along $\tilde\gamma_-$. In terms of the
solutions $h_+$ and $h_-^o$, formula~\eqref{f:hpm} takes the simplest
form
\begin{equation}
  \label{eq:13}
  f=h_+(1+o(1)) + [A-1 +o(1)]\,h_-^o\, (1+o(1)).
\end{equation}
Note that, for small $\varepsilon$, the absolute values of $h_+$
and $h_-^o$ are essentially determined by the factors
\begin{equation}
  \label{exp_pm}
  \exp\left(\frac i\varepsilon\int_{\gamma_+(\zeta)}
    \kappa\,d\zeta\right)\quad\text{and}\quad \exp\left(\frac
    i\varepsilon\int_{\tilde \gamma_-(\zeta)} \kappa\,d\zeta\right),
\end{equation}
where $\gamma_+(\zeta)$ is shown in Fig.~\ref{FDP:1}, part e). The
definition of Stokes lines implies that, along the Stokes lines
beginning at $\zeta_2$, the moduli of these factors are equal.
\subsubsection{The coefficient $A$}
\label{sec:coefficient-a}
The coefficient $A$ is defined in $U$, a sufficiently small constant
neighborhood of $E_0$. Formula~\eqref{eq:13} shows that it is
important to compare the modulus of $A$ with $1$. For $\varepsilon$
small, the modulus of $A$ is essentially determined by the factor
$\exp\left(-\frac 1\varepsilon\,\im\,\oint_{\gamma_0}\kappa
  d\zeta\right)$. So, when $\varepsilon\to 0$, depending of $E\in U$,
the coefficient $A$ may become exponentially small or exponentially
large. However, for some $E\in U$, it always is of order
$O(1)$. Indeed, one proves
\begin{Le}
  \label{modA}
  Fix $E\in U$. Assume that the configuration of the Stokes lines
  corresponds Fig.~\ref{FDP:1}, part a). Then, one has
  $\im\,\oint_{\gamma_0}\kappa d\zeta=0$.
\end{Le}
\demo Fix $E$ as in Lemma~\ref{modA} and consider $\kappa$ as a
function of $\zeta\in V$ ($V$ is the neighborhood of
$\sigma_1\cup\sigma_2$ defined in section~\ref{sec:prel-descr-probl}).
Cut $V$ along $\sigma_1$. First, we check that the branch of $\kappa$
(defined in a neighborhood of $\zeta_0$) is analytic $V\setminus
\sigma_1$. Consider the curve $\gamma$ beginning at $\zeta_0$ and
going to $\sigma_1$ along a straight line, then, going around
$\sigma_1$ just along it (infinitesimally close to it) and, finally,
coming back to $\zeta_0$ along the same straight line. Continue
$\kappa$ analytically along $\gamma$. Relation~\eqref{two-branches}
implies that, near $\zeta_0$, the values of $\kappa$ and of its
analytic continuation differ by the additive constant
$2(\kappa(\zeta_1)- \kappa(\zeta_2))$. But, as $\sigma_1$ is a Stokes
line for both $\zeta_1$ and $\zeta_2$, one has $\kappa(\zeta_1)=
\kappa(\zeta_2)$. This implies the analyticity of $\kappa$.
\smallpagebreak As $\kappa$ is single valued in $V\setminus\sigma_1$,
we can deform the integration contour $\gamma_0$ from the definition
of $A$ so that it go around $\sigma_1$ just along it.  Now, it follows
from the definition of the Stokes lines that $\im\oint_{\gamma_0}
\kappa d\zeta=\im\oint_{\gamma_0}(\kappa(\zeta)-\kappa(\zeta_1))
d\zeta=0$. \qed
\subsubsection{Generalizations of the Two-Waves Principle}
\label{sec:gener-two-waves}
\begin{figure}[h]
  \includegraphics[bbllx=71,bblly=620,bburx=389,bbury=721,width=11cm]{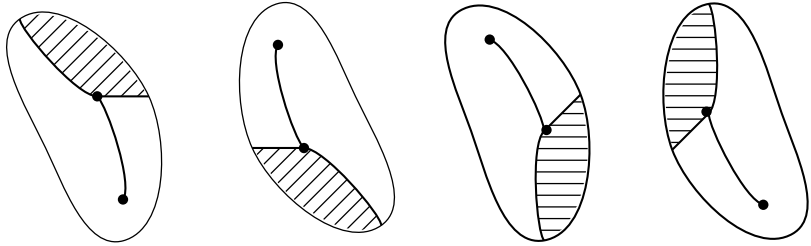}
  \caption{All the possible  geometric situations}\label{FDP:3}
\end{figure}

In the same way as we prove Lemma~\ref{tw:1}, one obtains analogous
statements for the ``symmetric'' geometries shown in Fig.~\ref{FDP:3}.
%
%
\subsection{How to use the Two-Waves Principle: an example}
\label{ex:tw}
Consider the solution $f$ studied in section~\ref{ex:cont-diag}.  We
now apply the Two-Waves Principle to obtain the asymptotics of $f$ to
the right of the Stokes line ``c''.
\subsubsection{Comparing the notations}
\label{sec:comparing-notations}
The points $\zeta_1$ and $\zeta_2$ are the branch points
$\zeta_{2n+2}$ and $\zeta_{2n+3}$; the Stokes lines $\sigma_1$ and
$\sigma_2$ are the Stokes lines $[\zeta_{2n+2},\zeta_{2n+3}]$ and
``c'' (more precisely, its segment below the line $\im\zeta=Y$,
$Y>\im\zeta_{2n+3}$). The domain $F$ situated above the line
$\im\zeta=\im\zeta_{2n+3}$ and to the right of ``c''.
\subsubsection{Checking the assumptions of the Two-Waves Principle}
\label{sec:check-assumpt-princ}
The assumptions of Lemma~\ref{tw:1} are satisfied: $\sigma_1$ is a
Stokes line both for $\zeta_{2n+2}$ and $\zeta_{2n+3}$; both
$\sigma_1$ and $\sigma_2$ are vertical; $f$ has the standard behavior
in $V\setminus(F\cup [\zeta_{2n+2},\zeta_{2n+3}])$; and, to the left
of the Stokes lines ``c'' and $[\zeta_{2n+2},\zeta_{2n+3}]$, near
them, one has $\im\kappa>0$.
\subsubsection{The solution $h_+$}
\label{sec:solution-h_+}
%
\begin{figure}[htbp]
  \centering \subfigure[The construction of $h_+$]{
    \includegraphics[bbllx=68,bblly=617,bburx=212,bbury=721,width=6cm]{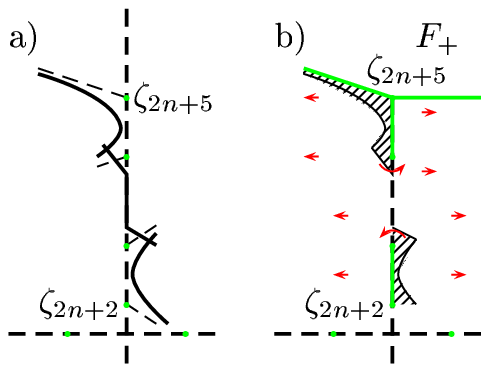}
      \label{fig:hp:cd}}
    \hskip2cm
    \subfigure[and of $h_-$]{
      \includegraphics[bbllx=71,bblly=617,bburx=144,bbury=721,width=3cm]{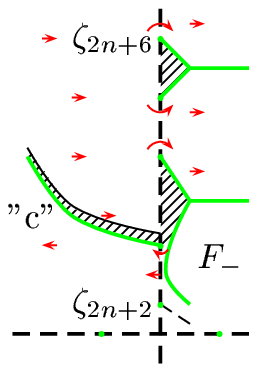}
      \label{fig:hm:cd}}
  \caption{Constructions of $h_\pm$}\label{fig:hpm:cd}
\end{figure}
%
To construct the solution $h_+$, we first build $\pi_+$, a
pre-canonical line, as shown in Fig.~\ref{fig:hp:cd}, part a). When
speaking about $\kappa$ on $\pi_+$, we mean the branch of the complex
momentum obtained by analytic continuation of the branch $\kappa$ from
the asymptotics of $f$ along $\pi_+$ from $D'$. Actually, $\pi_+$ is
pre-canonical with respect to $\kappa_1$, the branch of the complex
momentum equal to $2\pi-\kappa$. The construction of $\pi_+$ being
standard, we omit details and note only that $\pi_+$ consists of five
``elementary'' segments. The first (the lower) ``elementary'' segment
and the fourth one are segments of lines of Stokes type $\im\int^\zeta
(\kappa_1-\pi) d\zeta=\Const$.  The second and the fifth ones are
segments of lines of Stokes type $\im\int^\zeta\kappa_1d\zeta=
\Const$. The third elementary segment is a canonical line.\\
Having constructed $\pi_+$, we pick $\gamma_+$, a canonical line close
to $\pi_+$, and, by Theorem~\ref{T5.1}, construct the solution $g\sim
e^{-\frac i\varepsilon\int_\pi^{\zeta}\kappa_1 d\zeta}$ on $K_+$, a
canonical domain enclosing this canonical line.\\
We let $h_+= \exp\left(\frac{2\pi i(\zeta-\pi)}\varepsilon\right)g$.
On $K_+$, the function $h_+$ has the standard behavior $h_+\sim
e^{\frac i\varepsilon\int_\pi^{\zeta}\kappa d\zeta}\Psi_+$.\\
The computation of the continuation diagram of $h_+$ is explained in
Fig.~\ref{fig:hp:cd}, part b), where we show only what happens in the
domain $0<\re\zeta<2\pi$.\\
Note that we do not control the behavior of $h_+$ in the domain
denoted by $F_+$ in Fig.~\ref{fig:hp:cd}, part b).
\subsubsection{The solution $h_-$}
\label{sec:solution-h_-}
To construct $h_-$, we build $\pi_-$, a pre-canonical line similarly
to $\pi_+$. The only difference is that, above the point
$\zeta_{2n+3}$, instead of going along the line $\re\zeta=\pi$ to a
neighborhood of $\zeta_{2n+4}$ , the pre-canonical line $\pi_-$ goes
along a line of Stokes type $\im\int^\zeta(\kappa_1-\pi)
d\zeta=\Const$ which belongs to the same family as ``c'', is situated
to the right of ``c'' and is chosen sufficiently close to ``c''.\\
On $K_-$, a canonical domain enclosing $\gamma_-$, a canonical line
``approximating'' $\pi_-$, the solution $h_-$ has the standard
behavior $h_-\sim e^{-\frac i\varepsilon\int_\pi^{\zeta}\kappa
  d\zeta}\Psi_-$.\\
The analysis of the continuation diagram of $h_-$ is explained in
Fig.~\ref{fig:hm:cd}, where we show only what happens in the domain
$0<\re\zeta<2\pi$.\\
Note that we do not control the behavior of $h_-$ in the domain
denoted by $F_-$ in Fig.~\ref{fig:hm:cd}.
\subsubsection{Asymptotics of $f$}
\label{sec:asymptotics-f}
As in section~\ref{new-hm}, in terms of $h_-$, we define the
solution $h_-^o$. By the Two-Wave Principle, $f$ admits the
representation~\eqref{eq:13}. This yields the asymptotics of $f$
in $F$.\\
This leaves us with the following two questions:
\begin{itemize}
\item what is the asymptotics of $f$ in the domain $F\cap F_+$ (where
  the asymptotics $h_+$ is unknown)?
\item how to get the asymptotics of $f$ in the domain $F\cap F_-$
  (where the asymptotics $h_-$ is unknown)?
\end{itemize}
Denote by $\alpha$ the Stokes line beginning at $\zeta_{2n+5}$ and
going from it upwards to the left. To answer the first question, one
has to find the asymptotics of $h_+$ in the domain $F_+$ situated to
the right of $\alpha$ and above the line $\im\zeta=\im\zeta_{2n+5}$.
Therefore, one has just to apply the Two-Waves Principle once more
(now, to study $h_+$).
\smallpagebreak Denote by $\beta$ the Stokes line beginning at
$\zeta_{2n+3}$ going upwards to the right. Denote by $p$ the point
where it intersects the Stokes line beginning at $\zeta_{2n+4}$ going
downwards to the right. The answer to the second question is given by
\begin{Le}
  \label{le:t-w-ex}
  Let $D_1=\{\pi<\re\zeta<2\pi,\,\,\zeta_{2n+2}<\im\zeta<\im p\}$.
  Let $D_2$ be the part of $D_1$ situated to the right of $\beta$.
  Then, $D_2$ is in the continuation diagram of $f$ i.e. in $D_2$, $f$
  has standard asymptotics.
\end{Le}
\noindent We only explain the  idea guiding the proof of this
lemma and omit the technical details. The idea is the following.  Both
the solutions $h_+$ and $h_-^o$ have the standard behavior inside
$D_2$ near $l$, the left part of the boundary of $D_2$.  As $l$
consists of segments of Stokes lines, along $l$, the absolute values
of the exponentials~\eqref{exp_pm} determining the order of these
solutions coincide. To the right of $l$, the exponential term in the
asymptotics of $h_+$ becomes (exponentially) larger than the one in
the asymptotics of $h_-^o$. This and Lemma~\ref{modA} imply that, in
$D_2$, near $l$ (where both $h_+$ and $h_-^o$ have the standard
behavior), the second term in~\eqref{eq:13} is negligible with respect
to the first one so that, there, $f$ has the standard behavior
$f\sim\exp(\frac i\varepsilon\int_{\gamma_+(\zeta)}\kappa
\,d\zeta)\cdot \Psi_+$. This and the Rectangle Lemma, then, imply the
statement of Lemma~\ref{le:t-w-ex}.

%

%
\section{The proof of the Trapezium Lemma}
\label{proof:T-l}
\noindent In this section we prove the Trapezium Lemma,
Lemma~\ref{parallelogram-le}. We check only the first point. The
second one is proved similarly. We begin with studying the lines of
Stokes type inside $T$.
\smallpagebreak Pick $\zeta_0\in T$. By Corollary~\ref{le:s-l-2},
$\zeta_0$ belongs to exactly one line of Stokes type from the same
family as $\sigma_d$ and to exactly one line of Stokes type from the
same family as $\sigma_u$. We denote these lines by
$\sigma_d(\zeta_0)$ and $\sigma_u(\zeta_0)$ respectively. One has
\begin{Le}
  \label{le:4}
  One has
  \begin{itemize}
  \item The lines $\sigma_d(\zeta_0)$ and $\sigma_u(\zeta_0)$ stay
    vertical before leaving $U$.
  \item Above $\zeta_0$, the line $\sigma_u(\zeta_0)$ leaves $T$
    intersecting $\gamma_0$ at $\zeta_u$, an internal point of
    $\gamma_0$.
  \item Below $\zeta_0$, the line $\sigma_d(\zeta_0)$ leaves $T$
    intersecting $\gamma_0$ at $\zeta_d$, an internal point of
    $\gamma_0$.
  \end{itemize}
\end{Le}
\demo As $\im\kappa\ne 0$ in $U$, the first point of Lemma~\ref{le:4}
follows from Lemma~\ref{le:s-l-1}.\\
Let us check the second one. First, we note that $\sigma_u(\zeta_0)$
cannot leave $T$ by intersecting $\sigma_u$, its ``upper boundary'',
as both lines belong to one and the same family of line of Stokes
type. Second, show that it can not leave $T$ intersecting
$\tilde\gamma$ above $\zeta_0$. Consider the family $\{\sigma_u(t)\}
|_{t\in\tilde \gamma}$. It contains $\sigma_u$, the ``upper boundary''
of $T_j$, and, by assumption, $\sigma_u(\zeta_0)$.  By the second
point of Lemma~\ref{s-l:prop}, each line from this family intersects
$\tilde\gamma$ transversally. So, we can orient the tangent vectors at
the intersection points to the right. As, in $U$, one has
$\im\kappa\ne 0$, either all of these vectors are oriented upwards or
all of them are oriented downwards.  Therefore, all the tangent
vectors are directed to the right and downwards as does the tangent
vector to $\sigma_u$ (the ``upper boundary''). But, $\sigma_u
(\zeta_0)$ cannot go upward from $\zeta_0$, stay vertical and leave
$T$ intersecting $\tilde \gamma$ in this way.\\
Finally, show that $\sigma_u(\zeta_0)$ can not leave $T$ by
intersecting $\sigma_d$, the ``lower boundary'' of $T$. Therefore,
compare $t_d(\zeta_0)$ and $t_u(\zeta_0)$, the tangent vectors to
$\sigma_d(\zeta_0)$ and $\sigma_u(\zeta_0)$ at $\zeta_0$. As in $U$,
one has $\im\kappa\ne 0$, we orient both the vectors upwards. As
$\sigma_u(\zeta_0)$ and $\sigma_d(\zeta_u)$ belong to different
families of Stokes lines, either, for all $\zeta_0\in U$, the vector
$t_d(\zeta_0)$ is directed to the left with respect to $t_u(\zeta_0)$
or, for all $\zeta_0\in U$, it is directed to the right. Comparing the
tangent vectors at the point of intersection of $\sigma_d$ and
$\tilde\gamma$ (the lower and the right boundaries of $T$), we see
that we are in the second case i.e. for all $\zeta_0\in U$,
$t_d(\zeta_0)$ is directed to the right.\\
Assume that $\sigma_u(\zeta_0)$ leaves $T$ by intersecting $\sigma_d$,
the ``lower boundary'' of $T$. As $t_u$ is oriented to the left of
$t_d$, we conclude that either $\sigma_u(\zeta_0)$ intersects
$\sigma_d(\zeta_0)$ twice, or $\sigma_d(\zeta_0)$ intersects
$\sigma_d$. By Lemma~\ref{le:intersections}, both these events are
impossible. \\ So, we see that, above $\zeta_0$, $\sigma_u(\zeta_0)$
leaves $T$ intersecting $\gamma_0$. This is the second point of
Lemma~\ref{le:4}.\\
The third point is proved similarly.\qed
\smallpagebreak To complete the proof of Lemma~\ref{parallelogram-le},
we use Proposition~\ref{pro:pcl:2}.  First, assume that $\im\kappa\ne
0$ along $\gamma_0$.  For $\zeta_0\in T$, consider the line $\alpha$
which consists of the segment of $\sigma_u(\zeta_0)$ above $\zeta_0$
between $\gamma_0$ and $\zeta_0$ and of the segment of
$\sigma_d(\zeta_0)$ below $\zeta_0$ between $\zeta_0$ and $\gamma_0$.
This line is a pre-canonical line containing $\zeta_0$ and connecting
two internal point of the canonical line $\gamma_0$. As this line
exists for any $\zeta_0\in T$, Proposition~\ref{pro:pcl:2} implies
that $T$ is part of a canonical domain enclosing $\gamma_0$.  This
completes the proof in the case under consideration. In general case,
the line $\alpha$ may become horizontal (i.e. not vertical) at its end
points (recall that pre-canonical lines are supposed to be vertical).
If this is the case, one ``corrects'' $\alpha$ near its ends. For
example, near the upper end, one replaces a small segment of $\alpha$
by a small segment of a canonical line connecting an internal point of
the ``old'' $\alpha$ to an internal point of $\gamma_0$ situated above
the end of the ``old'' $\alpha$. The required canonical line is
obtained by a small $C^1$ deformation of $\gamma$ (as small $C^1$
deformations preserve the property of being canonical). In result, the
``new'' $\alpha$ becomes vertical. So, one again can apply
Proposition~\ref{pro:pcl:2}. This completes the proof of
Lemma~\ref{parallelogram-le}. \qed
%


%
\section{The proof of the Stokes Lemma}
\label{Principles-proof}
\noindent In this section we prove Lemma~\ref{st-lm}.
\subsection{Preliminaries}
For sake of definiteness, we assume that $\sigma_1$ is going downwards
from $\zeta_0$ and that the sector $S_1$ is adjacent to $\sigma_1$
from the left. All the other geometric situations are analyzed
similarly. Note that, by assumptions of the Stokes Lemma in the case
we consider, in $S_1$, near $\sigma_1$, one has $\im\kappa>0$.\\
For sake of briefness, we shall justify only the uniform asymptotics
of $f$ on $V':=V\setminus\sigma_1$, see Fig.~\ref{stokes:fig:1}. The
term ``standard behavior'' actually means more (see
section~\ref{sec:standard-asymptotics}). But, as our construction is
based on the analysis of solutions having standard behavior, reading
the proof, one easily checks that, in $V'$, the solution $f$ has
standard behavior.\\
Recall that one can always choose $\kappa_0$, a branch of the complex
momentum analytic on $V'$ and such that either $\kappa_0(\zeta_0)=0$
or $\kappa_0(\zeta_0)=\pi$ (natural branch).  Below, we assume that
$\kappa_0(\zeta_0)=0$; the second case is studied in a similar way.
\subsubsection{The plan of the proof}
\label{sec:plan-proof}
Our plan is roughly the following. First, we find $\varkappa$, a
canonical line in $V$ going to the right of $\zeta_0$ and $\sigma_1$,
and staying close to $\sigma_1$. The line $\varkappa$ will be
canonical with respect to $\kappa_0$, the natural branch of the
complex momentum. Then, by Lemma~\ref{LCD}, we construct $K$, a local
canonical domain containing $\varkappa$; Theorem~\ref{T5.1} then,
gives us $f_\pm $, two solutions having standard behavior
\begin{equation}
  \label{fpm}
  f_\pm \sim e^{\pm \frac i\varepsilon \int_{\zeta_*}^\zeta \kappa_0
    d\zeta} \Psi_\pm (x,\zeta,\zeta_*),\quad \zeta\in K.
\end{equation}
Here, $\kappa_0$ is the branch of the complex momentum with respect to
which $K$ is canonical, and $\zeta_*\in V'$ is a normalization point
(we assume that ${\mathcal E}(\zeta_*)\not\in P\cup Q$).\\
Recall that $f_\pm $ are analytic in $\zeta$ in the strip
$\{Y_1<\im\zeta<Y_2\}$, the smallest ``horizontal'' strip containing
$K$ (see Theorem~\ref{T5.1}).
\smallpagebreak Next, we express $f$ in the basis $f_\pm $
\begin{equation}
  \label{f:fpm}
  f(x,\zeta)=a(\zeta) f_+(x,\zeta)+b(\zeta)f_-(x,\zeta).
\end{equation}
The coefficients $a$ and $b$ are independent of $x$; they can be
expressed as
\begin{equation}
  \label{f:ab}
  a(\zeta)=\frac{w( f,f_-)}{w(f_+,f_-)}\quad\text{ and
    }\quad b(\zeta)=\frac{w(f_+,f)}{w(f_+,f_-)}.
\end{equation}
The Wronskians in this formula are analytic in the strip
$\{Y_1<\im\zeta< Y_2\}$ as the solutions $f$ and $f_\pm $ are.
Moreover, as $f$ and $f_\pm $ satisfy the
condition~\eqref{consistency}, the Wronskians are
$\varepsilon$-periodic in $\zeta$. Fix $\nu$ positive. For
sufficiently small $\varepsilon$, \ $|w(f_+,f_-)|$ is bounded away
from zero uniformly in the strip $\{Y_1+\nu<\im\zeta< Y_2-\nu\}$,
see~\eqref{W_of_f_pm}.  Returning to $a$ and $b$, we conclude that,
first, they are analytic in this strip, second, they are
$\varepsilon$-periodic in $\zeta$.
\smallpagebreak Lemma~\ref{st-lm} then, follows from the analysis of
the coefficients $a$ and $b$.
\subsubsection{Choice of the branch $\kappa_0$}
\label{sec:choice-branch-kapp}
Assume that $V$ is so small that it contains only one branch point
$\zeta_0$.  Consider $\kappa$, the branch of the complex momentum from
the asymptotics of $f$ in $S_1$; continue it analytically from $S_1$
to $V'$. Note that $\kappa_0$, the natural branch, is defined up to
the sign. We choose it so that $\im\kappa$ and $\im\kappa_0$ have the
same sign. We get
\begin{equation}
  \label{kappa0kappa}
  \kappa(\zeta)=\kappa_0(\zeta)+2\pi n_0,\quad \zeta\in V',
\end{equation}
where $n_0$ is a natural number independent of $\zeta$.
\subsubsection{Normalization of the solution $f$}
\label{sec:norm-solut-f}
As we express $f$ in terms of $f_\pm $ described by~\eqref{fpm}, it is
convenient to assume that the solution $f$ itself is normalized at
$\zeta_*$ and that, in $S_1$ and $S_2$, it has standard behavior
\begin{equation}
  \label{f:as}
  f\sim e^{\frac i\varepsilon \int_{\zeta_*}^\zeta \kappa_0
    d\zeta} \Psi_+(x,\zeta,\zeta_*).
\end{equation}
Note that, in~\eqref{f:as} (as in~\eqref{fpm}), we integrate
$\kappa_0$ but not $\kappa$. It is sufficient to consider this case.
Indeed, in view of~\eqref{kappa0kappa}, the solution $f$ can always be
represented in the form
\begin{equation}
  \label{f}
  f=f_0e^{2\pi i n_0(\zeta-\zeta_*)/\varepsilon} \tilde f,
\end{equation}
where $f_0$ is constant, and $\tilde f$ has the standard
behavior~\eqref{f:as}. Hence, it is sufficient to prove the Stokes
Lemma for $\tilde f$. So that, from now on, we simply assume that
$f=\tilde f$, i.e. $f$ is normalized at $\zeta_*$ and
$\kappa_0=\kappa$ (i.e. $n_0=0$).
\subsubsection{Three cases}
\label{sec:three-cases}
Consider the angle $\alpha$ between the Stokes line $\sigma_1$ and
the line $\{\im\zeta=\im\zeta_0,\ \re\zeta\ge \re\zeta_0\}$ at the
point $\zeta_0$. We measure this angle clockwise.  As we consider
the case where $\sigma_1$ is going downwards from $\zeta_0$, one
has $0<\alpha<\pi$.
\smallpagebreak When constructing the canonical line $\varkappa$, we
have to treat differently three cases:
\begin{description}
\item[a)] $0<\alpha<2\pi/3$ (see Fig.~\ref{fig:mat-mono1});
\item[b)] $\alpha=2\pi/3$ (see Fig.~\ref{fig:mat-mono2});
\item[c)] $2\pi/3<\alpha<\pi$ (see Fig.~\ref{fig:mat-mono3}).
\end{description}
However, having found the canonical line, in each of these cases, one
completes the proof by doing almost one and the same computation.
Thus, we only give a detailed proof of the Stokes Lemma in the case
a). For the two remaining cases, we describe with detail only the
construction of the canonical line.
\subsection{The proof of the Stokes Lemma in the case a)}
\label{sec:proof-stokes-lemma}
\subsubsection{Constructing the local canonical domain}
\label{sec:constr-local-canon}
Recall that the angle between $\sigma_1$ and $\sigma_3$ at $\zeta_0$
is equal to $2\pi/3$. So, the Stokes line $\sigma_3$ goes upwards from
$\zeta_0$. We assume that $V$ is sufficiently small so that
$\sigma_3\cap V$ is vertical.
\smallpagebreak When constructing the canonical line, we shall need
\begin{Le}
  \label{z1}
  If $V$ is sufficiently small and $0<\alpha<2\pi/3$, then, in $V'$,
  $\im\kappa$ vanishes only along $Z_0$, an analytic curve connecting
  $\zeta_0$ to a point of the boundary of $V$; this curve
  goes  inside the sector $S_1\cup\sigma_2\cup S_2$ of $V$.
  In the part of this sector situated between $\sigma_1$
  and $Z_0$, one has $\im\kappa>0$. In the rest of $V'\setminus Z_0$,
  one has $\im\kappa<0$.
\end{Le}
\demo The points where $\im\kappa=0$ are points of $Z$, the pre-image
of the set of the spectral bands of the periodic Schr{\"o}dinger
operator~\eqref{PSO} with respect to the mapping ${\mathcal E}:
\zeta\to E-W(\zeta)$. The ends of the connected components of $Z$ are
exactly the branch points of $\kappa$. So, there exists a connected
component of $Z$ beginning at $\zeta_0$, say $Z_0$.  Assume that
$0<\alpha<\pi/3$. Then, $\sigma_2$ goes downward from $\zeta_0$.  By
means of~\eqref{k:sqrt}, one easily checks that, in a sufficiently
small neighborhood of $\zeta_0$, $Z_0$ goes downward from $\zeta_0$
staying between $\sigma_1$ and $\sigma_2$. If $\alpha=\pi/3$, the
vectors tangent to $\sigma_2$ and to $Z_0$ at $\zeta_0$ are
horizontal. In this case, $Z_0$ and $\sigma_2$ go to the left from
$\zeta_0$. If $\pi/3<\alpha<2\pi/3$, then, in a sufficiently small
neighborhood of $\zeta_0$, $\sigma_2$ and $Z_0$ are going upwards from
$\zeta_0$, and $Z_0$ stays between $\sigma_2$ and $\sigma_3$.
%
%
\begin{floatingfigure}{4cm}
  \begin{center}
    \includegraphics[bbllx=71,bblly=577,bburx=219,bbury=721,width=4cm]{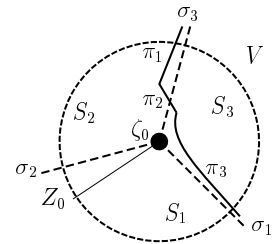}
  \end{center}
    \caption{The geometry in case a)}\label{fig:mat-mono1}
\end{floatingfigure}
%
%
\smallpagebreak The lines of Stokes type $\im\int^{\zeta}\kappa
d\zeta=\Const$ are tangent to the vector field
$\overline{\kappa(\zeta)}$ (as usual, we identify complex numbers with
vectors in $\R^2$). As $\sigma_1$ is vertical in $V$, it intersects
$Z$ (the set where $\im\kappa=0$) in $V$ only at $\zeta_0$. So, all
the connected components of $Z$ except $Z_0$ stay at a finite distance
from $\sigma_1$ (in $V$). Therefore, if $V$ is sufficiently small,
$Z_0$ is the only connected component of $Z$ in $V$.  Furthermore, as
$\sigma_1$ and $\sigma_3$ are vertical in $V$, $Z_0$ stays inside the
sector $S_1\cup \sigma_2\cup S_2$.
\smallpagebreak In a neighborhood of $\sigma_1$ to the left of
$\sigma_1$, the assumptions of the Stokes Lemma guaranty that
$\im\kappa>0$. So, we see that $\im\kappa$ remains positive in the
part of $V$ situated between $\sigma_1$ and $Z_0$ and adjacent to
$\sigma_1$ from the left. Also, $\im\kappa$ does not vanish in the
part of $V$ situated between $Z_0$ and $\sigma_1$ and adjacent to
$\sigma_1$ from the right. But as
$\kappa\sim\kappa_1\sqrt{\zeta-\zeta_0}$ for $\zeta\sim\zeta_0$, in
this sector, $\im\kappa<0$. This completes the proof of
Lemma~\ref{z1}. \qed
\vskip.2cm\noindent Now, we construct a pre-canonical curve $\pi$, and
use Proposition~\ref{pro:pcl:1} to find a canonical line $\varkappa$
close to $\pi$. The line $\pi$ is situated in $V'$ and is
pre-canonical with respect to the branch $\kappa$. It consists of
$\pi_1$, $\pi_2$ and $\pi_3$, three segments of lines of Stokes type.
\smallpagebreak Begin with describing $\pi_1$. Fix $a_1$, a point on
the boundary of $V$ between $Z_0$ and $\sigma_3$ (see
Fig.~\ref{fig:mat-mono1}). Consider $l_1$, the line of Stokes type
$\im\int_{a_1}^\zeta\kappa d\zeta=0$ passing through $a_1$. Recall
that $\sigma_2$ and $\sigma_3$ also are the lines of Stokes type
$\im\int^\zeta\kappa d\zeta=\Const$. As this family fibrates $S_2$, by
making $a_1$ close enough to $\sigma_3$, $l_1$ can be made arbitrarily
close to $\sigma_3\cup \sigma_2$. In addition, $l_1$ does not
intersect $\sigma_3\cup \sigma_2$.  We assume that $a_1$ is so close
to $\sigma_3$ that $l_1$ enters in $V$ at $a_1$ and goes downwards
from $a_1$. On $l_1$, we pick a point $a_2$ so that $\im \zeta_0<\im
a_2<\im a_1$ and so that the segment of $l_1$ between $a_1$ and $a_2$
is between $\sigma_3$ and $Z_0$.  This segment is the segment $\pi_1$.
Note that $\im\kappa<0$ along $\pi_1$ and that, as the line $l_1$ is
tangent to the vector field $\overline{\kappa}$, $\pi_1$ is vertical.
Let us underline also that, taking $a_1$ close enough to $\sigma_3$,
one can get $a_2$ arbitrarily close to $\zeta_0$.
\vskip.1cm\noindent To describe $\pi_2$, the second segment of $\pi$,
consider $l_2$, the line of Stokes type $\im\int_{a_2}^\zeta
(\kappa-\pi)d\zeta=0$ containing $a_2$. As $l_2$ is tangent to the
vector field $\overline{\kappa(\zeta)}-\pi$, it transversally
intersects $l_1$ at $a_2$. As $\im\kappa(a_2)<0$, the line $l_2$
intersects $\sigma_3$ from the left to the right and going downwards.
We make $a_1$ and $a_2$ so close to $\sigma_3$ that $l_2$ intersects
$\sigma_3$ in the same manner and staying vertical between $a_2$ and
$\sigma_3$. The segment $\pi_2$ is just a segment of $l_2$ connecting
the point $a_2$ to some point, say $a_3$, in $S_3$.  Clearly, $\pi_2$
is vertical, and $\im a_3<\im a_2$ (see Fig.~\ref{fig:mat-mono1}).
Note that $a_3$ can be taken arbitrarily close to $\sigma_3$.
\smallpagebreak The last segment of the pre-canonical line is a
segment of $l_3$, the line of Stokes type $\im\int_{a_3}^\zeta\kappa
d\zeta=0$ containing $a_3$. This line is tangent to the vector field
$\overline{\kappa}$. As $\im\kappa\ne0$ in $S_3$, $l_3$ is vertical in
$S_3$. The segment $\pi_3$ is the connected component of $l_3\cap S_3$
beginning at $a_3$ and going downwards. As the lines of Stokes type
$\im\int^\zeta\kappa d\zeta=\Const$ fibrate $S_3$, the line $\pi_3$
does not intersect neither $\sigma_3$ nor $\sigma_1$, and, choosing
$a_3$ close enough to $\sigma_3$, we can make $\pi_3$ arbitrarily
close to $\sigma_1\cup \sigma_3$. The segment $\pi_3$ is shown in
Fig.~\ref{fig:mat-mono1}.
\smallpagebreak The line $\pi$ being pre-canonical, by
Proposition~\ref{pro:pcl:1}, there exists a canonical line arbitrarily
close to $\pi$, say $\varkappa$. We can and do assume that the line
$\varkappa$ begins at $a_1$ and that $\zeta_0$ and $\sigma_1$ stay to
the left of $\varkappa$. Fix $\delta$ positive.  Choosing $\pi$ close
enough to $\sigma_1\cup\sigma_3$, we can assume that $\varkappa$ is in
the $\delta$-neighborhood of $\sigma_1\cup\sigma_3$.
\smallpagebreak Let $Y_1$ and $Y_2$ denote the imaginary parts of
the ends of $\varkappa$ so that $Y_1<Y_2$.
\smallpagebreak By Lemma~\ref{LCD}, there is a canonical domain $K$
enclosing $\varkappa$. We can (and do) assume that $K$ is situated in
$V'$ and in the $\delta$-neighborhood of $\sigma_1\cup\sigma_3$. Note
that, by construction, the point $\zeta_0$ and the Stokes line
$\sigma_1$ are to the left of $K$.
\smallpagebreak The strip $\{Y_1<\im\zeta<Y_2\}$ is the smallest
``horizontal'' strip containing $K$. Consider also the smallest
``horizontal'' strip $\{\tilde Y_2<\im\zeta< Y_2\}$ containing $K\cap
S_2$. As $a_2$ can be made arbitrarily close to $\zeta_0$ in the
construction of the pre-canonical line $\pi$, $\tilde Y_2$ can also be
made arbitrarily close to $\im\zeta_0$.
\subsubsection{Asymptotics of $a$ and $b$}
\label{sec:asymptotics-a-b}
Let $z_1$ be the lower end of $\sigma_1\cap V$, and let $z_2$ be the
upper end of $\sigma_3\cap V$. Fix $\delta_1>0$. If $\delta$ is
sufficiently small, then, $ Y_1<\im z_1+\delta_1$ and $Y_2>\im z_2-
\delta_1$. We prove
\begin{Le}
  \label{le:a}
  Fix $\delta_1$ positive. If $\delta$ is sufficiently small, then,
  for $\varepsilon\to 0$,
  \begin{equation}
    \label{a,b}
    a=1+o(1),\quad{\rm and}\quad \quad b=O\left(e^{-\eta/\varepsilon}\,
    e^{\frac{2i}{\varepsilon}\int_{\zeta_*}^{\zeta_0}\kappa d\zeta}\right),\quad
    \im z_1+\delta_1<\im\zeta<\im z_2-\delta_1,
  \end{equation}
  where $\eta$ is a positive constant (independent of $\varepsilon$).
  The estimates~\eqref{a,b} are uniform in $\zeta$.
\end{Le}
\demo In the proof of Lemma~\ref{le:a}, $C$ denotes different positive
constants independent of $\varepsilon$ and $\delta$. The proof of the
asymptotics of $a$ consists of three steps.
\smallpagebreak {\bf 1.} \ Recall that $a$ is given by~\eqref{f:ab}.
So, we need to compute $w(f,f_-)$. Above $\zeta_0$, in the domain
$K\cap S_2$, all the solutions $f$ and $f_\pm $ have standard
asymptotic behavior.  Moreover, in this region, the asymptotics of $f$
and of $f_+$ coincide.  Therefore, here, one has
$w(f,f_-)=w(f_+,f_-)(1+o(1))$, and $a$ admits the asymptotics
\begin{equation}
  \label{a:up}
  a=1+o(1).
\end{equation}
It is locally uniform. As $a$ is $\varepsilon$-periodic, this
asymptotics remains true in the strip $\{\tilde Y_2<\im\zeta< Y_2\}$.
\smallpagebreak {\bf 2.} \ Below $\zeta_0$, we can only estimate
$a$. We use Lemma~\ref{lempro}. To apply this lemma, we pick the
points $\zeta_1$ and $\zeta_2$ so that $\zeta_1\in S_1$ and
$\zeta\in K$ ($\im\zeta_1=\im\zeta_2<\im\zeta_0$). Then,
\begin{equation*}
|f(x,\zeta)|\leq C \left|e^{\dsize\frac
i{\varepsilon}\int_{\zeta_*}^{\zeta_1} \kappa d\zeta}\right|\cdot
e^{\dsize\frac1{\varepsilon}\int_{\zeta_1}^{\zeta}
      |\im\kappa| d\zeta}, \quad \zeta\in
      [\zeta_1,\zeta_2].
\end{equation*}
Here, the first integral is taken along a curve in $V'$, and the
second one is taken along $[\zeta_1,\zeta_2]$. Assume that
$\zeta_1$ and $\zeta_2$ are in the $\delta$-neighborhood of
$\sigma_1$.  Let $\zeta_b=[\zeta_1,\zeta_2]\cap \sigma_1$. Then
\begin{equation*}
|f(x,\zeta)|\leq C \left|e^{\dsize\frac
i{\varepsilon}\int_{\zeta_*}^{\zeta_b} \kappa
d\zeta}\right|\,\cdot e^{\dsize\frac{C\delta}{\varepsilon}}, \quad
\zeta\in [\zeta_1,\zeta_2].
\end{equation*}
Using the Stokes line definition,  we get finally
\begin{equation*}
|f(x,\zeta)|\leq C \left|e^{\dsize\frac
i{\varepsilon}\int_{\zeta_*}^{\zeta_0} \kappa d\zeta}\right|\cdot
e^{\dsize\frac{C\delta}{\varepsilon}}, \quad \zeta\in
      [\zeta_1,\zeta_2].
\end{equation*}
The derivative $\D\frac{\partial f}{\partial x}$ satisfies an
analogous estimate. Using the asymptotics of $f_-$, we get also
\begin{equation*}
|f_-(x,\zeta)|\leq
C\,\left|e^{\dsize-\frac{i}{\varepsilon}\int_{\zeta_*}^{\zeta_b}
\kappa d\zeta}\right|\,\,\left|
e^{\dsize-\frac{i}{\varepsilon}\int_{\zeta_b}^{\zeta} \kappa
d\zeta}\right|\le
C\,\left|e^{\dsize-\frac{i}{\varepsilon}\int_{\zeta_*}^{\zeta_0}
\kappa d\zeta}\right|\,\,e^{\dsize \frac{C\delta}{\varepsilon}},
\quad \zeta\in [\zeta_1,\zeta_2]\cap K,
\end{equation*}
where the integral is taken along a curve in $V'$.  Again, an
analogous estimate holds for $\D\frac{\partial f_-}{\partial x}$.  The
estimates for $f$ and $f_-$ allow to estimate their Wronskian, and to
get
\begin{equation}\label{a:down}
|a|\le Ce^{\dsize\frac {C\delta} {\varepsilon}}.
\end{equation}
As $a$ is $\varepsilon$-periodic, this estimate is valid and uniform
along any fixed line $\im \zeta=\Const$ in the strip $\{Y_1<\im
\zeta<\im\zeta_0\}$.
\smallpagebreak {\bf 3.} \ Now, the statement of Lemma~\ref{le:a}
concerning $a$ follows from estimates of the Fourier coefficients of
$a$. Fix $\nu>0$ sufficiently small. Then, for sufficiently small
$\varepsilon$, $a$ is analytic in a strip $\{Y_1+\nu\le \im\zeta\le
Y_2-\nu\}$. So, here, we can expand $a$ in a Fourier series with
exponentially decreasing coefficients; for any $\zeta'\in\{Y_1+\nu\le
\text{Im}\zeta\le Y_2- \nu\}$, one has
\begin{equation}
  \label{a:F}
  a(\zeta)=\sum_{n\in\Z}a_n e^{2\pi i
  n\frac{\zeta-\zeta_0}\varepsilon} \text{ where
  }a_n=\frac1{\varepsilon}\int_{\zeta'}^{\zeta'+\varepsilon}
  a(\zeta)e^{-2i\pi n\frac{\zeta-\zeta_0}\varepsilon}d\zeta.
\end{equation}
To estimate Fourier coefficients $(a_n)_{n\le 0}$, one uses the
estimate~\eqref{a:up} and~\eqref{a:F} with $\im \zeta'=Y_2-\nu$. This
gives
\begin{equation}
  \label{eq:1}
  a_0=1+o(1),\quad
  |a_n|\leq C e^{-2\pi |n|\, |Y_2-\nu-\im \zeta_0|/\varepsilon}.
\end{equation}
To estimate $(a_n)_{n>0}$, one uses~\eqref{a:down} and~\eqref{a:F}
assuming that $Y_1+\nu=\im\zeta'$. This yields
\begin{equation}
  \label{eq:2}
  |a_n|\leq C e^{C\delta/\varepsilon}
  e^{-2\pi |n|\,|\im \zeta_0-Y_1-\nu|/\varepsilon}.
\end{equation}
The estimates~\eqref{eq:1} and~\eqref{eq:2} are valid for sufficiently
small $\varepsilon$.  They imply the statement of Lemma~\ref{le:a}
concerning $a$.
\smallpagebreak The analysis of $b$ is also done in three steps.
Recall that $b$ is given by~\eqref{f:ab}. So, we need to study the
Wronskian $w(f,f_+)$.
\smallpagebreak {\bf 1.} \ First, we study $b$ above $\zeta_0$.  We
choose $\zeta\in K\cap S_2$.  Then, both $f$ and $f_+$ have the same
asymptotics.  So, we get
\begin{equation}
  \label{b:1}
  |w(f,f_+)|\le C\left|e^{\frac{2i}{\varepsilon}\int_{\zeta_b}^{\zeta}
      \kappa d\zeta}\cdot 
    e^{\frac{2i}{\varepsilon}\int_{\zeta_0}^{\zeta_b}\kappa d\zeta}\cdot 
    e^{\frac{2i}{\varepsilon}\int_{\zeta_*}^{\zeta_0}\kappa d\zeta}\right|,
\end{equation}
where $\zeta_b\in \sigma_3$ has the same imaginary part as $\zeta$. In
the first and the last integral, we integrate along curves in $V'$; in
the second integral we can integrate along the Stokes line $\sigma_3$,
hence, $\left|e^{\frac{2i}{\varepsilon}\int_{\zeta_0}^{\zeta_b}\kappa
    d\zeta}\right|=1$.\\
Consider the first integral. Let $D_0$ be the domain situated between
$Z_0$ and $\sigma_3$ where $\im\kappa<0$. For $c>0$, let $D_c$ be the
domain $D_0$ without the $c$-neighborhood of its boundary. In $D_c$,
one has $\left|e^{\frac{2i}\varepsilon \int_{\zeta_b}^\zeta\kappa
    d\zeta}\right|\le e^{-\eta/\varepsilon}$, where $\eta=\eta(c)$ is
positive. This implies that, in $D_c$ , we have $w(f,f_+)=O(
e^{-\eta/\varepsilon}\,e^{\frac{2i}\varepsilon
  \int_{\zeta_*}^{\zeta_0} \kappa d\zeta})$, and
\begin{equation}
  \label{st-lm:b}
  b=O(e^{-\eta/\varepsilon}\,
  e^{\frac{2i}{\varepsilon}\int_{\zeta_*}^{\zeta_0}\kappa d\zeta} ).
\end{equation}
Recall that $b$ is $\varepsilon$-periodic. Therefore, this estimate
holds in $S(D_c):=\{y_1<\im\zeta<y_2\}$, the smallest strip containing
$D_c$. This and the construction of the domain $K$ imply that, for any
fixed $\delta_3$, and sufficiently small $\varepsilon$, there is an
$\eta>0$ such that estimate~\eqref{st-lm:b} is uniform in the strip
$\{\tilde Y_2+\delta_3<\im\zeta< Y_2-\delta_3\}$.
\smallpagebreak {\bf 2.} \ To get an estimate below $\zeta_0$, we
proceed in the same way as for $a$ and  get $|b|\le
C\,\left|e^{\frac{2i}{\varepsilon}\int_{\zeta_*}^{\zeta_0}\kappa
d\zeta} \right|\cdot e^{\dsize\frac {C\delta} {\varepsilon}}$.
This estimate is valid and uniform along any fixed line $\im
\zeta=\Const$ in the strip $\{ Y_1<\im \zeta<\im\zeta_0\}$ for
sufficiently small $\varepsilon$.
\smallpagebreak {\bf 3.} \ The estimate for $b$ given in
Lemma~\ref{a,b} then follows from the analysis of the Fourier
coefficients of $b$ and the estimates obtained in the steps 1. and 2.
As it is similar to the analysis of $a$, we omit it.  \qed
\smallpagebreak {\it The asymptotics of $f$.} \ We know the
asymptotics of $f_\pm $, of $a$ and of $b$ in the domain $K\cap \{\im
z_1+\delta_1\le \im\zeta\le \im z_2-\delta_1\}$.  Substituting them
into~\eqref{f:fpm}, in $K\cap \{\im z_1+\delta_1\le \im\zeta\le \im
z_2-\delta_1\}$, we get
\begin{equation}\label{f:estimate}
  f=e^{\frac i\varepsilon \int_{\zeta_*}^\zeta \kappa d\zeta}
  \left(\Psi_+(x,\zeta,\zeta_*)+o(1)+O\left[
      e^{-\eta/\varepsilon-\frac{2i}\varepsilon \int_{\zeta_0}^\zeta \kappa
        d\zeta}\right]\right).
\end{equation}
The term $T=\im\int_{\zeta_0}^\zeta \kappa d\zeta$ is negative inside
the sector $S_3$ bounded by $\sigma_1$ and $\sigma_3$.  Indeed, we can
integrate on the curve going first along either $\sigma_1$ or
$\sigma_3$ to the point $\zeta_b$ with the same imaginary part as
$\zeta$, and then, along the line $\im\zeta=\Const$ to the point
$\zeta$. Hence, $T=\im \int_{\zeta_b}^\zeta \kappa d\zeta$. As $\im
\kappa<0$ in $S_3$, the term $T$ is negative. This implies that $f\sim
e^{\frac i\varepsilon \int_{\zeta_*}^\zeta\kappa d\zeta}
\Psi_+(x,\zeta,\zeta_*)$ both inside $K\cap S_3$ and, even, in the
part of a constant neighborhood of $\sigma_3$ situated in $K$ (because
of the factor $e^{-\eta/\varepsilon}$ in~\eqref{f:estimate}).
\smallpagebreak By assumption, in $S_1\cup\sigma_2\cup S_2$, one has
$f\sim e^{\frac i\varepsilon \int_{\zeta_*}^\zeta \kappa d\zeta}
\Psi_+(x,\zeta,\zeta_*)$. So, we see that this asymptotics is valid
locally uniformly in the whole domain $K\cap \{\im z_1+\delta_1\le
\im\zeta\}$.
\smallpagebreak Let us discuss the behavior of $f$ in $V$ outside $K$.
Both in $K$ and to the right of it (inside $V$), one has
$\im\kappa<0$. Fix $\delta_2>\delta_1$. Applying the Rectangle Lemma,
one sees that the standard asymptotics holds in the part of $V'$
situated in the strip $\{\im z_1+\delta_2\le \im\zeta\le \im
z_2-\delta_2\}$ to the right of $K$.
\smallpagebreak We have to justify the standard behavior of $f$ in the
rest of $S_3\cup(\sigma_3\cap V')$. Therefore, instead of $K$, we can
consider a similar canonical domain constructed for a smaller value of
the constant $\delta$ and for $\tilde Y_2$ closer to $\im\zeta_0$.  As
the constant $\delta$ (and, thus $\delta_1$ and $\delta_2$) can be
made arbitrarily small and as $\tilde Y_2$ can be made arbitrarily
close to $\im\zeta_0$, we conclude that, locally uniformly, $f$ has
the standard asymptotics in $S_3\cup(\sigma_3\cap V')$ and, therefore
in the whole domain $V'$. This completes the proof of The Stokes Lemma
in the case a).
\subsubsection{The proof of the Stokes Lemma in the case b)}
\label{sec:proof-stokes-lemma-1}
\smallpagebreak{\it Constructing the local canonical domain.\/} \ In
the case b), the Stokes line $\sigma_3$ goes to the right of
$\zeta_0$; the tangent vector to $\sigma_3$ at $\zeta_0$ is
horizontal. The tangent vector to $\sigma_2$ at $\zeta_0$ is oriented
upwards (see Fig.~\ref{fig:mat-mono2}). We assume that $V$ is
sufficiently small so that $\sigma_2\cap V$ is vertical.
\smallpagebreak Now, instead of Lemma~\ref{z1}, we get
\begin{Le}
  \label{z2}
  If $V$ is sufficiently small, and $\alpha=2\pi/3$, then, in $V'$,
  $\im\kappa$ vanishes only along $Z_0$, an analytic curve beginning
  at $\zeta_0$. The tangent vector to $Z_0$ at $\zeta_0$ is
  horizontal; $Z_0$ is going to the right from $\zeta_0$.  In the
  sector of $V'$ bounded by $\sigma_1$ and $Z_0$ and to the left of
  $\sigma_1$, one has $\im\kappa>0$. In the rest of $V'\setminus Z_0$,
  one has $\im\kappa<0$.
\end{Le}
\noindent  Being similar to that of Lemma~\ref{z1}, the proof of
Lemma~\ref{z2} is omitted.
\smallpagebreak Now, we construct the pre-canonical curve $\pi$. It is
situated in $V'$ and consists of three segments of lines of Stokes
type, say $\pi_1$, $\pi_2$ and $\pi_3$.
\smallpagebreak The segment $\pi_2$ is a segment of the line
$\re\zeta=\Const$ intersecting $Z_0$ close enough to $\zeta_0$. The
upper end of $\pi_2$, say $a_2$, belongs to $S_2$; $a_3$, the other
end of $\pi_2$, is in $S_3$. Choosing the intersection point close
enough to $\zeta_0$, we can make $\pi_2$ arbitrarily small. If $\pi_2$
is in a sufficiently small neighborhood of $\zeta_0$, then, it is a
canonical line. To justify this, one uses the fact that, in a
neighborhood of $\zeta_0$, $\kappa$ is analytic in
$\sqrt{\zeta-\zeta_0}$ and admits the representation~\eqref{k:sqrt}
with a non-zero constant $\kappa_1$. Omitting the elementary details,
we only make a remark on the sign of this constant. Choose the branch
of the square root in~\eqref{k:sqrt} so that $\sqrt{\zeta-\zeta_0}>0$
when $\im(\zeta-\zeta_0)=0$ and $\re(\zeta-\zeta_0)>0$. Then,
$\kappa_1$ is positive (as $\im\kappa>0$ above $Z_0$, and
$\im\kappa=0$ along $Z_0$).
%
%
\begin{floatingfigure}{4cm}
  \begin{center}
    \includegraphics[bbllx=71,bblly=577,bburx=219,bbury=721,width=4cm]{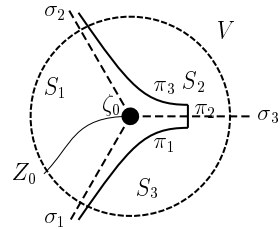}
  \end{center}
    \caption{The geometry in case b).}\label{fig:mat-mono2}
\end{floatingfigure}
%
%
\smallpagebreak Consider $l_1$, the line of Stokes type
$\im\int_{a_3}^\zeta\kappa d\zeta=0$ containing $a_3$. As the lines of
Stokes type $\im\int^\zeta\kappa d\zeta=\Const$ fibrate $S_3$,
choosing $\pi_2$ so that $a_3$ be close enough to $\sigma_3$, we can
make $l_1$ arbitrarily close to $\sigma_3\cup\sigma_1$.  Clearly,
$l_1$ does not intersect $\sigma_3\cup\sigma_1$. Recall that
$\im\kappa\ne 0$ in $V'$ below $Z_0$. Therefore $l_1$ is vertical at
$a_3$. If $a_3$ is close enough to $\zeta_0$ (and, thus, to
$\sigma_3\cup \sigma_1$), then, below $a_3$, the line $l_1$ stays
below $Z_0$.  Then, $\im\kappa\ne 0$ along $l_1$, and $l_1$ is
vertical in $V'$ also below $a_3$. We assume that this is the case.
The segment $\pi_1$ is the segment of $l_1$ going downwards from $a_3$
in $S_3$ to a point of the boundary of $V'$.
\smallpagebreak Let $l_2$ be the line of Stokes type
$\im\int_{a_2}^\zeta\kappa d\zeta=0$ containing $a_2$.  If $a_2$ is
close enough to $\zeta_0$, then, this line is arbitrarily close to
$\sigma_3\cup\sigma_2$. It does not intersect neither $\sigma_3$ nor
$\sigma_2$ and is vertical in $S_2$. It goes from $a_2$ upwards to
$a_1$, a point of the boundary of $V$.  The segment $\pi_3$ is just
the segment of this line between $a_2$ and $a_1$.
\vskip.2cm\noindent The line $\pi$ being pre-canonical, by
Proposition~\ref{pro:pcl:1}, arbitrarily close to $\pi$ , there exists
$\varkappa\subset S_2\cup \sigma_3\cup S_3$, a canonical line. Fix
$\delta$ positive. Choosing $\pi$ close enough to
$\sigma_1\cup\sigma_2$, we can assume that $\varkappa$ is in the
$\delta$-neighborhood of $\sigma_1\cup\sigma_2$.  By construction,
$\sigma_1\cup\sigma_2$ stays to the left of $\varkappa$. We denote by
$Y_1$ and $Y_2$ the imaginary parts of the ends of $\varkappa$ in $V$
so that $Y_1<Y_2$ (see Fig.~\ref{fig:mat-mono2}).
\smallpagebreak By Lemma~\ref{LCD}, there exists $K\subset S_2\cup
\sigma_3\cup S_3$, a canonical domain enclosing $\varkappa$
situated in the $\delta$-neighborhood of $\sigma_1\cup\sigma_2$.
By construction,  $\sigma_1\cup\sigma_2$ is to the left of $K$.
The strip $\{Y_1<\im\zeta<Y_2\}$ is the smallest ``horizontal''
strip containing $K$.
\smallpagebreak {\it Asymptotics of $a$ and $b$.\/} \ Let $z_1$ be the
lower end of $\sigma_1\cap V'$, and let $z_2$ be the upper end of
$\sigma_2\cap V'$. Fix $\delta_1>0$.  With these notations, the
``new'' coefficients $a$ and $b$ are described by Lemma~\ref{le:a}.
Let us discuss how the proof of Lemma~\ref{le:a} is modified.
\smallpagebreak The proof of the asymptotics of $a$ remains the same.
As about the asymptotics of $b$, only the step 1 (describing the
asymptotics of $b$ above $\zeta_0$) has to be modified. Let us give
the details.
\smallpagebreak {\it New step 1.} \ To get estimate~\eqref{a,b} for
$b$, we choose $\zeta$ in $K\cap S_2$.  There, both $f$ and $f_+$ have
the same asymptotics. Assuming in addition that $\im \zeta>\im
\zeta_0$, we again get~\eqref{b:1} where $\zeta_b\in \sigma_2$ has the
same imaginary part as $\zeta$, and in the second integral we
integrate along the Stokes line $\sigma_2$.  Let us discuss the
exponentials in~\eqref{b:1}.  As $\sigma_2$ is a Stokes line,
$\left|e^{\frac{2i}{\varepsilon}\int_{\zeta_0}^{\zeta_b}\kappa
    d\zeta}\right|=1$.  Assume that $\zeta$ is above $Z_0$. Then, $\im
\kappa>0$. Consider $\zeta$ such that, between the points $\zeta$ and
$\zeta_b$ (along the horizontal segment connecting them), one has
$\im\kappa>C>0$. Then $\left|e^{\frac{2i}{\varepsilon}
    \int_{\zeta_b}^{\zeta}\kappa d\zeta}\right|\le
e^{-2Cd/\varepsilon}$, where $d=|\zeta-\zeta_b|$.  In result, we see
that, in $K\cap S_2$, above any fixed constant neighborhood of $Z_0$,
\begin{equation*}
  b=O\left(e^{-\eta/\varepsilon}\,
    e^{\frac{2i}{\varepsilon}\int_{\zeta_*}^{\zeta_0}\kappa
      d\zeta}\right)
\end{equation*}
with a positive constant $\eta$ independent of $\varepsilon$.\\
Pick a $\delta_3>0$. Making $\delta$ smaller if necessary, we can get
that $Z_0$ is below the line $\im\zeta=\im\zeta_0+\delta_3$.  As $b$
is $\varepsilon$-periodic, we can conclude that, for sufficiently
small $\varepsilon$, there exists $\eta>0$ such that the last estimate
for $b$ holds locally uniformly in the strip $\{\im\zeta_0+\delta_3<
\im\zeta<Y_2\}$.
\vskip.1cm\noindent{\it The asymptotics of $f$.}  After having proved
Lemma~\ref{le:a}, the asymptotic of $f$ is derived almost in the same
way as in case a). In the domain $K\cap\{\im z_1+\delta_1\le\im\zeta
\le\im z_2-\delta_1\}$, we again get the
representation~\eqref{f:estimate}.  The new element is that the line
$Z_0$ (or a part of it) can now be situated in $S_3$.  This requires a
minor modification of the analysis.
\smallpagebreak Again one proves that, in~\eqref{f:estimate}, the
term $T(\zeta)=\im \int_{\zeta_0}^\zeta\kappa d\zeta$ is negative
in the sector $S_3$. If $Z_0$ does not enter the sector $S_3$, the
proof remains the same as in case a). Otherwise, arguing as in the
case a), one sees only that $T$ is negative in $S_3$ below $Z_0$.
Then, we note that in $V'$ (if $V$ is chosen sufficiently small),
$T(\zeta)$ vanishes only on the Stokes lines $\sigma_1$,
$\sigma_2$ and $\sigma_3$. These two observations imply that
$T(\zeta)< 0$ in the whole sector $S_3$. In result, as in case a),
we again conclude that, in the domain $K\cap\{\im
z_1+\delta_1\le\im\zeta\le \im z_2-\delta_1\}$, below $\sigma_3$
and in a constant neighborhood of $\sigma_3$, the solution $f$ has
the asymptotics $f\sim e^{\frac
  i\varepsilon\int_{\zeta_*}^\zeta\kappa d\zeta}
\Psi_+(x,\zeta,\zeta_*)$.
\smallpagebreak If $Z_0$ is outside the sector $S_3$, one completes
the proof as in case a).  Otherwise, arguing as in case a),
one sees only that $f$ has the desired asymptotics
\begin{itemize}
\item (1) in $S_1\cup S_2$ (by the assumptions of the Stokes Lemma);
\item (2) in the whole domain $K\cap\{\im\zeta_1+\delta_1\le
  \im\zeta\}$ (by the previous analysis and by (1));
\item (3) to the right of $K$ below $Z_0$ (by the Rectangle Lemma as
  in the case a)).
\item (4) to the left of $K$ and below the line
  $\im\zeta=\im\zeta_1+\delta_1$ (as in case a)).
\end{itemize}
This is sufficient. Indeed, one can reduce the size of $V$ so that the
new smaller $V'$ be contained in the union of the domains mentioned in
the above list. Then, for this new $V'$, the statement of the Stokes
Lemma has been proved.
\subsubsection{The proof of the Stokes Lemma in the case c)}
\label{sec:proof-stokes-lemma-2}
\smallpagebreak{\it Constructing the local canonical domain.\/} \ Now,
starting from $\zeta_0$, the Stokes lines $\sigma_1$ and $\sigma_3$ go
downwards, and $\sigma_2$ goes upwards (see Fig.~\ref{fig:mat-mono3}).
We assume that $V$ is so small that all three Stokes lines be vertical
in $V$. We use
\begin{Le}
  \label{z3}
  If $V$ is sufficiently small, and $2\pi/3<\alpha<\pi$, then, in
  $V'$, $\im\kappa$ vanishes only along $Z_0$, an analytic curve
  beginning at $\zeta_0$. The line $Z_0$ is vertical; starting from
  $\zeta_0$, it goes downwards staying in the sector $S_3$. In the
  sector of $V'$ bounded by $\sigma_1$ and $Z_0$ and to the left of
  $\sigma_1$, one has $\im\kappa>0$. In the rest of $V'\setminus Z_0$,
  one has $\im\kappa<0$.
\end{Le}
\noindent The proof of Lemma~\ref{z3} is similar to the one of
Lemma~\ref{z1} and is omitted.
\smallpagebreak To construct the pre-canonical line $\pi$, first
consider the line $\tilde\pi$ made of four segments of
``elementary'' lines $\pi_1$, $\pi_2$, $\pi_3$ and $\pi_4$, see
Fig.~\ref{fig:mat-mono3}. Let us briefly describe these segments
and their properties (the detailed analysis is similar to the one
done in the cases a) and b)).
%
%
\begin{floatingfigure}{4cm}
  \begin{center}
  \includegraphics[bbllx=71,bblly=577,bburx=219,bbury=721,width=4cm]{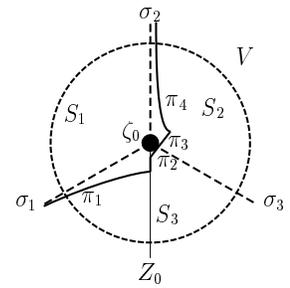}
  \end{center}
    \caption{The geometry in case c).}\label{fig:mat-mono3}
\end{floatingfigure}
%
%
\smallpagebreak The segment $\pi_1$ begins at $a_1$, a point of the
common part of the boundary of $V$ and $S_3$ situated strictly between
$\sigma_1$ and $Z_0$ (close enough to $\sigma_1$). This segment is a
segment of the line of Stokes type $\im\int_{a_1}^\zeta\kappa
d\zeta=0$. It stays in $S_3$ and connects the point $a_1$ to $a_2$, a
point of $Z_0$. Below $a_2$, it stays to the left of $Z_0$ and is
vertical. Taking $a_1$ close enough to $\sigma_1$, we can make $\pi_1$
arbitrarily close to $\sigma_1$.
\smallpagebreak The segment $\pi_2$ is a segment of $Z_0$ between
$a_2$ and $a_3$, an internal point of $Z_0$ such that $\im a_2<\im
a_3<\im\zeta_0$. We assume that $a_2$ and $a_3$ are close enough to
$\zeta_0$. Then, $Z_0$ is vertical above $a_2$, and, $0<\kappa<\pi$ on
$Z_0$. This implies that the segment $\pi_2$ is a canonical line.
\smallpagebreak We choose $a_3$ close enough to $\zeta_0$ and
construct the segment $\pi_3$ in a sufficiently small neighborhood of
$\zeta_0$.  It is a segment of $l_3$, the line of Stokes type
$\im\int_{a_3}^\zeta (\kappa-\pi)d\zeta=0$.  Beginning at $a_3$, it
goes to the right of $Z_0$. To the right of $a_3$, it is vertical and
goes upward, at least, while staying in $V '$ to the right of $Z_0$.
Above $a_3$, it can not come back to $Z_0$ without leaving $V'$ (this
follows from the analysis of the vector field $\overline{\kappa}-\pi$
near $Z_0$ to the right of it).  Therefore, in $V'$, $l_1$ stays
vertical above $a_3$. Moreover, if $a_3$ is close enough to $\zeta_0$,
then, $l_1$ intersects $\sigma_3$ above $a_3$. The segment $\pi_3$ is
the segment of $l_1$ between $a_3$ and $a_4$, a point of $S_2$. We
underline that, above $a_3$, \ $\pi_3$ is vertical, goes inside
$S_3\cup\sigma_3\cup S_2$ staying to the right of $\zeta_0$, and that
it can be constructed in an arbitrarily small neighborhood of
$\zeta_0$.
\vskip.1cm\noindent The segment $\pi_4$ is a segment of the line of
Stokes type $\im\int_{a_4}^\zeta\kappa d\zeta=0$. It goes upward from
$a_4$, is vertical above $a_4$ and, without intersecting
$\sigma_3\cup\sigma_2$, connects the point $a_4$ to $a_5$, a point of
the boundary of $S_2$ ($a_5\not\in\sigma_3\cup\sigma_2$). Taking $a_4$
close to $\sigma_3$, we can make $\pi_4$ arbitrarily close to
$\sigma_3\cup\sigma_2$.
\smallpagebreak The line $\tilde\pi$ is the union of the lines
$\pi_1$, $\pi_2$, $\pi_3$ and $\pi_4$. It is not pre-canonical as the
tangent vectors to $\pi_1$ and $\pi_3$ at the points of $Z_0$ are
horizontal. To get a pre-canonical line, we use the $C^1$-stability of
canonical lines and replace $\pi_2$ by a canonical line connecting an
internal point of $\pi_1$ to an internal point of $\pi_3$. This gives
us a pre-canonical line that we call $\pi$.
\smallpagebreak By Proposition~\ref{pro:pcl:1}, arbitrarily close
to $\pi$, there exists $\varkappa\subset S_2\cup \sigma_3\cup S_3$, a
canonical line. It stays to the right of $\sigma_1\cup\sigma_2$ and
can be constructed inside any given neighborhood of
$\sigma_1\cup\sigma_2$.  After having constructed the canonical line,
we complete the proof of the Stokes Lemma in the case c) exactly as in
the case b).
\smallpagebreak This completes the proof of Lemma~\ref{st-lm}, the
Stokes Lemma.\qed


%
\section{Proof of the Two-Waves Principle}
\label{tw:proof}
First, we note that the Wronskian of $h_\pm$ is non-zero. Recall
that $\zeta_0$ is to the right of $\sigma_1\cup\sigma_2$.
Computing the Wronskian at a point $\zeta\in D_+\cap D_-$ situated
to the right of $\sigma_1\cup\sigma_2$ (see Fig.~\ref{FDP:1}), we
get $w(h_+,h_-)= w(\Psi_+(x,\zeta),\Psi_-(x,\zeta))+o(1)$.
By~\eqref{Wcanonical}, the leading term in this formula equals to
$w(\Psi_+(x,\zeta_0),\Psi_-(x,\zeta_0))$, and, as $\zeta_0\not\in
P\cup Q$, the leading term is non-zero. This implies that, for
$\zeta$ in any compact set of $D_+\cap D_-$ and sufficiently small
$\varepsilon$, the solutions $h_\pm$ are linearly independent. So,
we can write~\eqref{f:hpm} with some coefficients $G$ and $g$
independent of $x$. These coefficients can be expressed in terms
of the Wronskians of the solutions:
\begin{equation}
  \label{eq:5}
  g(\zeta)=\frac{w(f,h_-)}{w(h_+,h_-)},\quad
  G(\zeta)=\frac{w(h_+,f)}{w(h_+,h_-)}.
\end{equation}
Recall that, the solutions having the standard behavior, they
satisfy the consistency condition. This implies that both $G$ and
$g$ are periodic (as Wronskians of consistent solutions). Now, to
get the asymptotics of $G$ and $g$, we have only to compute the
Wronskians defining these functions.
\smallpagebreak Begin with computing $g$. First, one assumes that
$\zeta$ is situated in $\zeta\in D\setminus F$ to the right of
$\sigma_1$. Here, the leading terms of the asymptotics of $f$ and
$h_+$ coincide and, as when computing $w(h_+,h_-)$, one gets
$w(f,h_-)= w(\Psi_+(x,\zeta_0),\Psi_-(x,\zeta_0))+o(1)
=w(h_+,h_-)+o(1)$. This, the representation for $g$
in~\eqref{eq:5} and the periodicity of $g$ imply that
\begin{equation}
  \label{eq:6}
  g=1+o(1), \quad \im \zeta_m < \im\zeta<\im\zeta_2,
\end{equation}
where $\zeta_m$ satisfies the inequality $\zeta_m<\zeta_0$ and is
determined by the position of the lower part of the boundary of $D_-$.
To estimate $g$ for $\im\zeta>\im\zeta_2$, we take a point $\zeta\in
D$ situated above the line $\im\zeta=\im\zeta_2$ in the
$\delta$-neighborhood of $\sigma_2$ to the left of $\sigma_2$. One has
\begin{equation}
  \label{eq:7}
  |w(f,h_-)|\le C\, \left|e^{\frac i\varepsilon \int_{\gamma(\zeta)}
      \kappa d\zeta}\,\,\, e^{-\frac i\varepsilon \int_{\gamma_-(\zeta)}
      \kappa d\zeta}\right|,
\end{equation}
where $C$ is a positive constant independent of $\varepsilon$,
$\gamma(\zeta)$ and $\gamma_-(\zeta)$ are two curves connecting
$\zeta_0$ to $\zeta$ in respectively $D\setminus (F\cup\sigma_1)$
and $D_-$, and we integrate the analytic continuations of $\kappa$
along the integration curves. Now, we deform these two curves
(without intersecting the branch points) so that each of them go
first from $\zeta_0$ to $\zeta_2$ (more precisely, to a point
infinitesimally close to $\zeta_2$) and then, along the Stokes
lines $\sigma_2$ and $\sigma_1$ (infinitesimally close to them),
to $\tilde\zeta$, the point of $\sigma_2$ (infinitesimally close
to $\sigma_2$) such that $\im\tilde \zeta=\im\zeta$, see
Fig.~\ref{FDP:4}. Now, discuss the right hand side
in~\eqref{eq:7}. First, consider the parts of the two integration
contours situated between $\zeta_0$ and $\zeta_2$. Their
contributions to the integrals are of opposite sign and, so, they
cancels one another. Furthermore, as
$\im\int^\zeta(\kappa-\kappa(\zeta_{1,2})) d\zeta$ is constant
along the Stokes lines, we see, that
\begin{equation}
  \label{eq:8a}
  \im \left( \int_{\gamma(\zeta)} \kappa
    d\zeta-\int_{\gamma_-(\zeta)} \kappa d\zeta \right)=
  \im \left(
    \int_{\tilde\zeta,\,\,\,\text{along \ }\gamma(\zeta)}^{\zeta}
    \kappa|_\gamma d\zeta
    -\int_{\tilde\zeta,\,\,\,\text{along \ }\gamma_-(\zeta)}^{\zeta}
    \kappa|_{\gamma_-} d\zeta
  \right).
\end{equation}
Here, $\kappa|_\beta(\zeta)$ denotes the value at $\zeta\in\beta$
of the analytic continuation of $\kappa$ along the curve $\beta$.
As $\kappa|_{\gamma(\zeta)}$ and $\kappa|_{\gamma_-(\zeta)}$ are
uniformly bounded, and as $|\tilde\zeta-\zeta|\le \delta$, we see
that the right hand side of~\eqref{eq:8a} is bounded by $C
\delta$. Therefore, $|w(f,h_-)|\le C\,e^{C\delta/\varepsilon}$,
and, so
%
%
\begin{floatingfigure}{4cm}
  \begin{center}
    \includegraphics[bbllx=69,bblly=577,bburx=190,bbury=721,width=4cm]{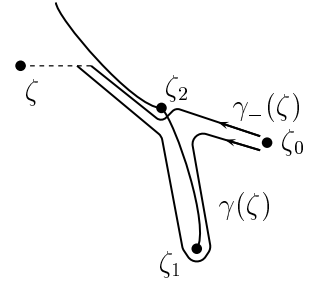}
  \end{center}
    \caption{$\gamma(\zeta)$ and $\gamma_-(\zeta)$}\label{FDP:4}
\end{floatingfigure}
%
%
\begin{equation}
  \label{eq:3}
  |g(\zeta)|\le C\,e^{C\delta/\varepsilon},\quad
  \im\zeta_2<\im\zeta<\im \zeta_M,
\end{equation}
where $\zeta_M>\zeta_2$ is defined by the position of the upper part
of the boundary of $D$. Let us underline that $\delta$ can be fixed
arbitrarily small.
\smallpagebreak Note that, being obtained using the standard
behavior of the solutions, each of the estimates for $g$ is
uniform in $\zeta$ in a given compact set (of the strip where the
estimate was obtained) and in $E$ provided $E$ stays in a
sufficiently small constant neighborhood of $E_0$.
\smallpagebreak Fix $\delta_1$ positive. As $g$ is analytic and
$\varepsilon$-periodic, from~\eqref{eq:6} and~\eqref{eq:3}, we
conclude that, for sufficiently small $\varepsilon$, in the strip
$\zeta_m+\delta_1\le \im\zeta\le \zeta_M-\delta_1$, \ $g$ admits the
asymptotics described in~\eqref{gG}.
\smallpagebreak Now, we compute the asymptotics of $G$. Assume
that $\zeta$ is situated in $D\cap D_+\cap D_-$ to the left of the
line $\sigma_1\cup\sigma_2$. Then, as we shall see, up to constant
factors, all three solutions $f$, $h_-$ and $h_+$ admit asymptotic
representations with one and the same leading term. So, to compute
$G$, one has just to compare the leading terms of the asymptotics
in the right and the left hand sides of~\eqref{f:hpm}.
\smallpagebreak First, we compare the asymptotics of $f$ and
$h_+$. Let $\gamma(\zeta)$ and $\gamma_+(\zeta)$ be curves
connecting $\zeta_0$ to $\zeta$ inside $D\setminus(F\cup\sigma_1)$
and $D_+$ respectively.  Define a curve $\gamma^0$ as shown in
Fig.~\ref{FDP:2}. We can write
\begin{equation}
  \label{eq:9}
  \gamma(\zeta)=\gamma^0+\gamma_+(\zeta),\quad
\end{equation}
\vskip.1cm\noindent Consider $\kappa$, $\omega_+$ and $\psi_+$ (the
functions defining the leading term of the asymptotics of $f$) along
$\gamma^0$. The curve $\gamma^0$ begins and ends at $\zeta_0$ and, so,
is closed.  But, as we are dealing with multi-valued functions, we
shall distinguish between its end and its beginning. We note that
\begin{itemize}
\item as the functions $\psi(x,\zeta)$ and $\omega(\zeta)$ are
  two-valued analytic functions, and as $\gamma^0$ goes around exactly
  two branch points, $\zeta_1$ and $\zeta_2$, the values of $\omega_+$
  and $\psi_+$ at the beginning and at the end of $\gamma^0$ coincide;
\item as, the branch points of $\kappa$ are of square root type, and,
  as $\kappa(\zeta_1)=\kappa(\zeta_2)$, the values of $\kappa$ at the
  beginning and at the end of $\gamma^0$ coincide.
\end{itemize}
The above observations and relation~\eqref{eq:9} show that
\begin{equation}
  \label{eq:15}
  \left.\left. q(\zeta)\,e^{\frac i\varepsilon\,\int_{\gamma(\zeta)}\kappa
        d\zeta+ \int_{\gamma(\zeta)}\omega_+ d\zeta}\psi_+(x,\zeta)
    \right|_{\gamma(\zeta)}=
    A\,q_+(\zeta)\, e^{\frac i\varepsilon\,\int_{\gamma_+(\zeta)}\kappa d\zeta+
      \int_{\gamma_+(\zeta)}\omega_+ d\zeta} \psi_+(x,\zeta)
  \right|_{\gamma_+(\zeta)}
\end{equation}
where $A$ given by~\eqref{tw:AB}, and $q$ and $q_+$ are the
branches of the function $\sqrt{k'({\mathcal E}(\zeta))}$ from the
formulas defining the canonical Bloch solutions from the
asymptotics of $f$ and $h_+$. Comparing this formula with the
asymptotics of $f$ and $h_+$, we see that, for the point $\zeta$
we consider, $f$ admits the representation
\begin{equation}
  \label{eq:10}
  f(x,\zeta)= A\,\tilde h_+(x,\zeta),
\end{equation}
where $\tilde h_+$ is a solution having the same asymptotic
representation as $h_+$ for the point $\zeta$ in consideration.
\smallpagebreak Second, we compare the asymptotics of $h_-$ and
$h_+$. Let $\gamma_\pm(\zeta)$ be curves connecting $\zeta_0$ to
$\zeta$ inside $D\pm$ respectively. Introduce the curve
$\gamma_-^0$  shown in Fig.~\ref{FDP:2}, part b). We can write
\begin{equation}
  \label{eq:11}
  \gamma_-(\zeta)=\gamma_-^0+\gamma_+(\zeta).
\end{equation}
\vskip.1cm\noindent Consider $\kappa$, $\omega_-$ and $\psi_-$ (the
functions in the asymptotics of $h_-$) along $\gamma_-^0$. Again, we
shall distinguish between the end and the beginning of this curve. We
note that
\begin{itemize}
\item as the functions $\psi(x,\zeta)$ and $\omega(\zeta)$ are two
  valued analytic functions, and as $\gamma_-^0$ goes around exactly
  one branch point, after analytic continuation along $\gamma_-^0$,
  the values of $\omega_-$ and $\psi_-$ at the end of $\gamma_-^0$
  coincide with $\omega_+$ and $\psi_+$ at its beginning;
\item as the branch points of $\kappa$ are of square root type, the
  values of $\kappa$ at the beginning and at the end of $\gamma_-^0$,
  say $\kappa_b$ and $\kappa_e$, are related by the formula
  $\kappa_b+\kappa_e=2\kappa(\zeta_2)$.
\end{itemize}
The above observations and relation~\eqref{eq:11} show that
\begin{multline}
  \label{eq:14}
  \left. q_-(\zeta)\,e^{-\frac i\varepsilon\,\int_{\gamma_-(\zeta)}\kappa
        d\zeta+ \int_{\gamma_-(\zeta)}\omega_- d\zeta}\psi_-(x,\zeta)
    \right|_{\gamma_-(\zeta)}=\\=\left.
    e^{-\frac{2i\kappa(\zeta_2)}\varepsilon\,(\zeta-\zeta_2)} B\, \,
    q_+(\zeta)\,e^{\frac i\varepsilon\,\int_{\gamma_+(\zeta)}\kappa d\zeta+
      \int_{\gamma_+(\zeta)}\omega_+ d\zeta} \psi_+(x,\zeta)
  \right|_{\gamma_+(\zeta)}
\end{multline}
with $B$ given by~\eqref{tw:AB}. Comparing this formula with the
asymptotics of $h_-$ and $h_+$, we see that, in a neighborhood of
$\zeta$, \ $h_-$ admits the representation
\begin{equation}
  \label{eq:12}
  h_-(x,\zeta)=
  e^{-\frac{2i\kappa(\zeta_2)}\varepsilon\,(\zeta-\zeta_2)}\, B\,
  \tilde{\tilde h}_+(x,\zeta).
\end{equation}
where $\tilde{\tilde h}_+$ is one more solution having the same
asymptotic representation as $h_+$ for the point $\zeta$ in
consideration.
\smallpagebreak Now, to compute the asymptotics of $G$, we
substitute into~\eqref{f:hpm} the asymptotic
representations~\eqref{eq:10} and~\eqref{eq:12} and the
asymptotic~\eqref{gG} for $g$. This leads to
\begin{equation}
  \label{eq:4}
  e^{-\frac{2i\kappa(\zeta_2)}\varepsilon\,(\zeta-\zeta_2)}\,G\,B=
  A(1+o(1))- 1+o(1).
\end{equation}
This implies formula~\eqref{gG} for $G$ and completes the proof of
Lemma~\ref{tw:1}. The uniformity properties of~\eqref{gG} follow from
the fact that $f$, $h_+$ and $h_-$ have the standard behavior. \qed
\smallpagebreak Note that the representation~\eqref{new-hm:as} follows
from~\eqref{eq:12}. Indeed, the solution $\tilde{\tilde h}_+$ has the
standard behavior in the same domain as $h_-$, i.e. in $D_-$. So, to
compute the leading term of the asymptotics of $\tilde{\tilde h}_+$ in
$F$, we have just to continue it analytically inside $D_-$ to $F$.
This leads to the desired representation.


%

%
\end{document}